%% file: main.tex
\pdfoutput=1
\documentclass[11pt,a4paper]{amsart}
\usepackage[foot]{amsaddr}
\usepackage[english]{babel}
\usepackage[utf8]{inputenc}
\usepackage{amssymb,amsmath}
\usepackage{mathtools}
\usepackage{graphicx}
\usepackage{bm,bbm}
\usepackage{color}
\usepackage{tikz}
\usepackage{ifthen}
\usetikzlibrary{snakes}
\usetikzlibrary{patterns}
\usepackage{algorithm}
\usepackage{setspace}
\usepackage[noend]{algpseudocode}
\usepackage{pgfplots}
\usepackage{enumitem}
\usepackage{array,blkarray}
\usepackage{bigdelim}
\usepackage{listings} 
\usepackage[hidelinks]{hyperref}
\usepackage{amssymb}
\usepackage{isotope}
\usepackage{stmaryrd}
\usepackage[]{placeins}
\usepackage[normalem]{ulem}
\usepackage[numbers]{natbib}
\usepackage{tikz}
\usepackage{pgfplots}
\usepackage{pgfplotstable}
\usepgfplotslibrary{groupplots}
\usetikzlibrary{shapes,arrows.meta}
\usetikzlibrary{external,spy}
\tikzset{external/system call={lualatex -shell-escape -halt-on-error -interaction=batchmode -jobname "\image" "\texsource"}}
\usepackage{ifthen} 
\newboolean{professormode}
\setboolean{professormode}{true}
\usepackage{caption}
\usepackage{subcaption}
\DeclareMathAlphabet{\mathbbold}{U}{bbold}{m}{n}

\algnewcommand{\LineComment}[1]{\State \(\triangleright\) #1}
\newcolumntype{L}[1]{>{\raggedright\let\newline\\\arraybackslash\hspace{0pt}}m{#1}}
\newcolumntype{C}[1]{>{\centering\let\newline\\\arraybackslash\hspace{0pt}}m{#1}}
\newcolumntype{R}[1]{>{\raggedleft\let\newline\\\arraybackslash\hspace{0pt}}m{#1}}
\theoremstyle{definition}

\theoremstyle{remark}

\numberwithin{theorem}{section}
\numberwithin{equation}{section}
\numberwithin{table}{section}
\numberwithin{figure}{section}

\DeclareMathOperator{\rank}{rank}
\input{def}

\definecolor{colorbalzani}{rgb}{0.0,0.0,1.0} 
\definecolor{colorcomment}{rgb}{1.0,0.0,0.0}

\newsavebox{\measurebox}

\begin{document}
\title[\(\mathcal{H}\)-sequence Rank-$1$ convexification]{Hierarchical Rank-One sequence convexification for the relaxation of variational problems with microstructures}

\author[]{M.~Köhler$^{*}$, T.~Neumeier$^{\dagger}$, M.~A.~Peter$^{\dagger, \ddagger}$,  D.~Peterseim$^{\dagger, \ddagger}$, D.~Balzani$^{*}$}
\address{${}^{*}$ Chair~of~Continuum~Mechanics, Ruhr-Universit\"at~Bochum, Universit\"atsstr.~150, 44801~Bochum, Germany}
\email{\{daniel.balzani, maximilian.koehler\}@rub.de}
\address{${}^{\dagger}$ Institute~of~Mathematics, University~of~Augsburg, Universit\"atsstr.~12a, 86159~Augsburg, Germany}
\email{timo.neumeier@uni-a.de}

\address{${}^{\ddagger}$ Centre~for~Advanced~Analytics~and~Predictive~Sciences (CAAPS), University~of~Augsburg, Universit\"atsstr.~12a, 86159~Augsburg, Germany}
\email{\{malte.peter, daniel.peterseim\}@uni-a.de}
\thanks{\textit{Acknowledgments.} The authors acknowledge funding by the Deutsche Forschungsgemeinschaft (DFG) within the Priority Program 2256 (“Variational Methods for Predicting Complex Phenomena in Engineering Structures and Materials”), Project ID 441154176, reference IDs BA2823/17-2, PE1464/7-2, and PE2143/5-2. Furthermore, the authors acknowledge the free and open source community of the Julia programming language, especially of Ferrite.jl.}

\date{\today}

\begin{abstract}
This paper presents an efficient algorithm for the approximation of the rank-one convex hull in the context of nonlinear solid mechanics.
It is based on hierarchical rank-one sequences and simultaneously provides first and second derivative information essential for the calculation of mechanical stresses and the computational minimization of discretized energies.
For materials, whose microstructure can be well approximated in terms of laminates and where each laminate stage achieves energetic optimality with respect to the current stage, the approximate envelope coincides with the rank-one convex envelope.
Although the proposed method provides only an upper bound for the rank-one convex hull, a careful examination of the resulting constraints shows a decent applicability in mechanical problems.
Various aspects of the algorithm are discussed, including the restoration of rotational invariance, microstructure reconstruction, comparisons with other semi-convexification algorithms, and mesh independency.
Overall, this paper demonstrates the efficiency of the algorithm for both, well-established mathematical benchmark problems as well as nonconvex isotropic finite-strain continuum damage models in two and three dimensions.
Thereby, for the first time, a feasible concurrent numerical relaxation is established for an incremental, dissipative large-strain model with relevant applications in engineering problems.
\end{abstract}

\maketitle

%
%
%

\input{sections/sec1.tex}

\input{sections/sec2.tex}

\input{sections/sec3.tex}

\input{sections/sec4.tex}

\input{sections/sec5.tex}

\input{sections/sec6.tex}


\bibliographystyle{alpha}
\bibliography{references}

\appendix

\end{document}

%% file: def.tex
\definecolor{myBlue}{RGB}{113,104,238} 
\definecolor{myGreen}{RGB}{154,205,50} 
\definecolor{myGreen2}{RGB}{114,175,30} 
\definecolor{myRed}{RGB}{180,50,50}  
\definecolor{myOrange}{RGB}{225,92,22} 
\definecolor{lgray}{RGB}{200,200,200} 
\definecolor{llgray}{RGB}{155,155,155} 
\definecolor{mycolor1}{rgb}{0.00000,0.44700,0.74100}%
\definecolor{mycolor2}{rgb}{0.85000,0.32500,0.09800}%
\definecolor{mycolor3}{rgb}{0.92900,0.69400,0.12500}%
\definecolor{mycolor4}{rgb}{0.49400,0.18400,0.55600}%
\definecolor{mycolor5}{rgb}{0.46600,0.67400,0.18800}%
\definecolor{mycolor6}{rgb}{0.30100,0.74500,0.93300}%
\definecolor{mycolor7}{rgb}{0.63500,0.07800,0.18400}%
\newcommand\N{\mathbb N}

\newcommand\R{\mathbb R}
\newcommand\Z{\mathbb Z}

%% file: sections/sec1.tex
%

\section{Introduction}\label{sec:Introduction}
The study of solids undergoing significant deformation reveals various dissipative behaviours, including plasticity, phase transformation and damage effects.
Accurate modelling of these phenomena is crucial for applications in materials engineering and structural analysis.
Typically, there are two broad classes of methodologies; namely, multiscale and phenomenological approaches.
In the phenomenological approach, internal variables \cite{Lub:1972:tfn,Lub:1973:sre} play a crucial role and evolve dynamically over time to reflect the intricate processes on the microstructural level.
Hence, the internal variables serve as a representative value of the microstructure.
This allows the model to achieve a higher level of accuracy in predicting stress and strain curves that closely match real-world observations without sacrificing computational complexity.
The inclusion of internal variables also facilitates the simulation of complex behaviours such as
strain-rate dependence and history-dependent effects.
The incremental stress potential framework, introduced by a series of papers, see e.g.~\cite{Hac:1997:gsm, OrtRep:1999:nem, OrtSta:1999:vfv, CarHacMie:2002:ncp, MieTheLev:2002:vfr}, is fundamental to the variationally and thermodynamically consistent discretisation of models with internal variables.
This framework transforms an inelastic problem into a sequence of elastic problems, where the current values of internal variables are obtained from minimisation,  while maintaining thermodynamic consistency throughout.
One of the advantages in comparison with classical material modeling approaches is that the material model can be fully specified by defining a suitable strain energy function and a dissipation potential.
Hence, only these two scalar-valued functions need to be implemented in a finite element context when using automatized differentiation techniques as in e.g.~\cite{TanBalSch:2016:ioi}.
However, its disadvantage lies in its often occurring inherent ill-posed nature due to the underlying structure of the dissipative part.
From a variational perspective, the loss of semi-convexity in the incremental stress potential poses a significant challenge, threatening the existence of minimisers crucial for mesh-independent (or mesh-objective) simulations.
Mechanically, this loss of semi-convexity may lead to material instability, characterised by a loss of ellipticity.
To correct this, regularisation techniques are applied.
Gradient-extended models, see e.g.~\cite{PlaBarMisTim:2021:med,RieBal:2023:sel,JunRieBal:2022:erna,JunSchJanHac:2019:afa,KieWafSprMen:2018:gdm,DimHac:2008:mge,DimHac:2011:rfd, ZuoHeAvrYanHac:2022:tfua} in the context of damage, emerge as a popular choice, introducing a length scale parameter related to the dissipative process.
While effective, the determination of an appropriate length-scale can be difficult, occasionally leading to deviations from observed physical phenomena, particularly in the context of crack curvatures \cite{LanKurMos:2023:cdg}.
Micromorphic, see e.g.~\cite{LanKurMos:2022:hrc,BhaLarJan:2023:mpm,YinZhaStoKal:2022:mdm,For:2019:maga}, and phase-field models, see e.g.~\cite{Wu:2017:upt,
WilLamMos:2021:pfm,MieAldRai:2016:pfm,Mil:2012:1pf}, fit into the broader category of gradient-extended models, demonstrating the versatility of this approach in capturing a spectrum of dissipative behaviours.
Although some few approaches have been published for this e.g.~in~\cite{RezRieMisBalPla:2024:apf} in the context of gradient elasticity, determining the optimal length scale remains a hurdle, and the introduction of a new coupled partial differential equation further complicates the computational situation.

An alternative regularisation approach is to include viscosity terms in the formulation, see e.g.~\cite{Nee:1988:mrd,FarOliCer:1998:spv,LanJunMos:2018:qdm,LanKurMos:2022:hrc}.
While this approach can effectively dominate material evolution, especially for sufficiently large viscosities, the downside is the need to select parameters that may appear unphysical.
This delicate balance between achieving regularisation dominance and maintaining physically relevant material parameter regimes underlines the ongoing challenges in this approach.
In cases where a direct multiscale description of inelastic processes is pursued, see e.g.~\cite{MasKouPeeGee:2010:chl,LiaRenLi:2018:mmq,CoeKouBosGee:2012:mab}, the computational complexity increases drastically.
The need for localization band transition schemes becomes crucial, especially when the deformation localizes and the separation of scales becomes invalid \cite{MasKouPeeGee:2010:chl,GitAskSlu:2007:rve}.
This requires a careful interplay between accuracy and computational efficiency, navigating the intricate interdependencies between scales.

Finally, relaxation introduces a unique perspective on regularisation.
By replacing the original non-convex generalised energy density with its (semi)convex envelope, a continuum of microstructures known as laminates is described.
Laminates capture oscillating gradients of the field of interest and provide a rich representation of material behaviour.
The main advantage is that existence of minimizers as well as mesh independency can be ensured (if the rank-one convex envelope coincides with the quasiconvex envelope and suitable growth conditions are present), without the need for additional primary variables or the solution of additional partial differential equations (as in the case of gradient-extended or micro-morphic models or phase-field approaches) or cross-finite-element information (as in the case of integral-based regularisation).
However, it is important to note that while laminates provide a nuanced understanding of microstructural evolution, they lack a direct association with a specific length scale or period for the observed oscillations.
This aspect emphasises the broader homogenisation of a continuum of potential microstructures within the relaxation framework.
This can, on the other hand, also be interpreted as advantage as it avoids the choice of suitable values for the length scale.
While having several advantageous properties, relaxation is still not a popular choice for regularisation, due to the lack of efficient computational approaches to determine the (semi)convex envelope.
In specific cases, especially for non-convex elastic formulations, the energy density can be relaxed analytically, cf.~\cite{LeRao:1995:qes, DeSDol:2002:mrn}.
Analytic relaxation results have been carried out in e.g.~\cite{ConThe:2005:sem,ConHauOrt:2007:cmc,CarConOrl:2008:man} under suitable assumptions.
However, for a general setting, it is often difficult to formulate analytic results due to varying parts of the model.
In order to relax in the general setting, numerical convexification comes into play.
There, an adhoc relaxation is performed whenever needed.
Typically, in each integration point of the finite element discretisation, a computational algorithm is invoked such that the first and second derivative of the envelope can be returned to the assembly process.
While the procedure in terms of assembly is clear, computational methods for the concurrent (i.e.~embedded) numerical relaxation are sparse.
Only recently, methods for accurate relaxation for a single increment -- or alternatively an elastic potential -- were published in \cite{NeuPetPetWie:2023:cpi,BalKohNeuPetPet:2023:mrc,ConDol:2018:ara, ObeRua:2017:pde, KumVidKoc:2020:ant}.
Earlier works in numerical relaxation can be found in \cite{Bar:2005:rea,BarCarHacHop:2004:erm,Bar:2004:lca,AubFagOrt:2003:csa,OrtRepSta:2000:tsd,DolWal:2000:ena, Dol:1999:ncr}.
Alternatively, methods with incompatible microstructures were shown to be computationally effective, see \cite{SchJunHac:2020:vrd}.
Still, the methods are too computationally demanding to simulate realistic boundary value problems in three spatial dimensions while ensuring compatibility of the described microstructures (laminates).
Another reoccurring modelling in the domain of relaxation was the use of simplified models, such that e.g.~only a one-dimensional convexification was performed, see e.g.~\cite{KohNeuMelPetPetBal:2022:acma,KohBal:2023:emr,GurMie:2011:edm,BalOrt:2012:riv,SchBal:2016:riv,LamMieDet:2003:ern}.

In this paper, we present a novel algorithm based on certain mechanically motivated assumptions in the modelling of the incremental stress potential, which is able to relax mathematical benchmark problems in multiple dimensions and to compute relaxed continuum damage mechanics boundary value problems without loss of generality.
This work directly extends ideas from \cite{AubFagOrt:2003:csa} with computational geometry algorithms, namely the Graham's Scan \cite{Gra:1972:ead}.
Instead of discretising the full $d^2$-dimensional space of deformation-gradients and computing the successive lamination in each grid point, like done in e.g.~\cite{Bar:2004:lca, DolWal:2000:ena}, we utilize the hierarchical rank-one connected sequences (so called $\mathcal{H}$-sequences) characterization of the rank-one convex envelope, see e.g.~\cite[Section 4.1.1.3]{Dac:2008:dmca}.
With this characterization, the point evaluation of the rank-one convex envelope can be performed by evaluating the convex combination of the optimal $\mathcal{H}$-sequence.
We aim to find a candidate $\mathcal{H}$-sequence for this minimisation by successively choosing locally optimal laminates.
Computationally, each sequential lamination is performed by one dimensional convexifications along a set of discretised rank-one directions.
Starting from a given evaluation point, the subsequent lamination of already found laminates leads to a candidate for the global rank-one convexification problem, which, for a special class of functions, seems to be optimal and delivers an approximation to the rank-one convex envelope.
Since the maximum lamination depth is assumed to be given, the computational effort is dominated by the one-dimensional convexifications, i.e.~the discretisation of the one-dimensional convexifications.
This yields a magnificent computational advantage when compared to mesh-based successive lamination like e.g.~\cite{Bar:2004:lca, DolWal:2000:ena}.
In the general case, the algorithm fails to approximate the rank-one convex envelope, but in presence of structural properties, the computed $\mathcal{H}$-sequence is optimal and results are found to coincide with the actual rank-one convex envelope.

This contribution is structured as follows: Section~\ref{sec:VP} establishes the mathematical framework by introducing the general variational problem and the necessary notions for the algorithms.
Section~\ref{sec:HROC} describes the novel algorithm and the assumption made in order to have a convergent scheme, followed in Section~\ref{sec:benchmark} with convergence examples for mathe\-matical benchmark problems as well as a counterexample where the required underlying structure of the general energy density is not given.
In Section~\ref{sec:examples}, the algorithm is applied to a
practically relevant problem, i.e.~a continuum damage mechanics model at finite strains, which has an application in soft biological tissues such as arteries under supra-physiological loadings, cf.~e.g.~\cite{AntBalDesDeeMacKam:2019:mdc}.
Finally, a conclusion and outlook are given in Section~\ref{sec:conclusion}.

%% file: sections/sec2.tex

\section{Non-convex Variational Problems in Nonlinear Solid Mechanics}\label{sec:VP}
In solid mechanics, we describe deformable solids in terms of their reference configuration $\mathcal{B}$ and current configuration $\mathcal{B}_t$.
Points within the solid are represented by reference coordinates $\boldsymbol{X}\in\mathcal{B}$ and current coordinates $\boldsymbol{x}\in\mathcal{B}_t$.
The displacement field $\boldsymbol{u}(\boldsymbol{X},t)$ characterizes the movement of material points from the reference to the current configuration.
The deformation gradient $\boldsymbol{F}$ describes local deformation, relating infinitesimal line elements in the reference to the current configuration.
It is defined as $\boldsymbol{F}(\boldsymbol{u}) = \boldsymbol{I} + \text{Grad} \, \boldsymbol{u}$.
In variational problems in solid mechanics, we minimise energy functionals with respect to displacement fields to derive equations governing deformation behaviour, taking into account appropriate boundary conditions.

\subsection{Elastic Problems}

Elasticity describes the ability of a material to return to its original shape after deformation when external forces are removed.
In hyperelasticity, this behaviour can be described by the principle of minimising the potential energy, see e.g.~\cite[Section 8.2]{Hol:2000:nsm}, where the potential energy $\Pi$ of the solid is defined as the integral of the energy density $W$ over the body occupied by the solid and the external forces and body forces, if present, are gathered in a term $\Pi^{\text{ext}}$, i.e.
\begin{equation}
    \label{eq:principalenergy}
    \Pi(\boldsymbol{u}) = \int_{\mathcal{B}}W(\boldsymbol{F}(\boldsymbol{u})) \, \text{d}V + \Pi^{\text{ext}}(\boldsymbol{t},\boldsymbol{u}).
\end{equation}
Setting the first variation to zero yields
\begin{equation}
    \label{eq:firstvariation}
    \delta_{\boldsymbol{u}}
    \Pi(\boldsymbol{u})= \int_{\mathcal{B}} \boldsymbol{P}\left(\boldsymbol{F}\left(\boldsymbol{u}\right)\right) :
\text{Grad}\, \delta \boldsymbol{u} \, \text{d}V
    + \delta_{\boldsymbol{u}}\Pi^{\text{ext}} = 0,
\end{equation}
with the variation of the displacements~$\delta\boldsymbol{u}$.
Here, $\,:\,$ denotes the
inner product and $\boldsymbol{P}=\partial_F W$ the first derivative of $W$ with respect to the deformation gradient $\boldsymbol{F}$, i.e.~the first Piola--Kirchhoff stress tensor
\begin{equation}
	\boldsymbol{P}\left(\boldsymbol{F}\left(\boldsymbol{u}\right)\right) = \frac{\partial W\left(\boldsymbol{F}\left(\boldsymbol{u}\right)\right)}{\partial \boldsymbol{F}}.
\end{equation}
Often, Newton schemes, see e.g.~\cite[Algorithm 4.2]{Bar:2015:nmn}, are employed to solve the nonlinear equation~\eqref{eq:firstvariation}. Within these methods, the second derivative of the strain energy density, i.e.~the fourth-order tangent moduli
\begin{equation}
    \mathbb{A}\left(\boldsymbol{F}\left(\boldsymbol{u}\right)\right) = \frac{\partial^2 W\left(\boldsymbol{F}\left(\boldsymbol{u}\right)\right)}{\partial \boldsymbol{F} \partial \boldsymbol{F}},
\end{equation}
is required.
With this at hand, the problem can be discretized by finite elements in order to approximate solutions to \eqref{eq:firstvariation}.
However, these solutions
are only minimisers of \eqref{eq:principalenergy} if $W$ satisfies certain properties such as (semi)convexity.
Some well-known and widely used geometrically nonlinear but in specific stress-strain measures physically linear elastic material models exhibit a severely non-convex behaviour such as Biot or St.~Venant--Kirchhoff formulations \cite{BerBohSil:2007:rcs}.
In this case, numerical results often become mesh-dependent, meaning that, e.g.~deformation fields and material responses may be sensitive to changes in meshing and small variations in material parameters.

\subsection{Inelastic Problems}
While elasticity provides a foundational framework for understanding material behaviour, many real-world materials exhibit inelastic behaviour under certain conditions.
Inelasticity refers to irreversible effects of materials, which can arise due to various physical mechanisms such as plasticity, phase transformation, or damage.
Unlike elastic deformation, inelastic deformation results in permanent changes to the material's
microstructure.

Describing such material behaviour requires more advanced modelling approaches.
The starting point of the incremental formulation is the fact that a solid should minimise its work deformation energy density.
This work deformation energy density can be expressed as
\begin{equation} 
	\label{eq:workdeformation}
    \mathcal{W} = \int_{0}^T \boldsymbol{P} : \dot{\boldsymbol{F}} \, \text{d}t,
\end{equation}
where $\boldsymbol{P}$ denotes the first Piola--Kirchhoff tensor and $(\,\dot{}\,)$ the time derivative.
A strain
energy density $\psi(\boldsymbol{F},\boldsymbol{\alpha})$, dependent on $\boldsymbol{F}$ and a vector collecting all internal variables $\boldsymbol{\alpha}$, is postulated to exist, satisfying all required continuum mechanical properties such as objectivity and invariances.
Taking the time derivative of $\psi$ yields
\begin{equation}\label{eq:dotpsi}
    \dot{\psi}(\boldsymbol{F},\boldsymbol{\alpha}) = \underbrace{\frac{\partial \psi}{\partial \boldsymbol{F}}}_{\boldsymbol{P}} : \dot{\boldsymbol{F}} + \underbrace{\frac{\partial \psi}{\partial \boldsymbol{\alpha}}}_{\boldsymbol{-\beta}} \cdot \;\dot{\boldsymbol{\alpha}}.
\end{equation}
Defining the dissipation potential as $\phi \coloneqq \boldsymbol{\beta} \cdot \dot{\boldsymbol{\alpha}}$ and rearranging \eqref{eq:dotpsi} allows to rewrite \eqref{eq:workdeformation} as
\begin{equation}
    \label{eq:ISPstart}
    \mathcal{W}(\boldsymbol{F},\boldsymbol{\alpha}) = \int_{0}^T (\dot{\psi} + \phi)  \, \text{d}t.
\end{equation}
Now, discretizing the time interval $[0,T]$ into finite increments and integrating over a single increment, in combination with the minimisation w.r.t.~the collection of internal variables $\boldsymbol{\alpha}$ in \eqref{eq:ISPstart}, leads to
\begin{equation}
    W(\boldsymbol{F}_{k+1}) = \inf_{\alpha_{k+1}} \int_{t_k}^{t_{k+1}} (\dot{\psi} + \phi) \, \text{d}t.
\end{equation}
This minimisation problem can be solved
per increment
via the principle of minimum of potential energy, cf.~\eqref{eq:principalenergy}, and thus, renders per increment a pseudo-elastic solution process, where the internal variable is updated by
\begin{equation}
    \boldsymbol{\alpha}_{k+1} = \arg\inf \mathcal{W}(\boldsymbol{F}_{k+1},\boldsymbol{\alpha}_{k+1}).
\end{equation}
Careful mathematical considerations show that $W$ becomes non-convex at different condensed states for different deformation gradients in general.
Since the original model does not encode a length scale, oscillations in the deformation gradient occur, which become finer and finer w.r.t.~the discretization size, cf.~\cite[Section 2.1.8]{Bar:2015:nmn}, \cite[Example 4.4]{Bar:2015:nmn}.
Relaxation realises a scale separation of the oscillations from the macroscopic response.

\subsection{Relaxation by rank-one convexification} \label{sec:RankOne}

We aim to regularise the non-convex models by means of relaxation.
Here, we refer to the notion of relaxation by replacing the original non-convex generalized energy density by its (semi)convex envelope.
Possible choices of (semi)convex hulls are the convex, polyconvex, quasiconvex, and rank-one convex envelope, which are related by the following inequalities 
\begin{equation}
	W^{\text{c}} \leq W^{\text{pc}} \leq W^{\text{qc}} \leq W^{\text{rc}} \leq W.
\end{equation}
In this work, we focus on the rank-one convex envelope.
This hull typically offers a close approximation to the quasiconvex envelope while ensuring the preservation of physically relevant properties inherent to the energy density.
Later on, it will become evident that this choice enables the definition of an algorithm with sufficiently low complexity, rendering it suitable for concurrent relaxation within finite element simulations for nonlinear boundary value problems.

In what follows, we recall the basic definitions and results for rank-one convex functions.
A function $W\colon\R^{d \times d} \to \R$ is called \emph{rank-one convex}, if 
\begin{equation}
	W(\lambda \, \boldsymbol{A} + (1 - \lambda) \, \boldsymbol{B}) \leq \lambda \, W(\boldsymbol{A}) + (1 - \lambda) \, W(\boldsymbol{B})
\end{equation}
holds for all $\lambda \in [0, 1]$ and all matrices $\boldsymbol{A}, \boldsymbol{B} \in \R^{d \times d}$ that are rank-one connected, i.e.~$\rank(\boldsymbol{A} - \boldsymbol{B}) = 1$.
This condition is equivalent to the property that $W$ is convex along rank-one directions, i.e.~the function $g\colon\eta \mapsto W(\boldsymbol{F} + \eta \, \boldsymbol{a} \otimes \boldsymbol{b})$ is convex for all $\boldsymbol{a}, \boldsymbol{b} \in \R^{d}$ and $\boldsymbol{F} \in \R^{d \times d}$. 
The \emph{rank-one convex envelope} $W^{\text{rc}}$ of a function $W$ is the largest rank-one convex function below $W$, i.e.~in each $\boldsymbol{F}$ it is given by
\begin{equation}
	W^{\text{rc}}(\boldsymbol{F}) = \sup\{\tilde{W}(\boldsymbol{F}) \mid \tilde{W}\colon\R^{d \times d} \to \R \text{ is rank-one convex} \text{ with } \tilde{W} \leq W\}.
\end{equation}

Various characterizations for the rank-one convex envelope can be found for example in \cite[Chapter 5]{Dac:2008:dmca}.
One characterization is the construction by successive lamination.
This is done by setting $W^{0} = W$ and, for $k > 0$, each next laminate is obtained by
\begin{align}
	W^{k + 1} (\boldsymbol{F}) & = \inf_{\substack{\lambda \in [0, 1] \\ \boldsymbol{A}, \boldsymbol{B} \in \mathbb{R}^{d \times d}}}\left\{ \lambda \, W^{k}(\boldsymbol{A}) + (1 - \lambda) \, W^{k}(\boldsymbol{B}) \,  \bigg\vert 
	\begin{array}{c} 
	\lambda \, \boldsymbol{A} + (1 - \lambda) \, \boldsymbol{B} = \boldsymbol{F}, \\ \mathrm{rank}(\boldsymbol{A} - \boldsymbol{B}) = 1
	\end{array} \right\}
\end{align}
at all points $\boldsymbol{F} \in \R^{d \times d}$. The lamination-convex envelope $W^{\text{lc}}$ is defined in a pointwise manner through the limit 
\begin{equation}
	\label{eq:laminationlimit}
	W^{\text{lc}} (\boldsymbol{F}) \coloneqq \lim\limits_{k \to \infty} W^{k} (\boldsymbol{F}).
\end{equation}
In \cite[5.C]{KohStr:1986:odr}, it is shown that the rank-one convex envelope can be obtained by successive lamination, that is $W^{\text{rc}} = W^{\text{lc}}$.
While this approach can be used to formulate a computational algorithm, as done in \cite{Dol:1999:ncr,Bar:2004:lca,ConDol:2018:ara,BalKohNeuPetPet:2023:mrc}, all of these algorithms are too computationally demanding for simultaneous embedding in realistic finite element computations, which is mainly due to the fact that the high dimensional space of $d\times d$ matrices needs to be discretized in order to find the corresponding laminates in the deformation gradient space.
This discretization aspect implies a computational complexity scaling by $d^2$, without even considering the discretization of rank-one directions, one-dimensional convexifications or the required laminations.

Another equivalent characterization for the rank-one convex envelope can be found e.g.~in \cite[Proposition~5.16]{Dac:2008:dmca} and is based on hierarchical rank-one connected matrices.
These so-called $\mathcal{H}_M$-Sequences encode a condition on a set of $M \in \N$ matrices $\boldsymbol{F}_1, \dots, \boldsymbol{F}_M \in \R^{d \times d}$.
We use the notation of \cite{Dol:1999:ncr}.
For \(M = 1\) the pair $(\xi_1, \boldsymbol{F}_1)$ always satisfies the condition $\mathcal{H}_1$, i.e.~$(\xi_1, \boldsymbol{F}_1) \in \mathcal{H}_1$.
The set of pairs $(\xi_i, \boldsymbol{F}_i) \in [0, 1] \times \R^{d \times d}$ for $i =1, \ldots, M$ fulfils the condition $\mathcal{H}_M$ (notation: $(\xi_i, \boldsymbol{F}_i) \in \mathcal{H}_M$) if $\sum_{i = 1}^{M} \xi_i = 1$ and if, up to a permutation of the indices $\{1, \dots, M\}$, it holds {${\rank(\boldsymbol{F}_1 - \boldsymbol{F}_2) = 1}$} and defining
\begin{equation}
	\begin{aligned}
		\zeta_1 & = \xi_1 + \xi_2, &\qquad \hat{\boldsymbol{F}}_{1} & = \frac{1}{\zeta_1} (\xi_1 \boldsymbol{F}_1 + \xi_2 \boldsymbol{F}_2), \\
		\zeta_i & = \xi_{i+1}, &\qquad \hat{\boldsymbol{F}}_i & = \boldsymbol{F}_{i+1}
	\end{aligned}
\end{equation}
for all $i \in \{2, \dots, M-1\}$, it holds $(\zeta_i, \hat{\boldsymbol{F}}_i) \in \mathcal{H}_{M - 1}$. 

\begin{figure}
	\centering
	\begin{subfigure}[b]{0.3\textwidth}
		\centering
		\ifthenelse{\boolean{professormode}}
		{\includegraphics[width=\textwidth]{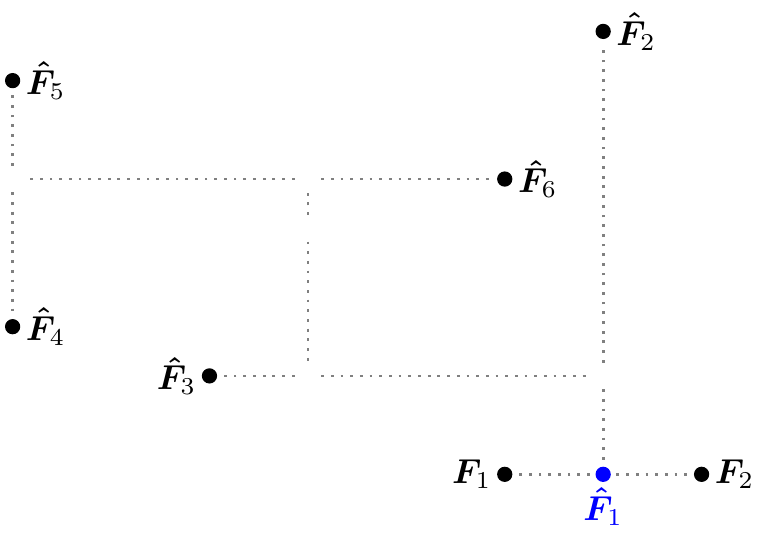}}
		{\resizebox{\textwidth}{!}{%
			\input{figures/tikz/h-set-construction.tex}
			}%
		}
		\caption{}
		\label{fig:Hsetconstruction}
	\end{subfigure}
	\begin{subfigure}[b]{0.3\textwidth}
		\centering
		\ifthenelse{\boolean{professormode}}
		{\includegraphics[width=\textwidth]{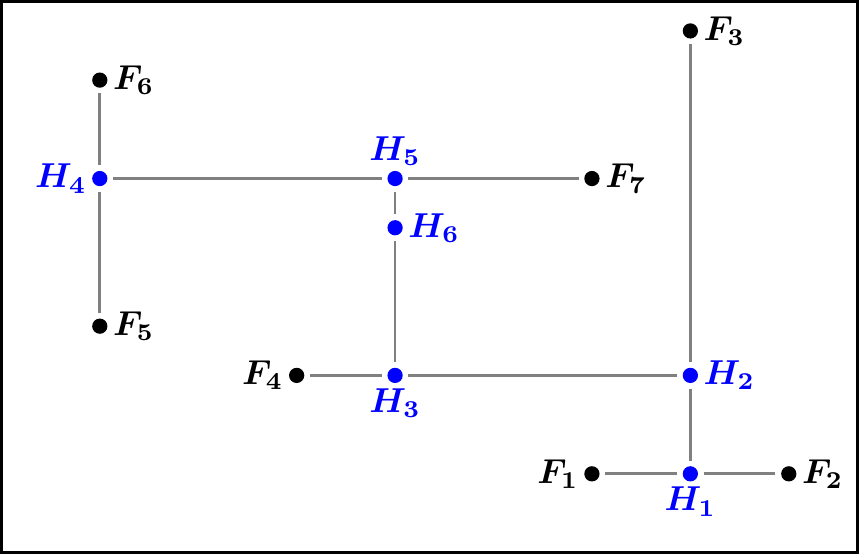}}
				{\resizebox{\textwidth}{!}{%
				\input{figures/tikz/h-set.tex}
			}%
		}
		\caption{}
		\label{fig:Hset}
	\end{subfigure}
	\begin{subfigure}[b]{0.3\textwidth}
		\centering
		\ifthenelse{\boolean{professormode}}
		{\includegraphics[width=\textwidth]{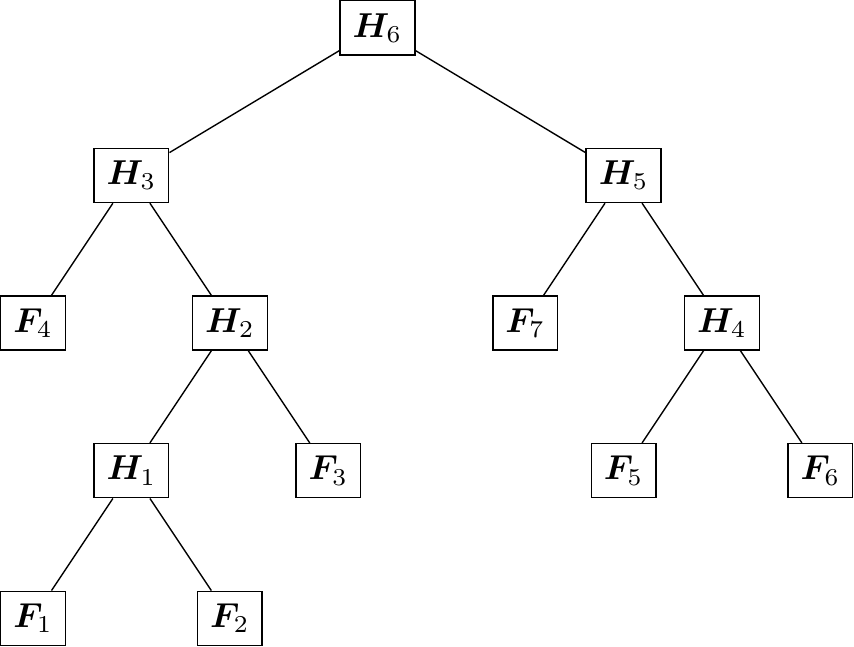}}
		{\resizebox{\textwidth}{!}{%
				\input{figures/tikz/tree.tex}
			}%
		}
		\caption{}
		\label{fig:tree}
	\end{subfigure}
	\caption{(\subref{fig:Hsetconstruction}) construction of $\mathcal{H}$ set, (\subref{fig:Hset}) $\mathcal{H}$ set, and (\subref{fig:tree}) Rank-one tree. The lines in (\subref{fig:Hsetconstruction}) and (\subref{fig:Hset}) encode rank-one directions. (\subref{fig:Hsetconstruction}) shows the first recursive check for the $\mathcal{H}_7$ condition.}
	\label{fig:Hsetvis}
\end{figure}

An illustration of this hierarchical rank-one connectivity condition is given in Figure~\ref{fig:Hsetvis} by an exemplary set of seven matrices. 
For the sake of simplicity, the representation is done in two dimensions, e.g.~the $F_{11}$--$F_{22}$ plane.
In Figure~\ref{fig:Hsetconstruction} the first check is illustrated exemplary for $M = 7$ by connecting $\boldsymbol{F}_1$ and $\boldsymbol{F}_2$ by a rank-one line resulting in $\hat{\boldsymbol{F}}_1$.
This is followed by a cascade of similar checks on hierarchical rank-one connectivity for the set $\hat{\boldsymbol{F}}_{i}$ for $i = 1, \dots, 6$.
The final set of rank-one connected matrices and their intermediate connectivity points as well as their centre of mass (the resulting convex combination) $\boldsymbol{H}_6$ is depicted in Figure~\ref{fig:Hset}.
The hierarchically connected matrices $\boldsymbol{F}_1, \dots, \boldsymbol{F}_M$ are exactly the leaves of the corresponding rank-one tree in Figure~\ref{fig:tree}, which in general might not be unique.

With this definition at hand, the rank-one convex envelope $W^{\text{rc}}$ can now be characterized by
\begin{equation} \label{eq:WrcHseq}
	W^{\text{rc}}(\boldsymbol{F}) = \inf_{} \left\{\sum_{i = 1}^{M} \xi_i \, W(\boldsymbol{F}_i) \, \Big\vert \, (\xi_i, \boldsymbol{F}_i) \in \mathcal{H}_M , \boldsymbol{F} = \sum_{i = 1}^{M} \xi_i \boldsymbol{F}_i, M \in \N\right\}.
\end{equation}
In general, the number of matrices required for the rank-one convex hull, $M \in \N$, cannot be bounded.
The main difference between successive lamination and hierarchical sequences can be seen in how the minimisation problems are structured to obtain the rank-one convex hull.
For the example in Figure~\ref{fig:Hsetvis}, basically, for successive lamination, a set of 13 recursive minimisation problems, each with two degrees of freedom, has to be solved, whereas for hierarchical sequences, one minimisation problem with 13 degrees of freedom has to be solved.
The resulting minimisation problems are marked as black boxes in Figure~\ref{fig:Hset} and Figure~\ref{fig:tree}.
Therefore, the characterisation in terms of hierarchical sequences allows,
theoretically, to find the rank-one convex envelope without the iterative process of laminations, due to the absence of any recursive definition in~\eqref{eq:WrcHseq}.
Instead, one has to minimise over possible hierarchical rank-one sequences, which, however, contain some hidden recursivity in their construction.
Despite the intricate construction of hierarchical rank-one sequences, the $\mathcal{H}$-sequence representation \eqref{eq:WrcHseq} provides a natural path towards a local representation of the rank-one convex envelope.

Assume a point $\boldsymbol{F}$ in deformation-gradient space is given. 
A single point evaluation of the rank-one convex envelope via the iterative lamination characterization \eqref{eq:laminationlimit} requires the knowledge of all deepest laminates of all points in the deformation gradient space, i.e.~successive laminates for all points have to be computed.
This is not the case for the $\mathcal{H}$-sequence characterization.
Instead, this characterization allows to sweep through the $d \times d$ space and search for suitable $\mathcal{H}$-sequence candidates for the minimisation problem in \eqref{eq:WrcHseq}.
This can be done by starting at the given point $\boldsymbol{F}$ and iteratively constructing hierarchical rank-one connected set candidates by following rank-one lines. 
Of course, this task is too complex in general. 
However, it is this aspect of constructing $\mathcal{H}$-sequence candidates for the minimisation problem of the rank-one convex envelope that motivates the following algorithmic approach.

%% file: figures/tikz/h-set-construction.tex
\tikzsetfigurename{h-set-construction}
\begin{tikzpicture}[]
	\pgfmathsetmacro\firstsize{3.5}
	\pgfmathsetmacro\secondsize{1.75}
	\pgfmathsetmacro\thirdsize{1.0}
	\pgfmathsetmacro\verticalsize{1.5}
	\node (H1) at (3,-2) {};
	\node (H2) at (3,-1) {};
	\node (H3) at (0,-1) {};
	\node (H4) at (-3,1) {};
	\node (H5) at (0,1) {};
	\node (H6) at (0,0.5) {};
	\node (F1) at (2,-2) {};
	\node (F2) at (4,-2) {};
	\node (F3) at (3,2.5) {};
	\node (F4) at (-1,-1) {};
	\node (F5) at (-3,-0.5) {};
	\node (F6) at (-3,2) {};
	\node (F7) at (2,1) {};
	\filldraw[blue] (H1) circle (2pt) node[anchor=north]{\small $\boldsymbol{\hat{F}}_1$};
	%
	\filldraw[black] (F1) circle (2pt) node[anchor=east]{\small $\boldsymbol{F}_1$};
	\filldraw[black] (F2) circle (2pt) node[anchor=west]{\small $\boldsymbol{F}_2$};
	\filldraw[black] (F3) circle (2pt) node[anchor=west]{\small $\boldsymbol{\hat{F}}_2$};
	\filldraw[black] (F4) circle (2pt) node[anchor=east]{\small $\boldsymbol{\hat{F}}_3$};
	\filldraw[black] (F5) circle (2pt) node[anchor=west]{\small $\boldsymbol{\hat{F}}_4$};
	\filldraw[black] (F6) circle (2pt) node[anchor=west]{\small $\boldsymbol{\hat{F}}_5$};
	\filldraw[black] (F7) circle (2pt) node[anchor=west]{\small $\boldsymbol{\hat{F}}_6$};
	\draw[thick, gray, dotted] (H6) -- (H5) {};	
	\draw[thick, gray, dotted] (H6) -- (H3) {};
	\draw[thick, gray, dotted] (H5) -- (H4) {};	
	\draw[thick, gray, dotted] (H5) -- (F7) {};
	\draw[thick, gray, dotted] (H3) -- (H2) {};
	\draw[thick, gray, dotted] (H3) -- (F4) {};
	\draw[thick, gray, dotted] (H4) -- (F5) {};
	\draw[thick, gray, dotted] (H4) -- (F6) {};
	\draw[thick, gray, dotted] (H2) -- (H1) {};
	\draw[thick, gray, dotted] (H2) -- (F3) {};
	\draw[thick, gray, dotted] (H1) -- (F1) {};
	\draw[thick, gray, dotted] (H1) -- (F2) {};
\end{tikzpicture}

%% file: figures/tikz/h-set.tex
\tikzsetfigurename{h-set}
\begin{tikzpicture}
    \draw[black, thick] (-4.0,-2.8) rectangle (4.7,2.8);
	\pgfmathsetmacro\firstsize{3.5}
	\pgfmathsetmacro\secondsize{1.75}
	\pgfmathsetmacro\thirdsize{1.0}
	\pgfmathsetmacro\verticalsize{1.5}
	\node (H1) at (3,-2) {};
	\node (H2) at (3,-1) {};
	\node (H3) at (0,-1) {};
	\node (H4) at (-3,1) {};
	\node (H5) at (0,1) {};
	\node (H6) at (0,0.5) {};
	\node (F1) at (2,-2) {};
	\node (F2) at (4,-2) {};
	\node (F3) at (3,2.5) {};
	\node (F4) at (-1,-1) {};
	\node (F5) at (-3,-0.5) {};
	\node (F6) at (-3,2) {};
	\node (F7) at (2,1) {};
	\filldraw[blue] (H1) circle (2pt) node[anchor=north]{\small $\boldsymbol{H_1}$};
	\filldraw[blue] (H2) circle (2pt) node[anchor=west]{\small $\boldsymbol{H_2}$};
	\filldraw[blue] (H3) circle (2pt) node[anchor=north]{\small $\boldsymbol{H_3}$};
	\filldraw[blue] (H4) circle (2pt) node[anchor=east]{\small $\boldsymbol{H_4}$};
	\filldraw[blue] (H5) circle (2pt) node[anchor=south]{\small $\boldsymbol{H_5}$};
	\filldraw[blue] (H6) circle (2pt) node[anchor=west]{\small $\boldsymbol{H_6}$};
	\filldraw[black] (F1) circle (2pt) node[anchor=east]{\small $\boldsymbol{F_1}$};
	\filldraw[black] (F2) circle (2pt) node[anchor=west]{\small $\boldsymbol{F_2}$};
	\filldraw[black] (F3) circle (2pt) node[anchor=west]{\small $\boldsymbol{F_3}$};
	\filldraw[black] (F4) circle (2pt) node[anchor=east]{\small $\boldsymbol{F_4}$};
	\filldraw[black] (F5) circle (2pt) node[anchor=west]{\small $\boldsymbol{F_5}$};
	\filldraw[black] (F6) circle (2pt) node[anchor=west]{\small $\boldsymbol{F_6}$};
	\filldraw[black] (F7) circle (2pt) node[anchor=west]{\small $\boldsymbol{F_7}$};
	\draw[thick, gray] (H6) -- (H5) {};	
	\draw[thick, gray] (H6) -- (H3) {};
	\draw[thick, gray] (H5) -- (H4) {};	
	\draw[thick, gray] (H5) -- (F7) {};
	\draw[thick, gray] (H3) -- (H2) {};
	\draw[thick, gray] (H3) -- (F4) {};
	\draw[thick, gray] (H4) -- (F5) {};
	\draw[thick, gray] (H4) -- (F6) {};
	\draw[thick, gray] (H2) -- (H1) {};
	\draw[thick, gray] (H2) -- (F3) {};
	\draw[thick, gray] (H1) -- (F1) {};
	\draw[thick, gray] (H1) -- (F2) {};
\end{tikzpicture}

%% file: figures/tikz/tree.tex
\tikzsetfigurename{tree}
\begin{tikzpicture}
	\pgfmathsetmacro\verticalsize{1.5} %
	\pgfmathsetmacro\firstsize{2.5} %
	\pgfmathsetmacro\secondsize{1.} %
	\pgfmathsetmacro\thirdsize{1.0} %
	\pgfmathsetmacro\fourthsize{1.0} %
	\node[draw] (F) at (0,0) {\small $\boldsymbol{H}_6$};
	\node[draw] (F1) at (-\firstsize,-\verticalsize) {\small $\boldsymbol{H}_3$};
	\node[draw] (F2) at (\firstsize,-\verticalsize) {\small $\boldsymbol{H}_5$};
	\draw (F) -- (F1) node[midway,left,yshift=1em] {}; 
	\draw (F) -- (F2) node[midway,right,yshift=1em] {}; 
	\node[draw] (F11) at (-\firstsize-\secondsize,-2*\verticalsize) {\small $\boldsymbol{F}_{4}$};
	\node[draw] (F12) at (-\firstsize+\secondsize,-2*\verticalsize) {\small $\boldsymbol{H}_{2}$};
	\draw (F1) -- (F11) node[midway,left] {}; 
	\draw (F1) -- (F12) node[midway,right] {}; 
	\node[draw] (F21) at (\firstsize-\secondsize,-2*\verticalsize) {\small $\boldsymbol{F}_{7}$};
	\node[draw] (F22) at (\firstsize+\secondsize,-2*\verticalsize) {\small $\boldsymbol{H}_{4}$};
	\draw (F2) -- (F21) node[midway,left] {}; 
	\draw (F2) -- (F22) node[midway,right] {}; 
	%
	\node[draw] (F121) at (-\firstsize+\secondsize-\thirdsize,-3*\verticalsize) {\small
		$\boldsymbol{H}_{1}$};
	\node[draw] (F122) at (-\firstsize+\secondsize+\thirdsize,-3*\verticalsize) {\small
		$\boldsymbol{F}_{3}$};
	\draw (F12) -- (F121) node[midway,left] {}; 
	\draw (F12) -- (F122) node[midway,right] {}; 
	%
	%
	\node[draw] (F221) at (+\firstsize+\secondsize-\thirdsize,-3*\verticalsize) {\small
		$\boldsymbol{F}_{5}$};
	\node[draw] (F222) at (+\firstsize+\secondsize+\thirdsize,-3*\verticalsize) {\small
		$\boldsymbol{F}_{6}$};
	\draw (F22) -- (F221) node[midway,left] {}; 
	\draw (F22) -- (F222) node[midway,right] {}; 

	\node[draw] (F1211) at (-\firstsize+\secondsize-\thirdsize-\fourthsize,-4*\verticalsize) {\small
		$\boldsymbol{F}_{1}$};
	\node[draw] (F1212) at (-\firstsize+\secondsize-\thirdsize+\fourthsize,-4*\verticalsize) {\small
		$\boldsymbol{F}_{2}$};
	\draw[draw] (F121) -- (F1211) node[midway,left] {}; 
	\draw[draw] (F121) -- (F1212) node[midway,right] {}; 
	
\end{tikzpicture}

%% file: sections/sec3.tex
%

\section{Hierarchical rank-one sequence convexification algorithm} \label{sec:HROC}
We introduce the basic steps of the convexification algorithm motivated by the hierarchical rank-one sequence representation of the rank-one convex envelope.
In addition, we comment on the generation of derivative information in order to make the relaxation algorithm applicable to the simulation of boundary value problems.
Based on the rank-one tree information, we further address the visualisation of the microstructure associated with the relaxation process.

\subsection{Algorithm} \label{sec:algorithm}
\begin{figure}
	\centering
	\begin{subfigure}[b]{0.7\textwidth}
		\centering
		\ifthenelse{\boolean{professormode}}
		{\includegraphics[width=0.49\textwidth]{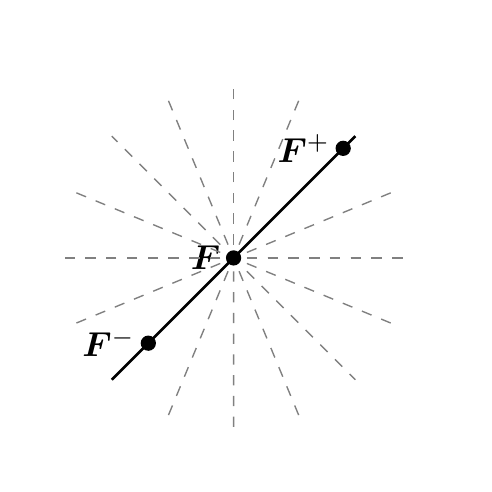} \includegraphics[width=0.49\textwidth]{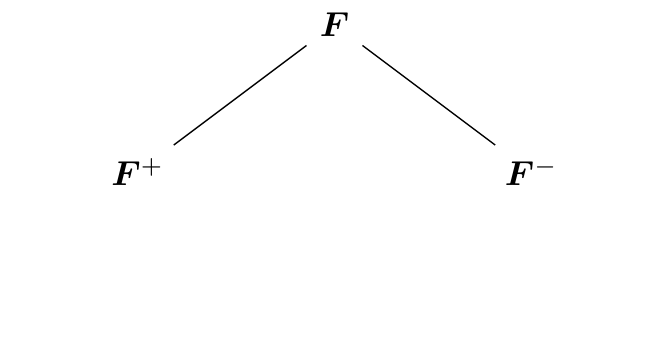}}
		{\resizebox{\textwidth}{!}{%
				\input{figures/tikz/h_search_begin.tex} \input{figures/tikz/h_tree_begin.tex}
			}%
		}
		\caption{}
		\label{fig:h_search_begin}
	\end{subfigure}
	\begin{subfigure}[b]{0.7\textwidth}
		\centering
		\ifthenelse{\boolean{professormode}}
		{\includegraphics[]{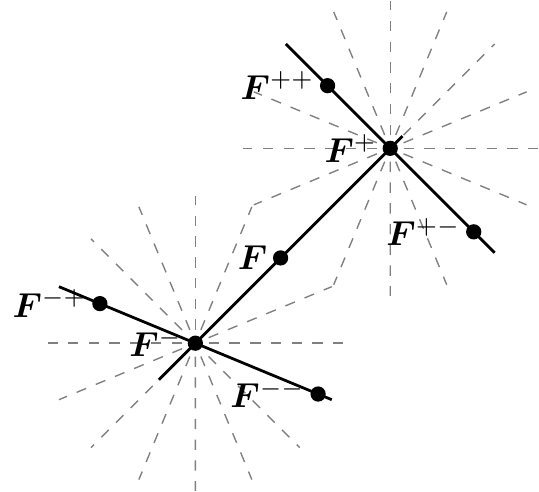}\includegraphics[]{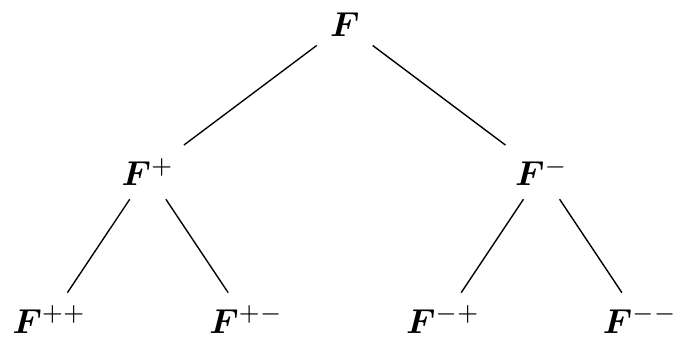}}
		{\resizebox{\textwidth}{!}{%
		    \input{figures/tikz/h_search_end.tex} \input{figures/tikz/h_tree_end.tex}
			}%
		}
		\caption{}
		\label{fig:h_search_end}
	\end{subfigure}
	\caption{Construction of the $\mathcal{H}$-sequence by convexifying first in the point $\boldsymbol{F}$ (\subref{fig:h_search_begin}) and afterwards in $\boldsymbol{F}^{+}$ and $\boldsymbol{F}^{-}$ (\subref{fig:h_search_end}).
        	 In (\subref{fig:h_search_begin}), the given macroscopic deformation gradient $\boldsymbol{F}$ is convexified along all the discretised rank-one lines contained in $\mathcal{R}$ (dashed lines). 
        	 $\boldsymbol{F}^{+}$ and $\boldsymbol{F}^{-}$ correspond to the supporting points of the line delivering the lowest convexified value of the energy at $\boldsymbol{F}$ (drawn as solid line).
	         The corresponding first-order laminate tree is visualised on the right-hand side in Figure~\ref{fig:h_search_begin}.
	         Next, in (\subref{fig:h_search_end}), all rank-one lines are checked again in both the points $\boldsymbol{F}^{+}$ and $\boldsymbol{F}^{-}$ if a lower convexified function value is possible in these two points, giving a second level laminate.
	         The leaves might then be checked for even further split ups possibly leading to level-three laminates.}
	\label{fig:h_search}
\end{figure}

With the $\mathcal{H}$-sequence representation at hand, one can potentially sweep through high-dimensional space and connect matrices by rank-one lines to obtain the hierarchical sequence that constitutes the rank-one convex envelope for a given $\boldsymbol{F} \in \R^{d \times d}$.
The critical point is how to construct the hierarchical set itself, or to be more precise, how to choose the connected matrices.
In general, this task is too complex in the sense that the set of possible hierarchical rank-one connected sets is infinite.
Apart from the fact that the $(\xi_i, F_i)$ are pairs of scalar values and matrices, the number $M$ in \eqref{eq:WrcHseq} is not bounded in general. 
That is why we aim for solving the minimisation problem \eqref{eq:WrcHseq} on a restricted set of hierarchical rank-one connected sets, i.e.~we reduce the number of possible candidates in the minimisation problem by a local construction.
As a first approximation, for a given point $\boldsymbol{F}$, we aim to find the two rank-one connected matrices which deliver locally the lowest convexified value, where the one-dimensional convexification is performed exactly at the connecting rank-one convex line. 
Therefore, a discretisation of rank-one directions $\mathcal{R}$ is used.
After that, the algorithm proceeds recursively with the supporting points of the convexified lines and searches again for possible rank-one connected matrices, further decreasing the convex combinations' function value.
Figure~\ref{fig:h_search} illustrates this recursive procedure.

The resulting algorithm, referred to as Hierarchical Rank-One Sequence Convexification (HROC), involves the construction of a binary lamination tree to solve efficiently multi-dimensional rank-one line convexification problems.
The parameters of the algorithm are the maximum tree depth \(k_{\max} \in \N\), the number of discretisation points in the one-dimensional convexifications \(N \in \N\), and the set of discretised rank-one directions \(\mathcal{R}\).
The algorithm aims to be as global as necessary while being as local as possible.
It does this by sweeping through the high-dimensional space of $d \times d$ matrices by rank-one lines.
This approach aims to refine iteratively the rank-one convex hull by recursively splitting the determined rank-one convex hull supporting points of the lowest convexified rank-one line.

The one-dimensional convexifications along the rank-one lines can be carried out by computational geometry algorithms, such as the Graham's Scan \cite{Gra:1972:ead}. 
A pseudo code of this procedure is given in Algorithm~\ref{alg:oneDimConvexification}.\\
\begin{algorithm}[H]
	\caption{One-dimensional convexification}
	\label{alg:oneDimConvexification}
	\begin{algorithmic}[1]
		\Function{convexify}{\texttt{x, w}} \Comment{Input arrays of length \texttt{L}}
		\State{\texttt{y[1] = w[1], y[2] = w[2], c[1] = w[1], c[2] = w[2]}}
		\State{\texttt{n = 2}}
		\For{\texttt{i = 3, 4, \ldots, L}}
		\While{\texttt{{(c[n] - c[n-1]) * (x[i] - y[n]) >=}}\par \texttt{{\qquad\qquad\qquad\qquad(w[i] - c[n]) * (y[n] - y[n-1])}} \textbf{and} \texttt{n >=  1}}
		\State{\texttt{n -= 1}}					
		\EndWhile{}
		\State{\texttt{n += 1}}
		\State{\texttt{y[n] = x[i], c[n] = w[i]}}
		\EndFor{}
		\State \Return \texttt{y, c} \Comment{Output arrays of length \texttt{n}}
		\EndFunction
	\end{algorithmic}
\end{algorithm}

This algorithm exploits the monotonicity property of a convex function's derivative.
It starts to sweep through the discretised interval and as soon as a non-convex regime is entered, points within this region of non-convexity are either deleted or stored in the array of supporting points of the convex hull.
For more information, see e.g.~\cite{KohNeuMelPetPetBal:2022:acma}.
A single one-dimensional convexification is of complexity $\mathcal{O}(N)$ which coincides with the overall complexity of the proposed algorithm for simple laminates.

This approach is now used to construct binary trees by pushing the lowest convex hull supporting points along rank-one lines into a queue.
The construction of the binary tree starts with a root node representing the macroscopic deformation gradient $\boldsymbol{F}$.
At each level of the tree, the leaf nodes are split into two parts if there is a lower rank-one convexified line, the resulting supporting points are labelled \textit{plus} and \textit{minus}.
The main steps in constructing the binary rank-one lamination tree are as follows:
\begin{itemize}
    \item \textbf{Initialization}: The root node of the tree is created to represent the macroscopic deformation gradient $\boldsymbol{F}$, e.g.~as predicted by the global finite element solver.
    \item \textbf{Splitting}: At each level of the tree, the one-dimensional convexifications are performed along all discretised rank-one directions contained in $\mathcal{R}$.     
 	The direction leading to the minimal convexified function value is accepted, and leads to a split into a simple laminate.
    \item \textbf{Evaluation}: The points of the possible, initial, simple laminate split into $\boldsymbol{F}^-$ and $\boldsymbol{F}^+$ are convexified along all discretised rank-one lines to determine if a direction leads to a nested laminate and therefore lower predicted hierarchical rank-one convex hull.
    \item \textbf{Recursion}: If necessary, the process is recursively applied to the determined laminate phases to refine the approximated rank-one convex hull further.
    \item \textbf{Termination}: The process terminates when a stopping criterion is met, e.g.~a maximum tree depth is reached or further convexifications do not result in a lower convexified energy density value.
\end{itemize}

The splitting is realised in the \textsc{hroc-kernel} function, which plays a crucial role in this process by computing lamination candidates. 
A pseudocode of this function is illustrated in Algorithm~\ref{alg:hroc-kernel}.
The input arguments are the energy density function $W$, the current state of the tree (given by its root node $root$), the evaluation point $\boldsymbol{F}$, and the convexification parameters $N, r, k_{\max}$ and $\mathcal{R}$.
The function iterates over possible rank-one directions contained in the discretisation set $\mathcal{R}$, computes the convex hull along these rank-one lines, and selects the direction that yields the lowest convexification result.
First, the scalar value $W_{\text{ref}}$ is initialized by
\begin{equation} \label{eq:Wevaluation}
	W_{\text{ref}} \gets \sum_{i = 1}^{M} \xi_i \, W(\boldsymbol{F}_i)
\end{equation}
using the current lamination tree. Here, $M$ denotes the number of leaves in the tree associated to $root$ and $\boldsymbol{F}_i$ the actual leaves with their volume fractions $\xi_i$.
Afterwards, the iteration through the discretised directions starts. 
For a given direction $\boldsymbol{R}$, the scaling factor $\delta$ has to be determined ensuring that the $N$ points along the rank-one line $\boldsymbol{F} + s \boldsymbol{R}$ for $s \in \R$ are located within the bounding box of radius $r$.
After scaling of the rank-one direction, the one dimensional convexification is performed by Algorithm~\ref{alg:oneDimConvexification}.
Our implementation uses a buffer for storing the associated function values of $W$ along the rank-one line $\boldsymbol{F} + i \, \mathbf{R}$ in order to avoid reallocation but any array representation can be utilized.
The index $i \in \Z$ is restricted ensuring that the rank-one line does not leave the bounding box of radius $r$.
If the checked rank-one line delivers a lower convex combination for the function evaluation by \eqref{eq:Wevaluation} involving the new laminate, the supporting points $\boldsymbol{F}^+$ and $\boldsymbol{F}^-$ as well as the rank-one direction $\boldsymbol{R}$ are stored by overwriting the variable $laminate$.
Afterwards, the returned $laminate$ can be added to the rank-one tree represented by $root$ in the \textsc{hroc} main function.
\begin{algorithm}[H]
	\caption{\textsc{hroc-kernel}}
	\label{alg:hroc-kernel}
	\begin{algorithmic}[1]
		\Function{hroc-kernel}{$root, N, r, \mathcal{R}, W, \boldsymbol{F}$}
		\State Initialize variables $W_{\text{ref}}$ and empty $laminate$
		\For{each rank-one direction $\mathbf{R}$ in $\mathcal{R}$}
		\State Calculate scaling factor $\delta$ based on $\mathbf{R}$, $N$ and $r$
		\State Scale $\mathbf{R}$ by $\delta$
		\State One-dimensional convexification along $\boldsymbol{F} + i \, \mathbf{R}$
		\If{convexification result is lower than $W_{\text{ref}}$}
		\State $W_{\text{ref}} \gets W(laminate)$
		\State $laminate \gets$ ($\boldsymbol{F}^+, \boldsymbol{F}^-, \mathbf{R}$)
		\EndIf
		\EndFor
		\State \Return $laminate$
		\EndFunction
	\end{algorithmic}
\end{algorithm}
With this subroutine at hand, the pseudocode for the overall algorithm can be given by Algorithm~\ref{alg:hrock}.
In this pseudocode, the HROC function orchestrates the construction of the binary rank-one lamination tree by iteratively splitting deformation-gradient values into laminates and adding both of the laminate phases to the queue.
This is done until the lamination candidate queue is empty or the maximum tree depth $k_{\max}$ is reached.
The \textsc{hroc-kernel} function is called on each computed lamination candidate and expresses the recursive nature of the algorithm. 
\begin{algorithm}[H]
\caption{Hierarchical Rank-One Convexification (HROC)}
\label{alg:hrock}
\begin{algorithmic}[1]
\Function{HROC}{$\text{convexification}, \text{buffer}, W, \boldsymbol{F}$}
    \State Initialize root node $root$ with $\boldsymbol{F}$
    \State $laminate \gets$ \Call{hroc-kernel}{$root, \text{convexification}, \text{buffer}, W, \boldsymbol{F}$}
    \State Initialize queue with $(root, laminate)$
    \While{queue is not empty}
        \State $(parent, lc) \gets$ pop queue
        \If{depth of $parent < k_{\max}$}
	        \State $\lambda \gets$ compute splitting ratio based on $parent$ and $lc$
	        \State attach $\boldsymbol{F}^+$ and $\boldsymbol{F}^-$ from $lc$ with volume fractions $\lambda$ and $1-\lambda$ to $parent$ 
	        \State $lc^+ \gets$ \Call{hroc-kernel}{$root, N, r, \mathcal{R}, W, \boldsymbol{F}^+$} 
	        \State $lc^- \gets$ \Call{hroc-kernel}{$root, N, r, \mathcal{R}, W, \boldsymbol{F}^-$} 
	        \State Push $(\boldsymbol{F}^+,lc^+)$ to queue
	        \State Push $(\boldsymbol{F}^-,lc^-)$ to queue
		\EndIf
    \EndWhile
    \State \Return $root$
\EndFunction
\end{algorithmic}
\end{algorithm}

We want to highlight one important aspect of the algorithm when it comes to the semi-convexification process.
Namely, we always choose the locally lowest possible laminate for each recursive application of \textsc{hroc-kernel}, i.e.~we choose the rank-one line which delivers the smallest function value in the overall convex combination.
The overall algorithm only minimises over a certain set of hierarchically connected rank-one sets in \eqref{eq:WrcHseq}.
For arbitrary energy densities, this local choice might not result in the global optimal hierarchical rank-one connected sequence.
That is why a suitable approximation of the rank-one convex envelope for general energy densities cannot be guaranteed.

Mechanically, this assumption is violated if a lower, nested laminate can be constructed from non-optimal intermediate level laminates.
A counter example in which the HROC algorithm fails to approximate the rank-one convex envelope is discussed in Section~\ref{sec:benchmark:counter_example}.
However, we assume that we are working with energy densities where the assumption of energetically optimal laminates per level is valid.
Thus, we propose a domain specific algorithm which is only valid for a certain class of functions, but allows for the efficient approximation of envelopes as well as their derivatives within this class.
It is worth mentioning that the derivatives of $W^{\text{rc}}$ are of fundamental interest as they are used by the finite element solver within the assembly procedure and they are linked to the computation of mechanical stress.
In the case of an energy-based line search, the function value of $W^{\text{rc}}$ becomes important in order to assemble the total energy over the domain.

\subsection{Derivative Information}
The response of the HROC algorithm is a tree of matrices which are hierarchically rank-one connected.
The leaves of the tree form the hierarchical rank-one connected set.
They are denoted by $\boldsymbol{F}_1, \dots, \boldsymbol{F}_M \in \R^{d \times d}$ and their associated volume fractions are $\xi_1, \dots, \xi_M \in \R$.
With this $\mathcal{H}$-sequence, the approximation of first- and second-order derivative can be obtained by
\begin{equation} \label{eq:deriv}
	\boldsymbol{P} = \partial_{\boldsymbol{F}}W^{\text{rc}}(\boldsymbol{F}) \approx \sum_{i = 1}^{M} \xi_i \, \partial_{\boldsymbol{F}}W(\boldsymbol{F}_i)
\end{equation}
and
\begin{equation}
	\mathbb{A} = \partial\partial_{\boldsymbol{F}}W^{\text{rc}}(\boldsymbol{F}) \approx \sum_{i = 1}^{M} \xi_i \, \partial\partial_{\boldsymbol{F}}W(\boldsymbol{F}_i),
\end{equation}
respectively. It is advantageous that the algorithm relies on the derivative evaluation of the original function which is often at least $C^2$, thus, giving a well-posed formula for the derivatives, except for the second derivative which is lost along rank-one paths, due to the definition of the rank-one convex hull.
For a mathematical discussion of the continuity of the rank-one convex envelope, see \cite[Section~4]{BalKirKri:2000:rqe}.

\subsection{Microstructure Reconstruction}
The computed rank-one tree for $\boldsymbol{F}$ represents the formation of microstructures associated to the relaxation process. 
However, the tree represents an $\mathcal{H}$-sequence for the deformation gradient $\boldsymbol{F}$, but there is no information about an actual microstructural deformation field.
We are interested in the characterisation of the microstructure based on the lamination tree obtained by the HROC algorithm on a reference volume element $\Omega = [0,1]^d$.
Given a binary lamination tree with leaves $\boldsymbol{F}_i$ for $i=1,\dots,M$, a recursive call of characteristic functions can be used to obtain a lamination coefficient $\boldsymbol{F}^{\pm}$ for any point $\boldsymbol{x} \in \Omega$.
This coefficient function $\boldsymbol{F}^{\pm} \colon \Omega \to \R^{d \times d}$ associates a leave of the rank-one tree with root node $\boldsymbol{F}$, i.e.~one $\boldsymbol{F}_i$ of the $\mathcal{H}$-sequence, to any given point $\boldsymbol{x} \in \Omega$. It is defined by
\begin{equation}
	\boldsymbol{F}^{\pm}(\boldsymbol{x}) = \boldsymbol{F} + \hat{\boldsymbol{F}}^{\pm}(\boldsymbol{x}, \boldsymbol{F}) 
\end{equation}
where the function $\hat{\boldsymbol{F}}^{\pm} \colon \R^{d} \times \R^{d \times d} \to \R^{d \times d}$ is given by
\begin{equation}
	\hat{\boldsymbol{F}}^{\pm}(\boldsymbol{x}, \boldsymbol{A}) = 
	\begin{cases}
		 \boldsymbol{0} \in \R^{d \times d} & \text{if } \boldsymbol{A} \text{ has no children}, \\
		 - \lambda \, \boldsymbol{R} + \hat{\boldsymbol{F}}^{\pm}(\boldsymbol{x}, \boldsymbol{A}^{+}) & \text{if }  \{\frac{\boldsymbol{x} \cdot \boldsymbol{n}}{\varepsilon}\} \in [0, 1 - \lambda),\\
		 (1 - \lambda) \, \boldsymbol{R} + \hat{\boldsymbol{F}}^{\pm}(\boldsymbol{x}, \boldsymbol{A}^{-}) & \text{if } \{\frac{\boldsymbol{x} \cdot \boldsymbol{n}}{\varepsilon}\} \in [1 - \lambda, 1).
	\end{cases}
\end{equation}
The recursive function call of $\hat{\boldsymbol{F}}^{\pm}$ traces down a binary tree until a leave $\boldsymbol{F}_i$ of the tree is reached.
Within this function, $\boldsymbol{A}^{+}$ and $\boldsymbol{A}^{-}$ denote the lamination split for a node $\boldsymbol{A}$, i.e.~the child nodes of $\boldsymbol{A}$ in the tree, $\lambda \in [0, 1]$ denotes the volume fraction of the phase $\boldsymbol{A}^{-}$ and the volume fraction of $\boldsymbol{A}^{+}$ is consequently given by $1 - \lambda$. 
The vector $\boldsymbol{n} \in \R^{d}$ is the normal of the rank-one direction $\boldsymbol{R}$ associated to the $\boldsymbol{A}^+$, $\boldsymbol{A}^-$ split, i.e.~$\boldsymbol{R} = \boldsymbol{a} \otimes \boldsymbol{n} = \boldsymbol{A}^- - \boldsymbol{A}^+$.
By $\{\, \bullet \,\}$, we denote the fractional part of a real number, i.e.~$\{s\} = s - \lfloor s \rfloor \in [0, 1)$ for $s \in \R$.
The parameter $\varepsilon \in [0, 1]$ denotes the length scale on which the laminate consisting of $\boldsymbol{A}^+$ and $\boldsymbol{A}^-$ oscillates.
While the phases, the associated volume fractions and the rank-one direction, are given by the binary lamination tree (e.g.~obtained by the output of the HROC algorithm), the oscillating length scale $\varepsilon$, or alternatively the frequency $1 / \varepsilon$, needs to be chosen in order to visualize a possible microstructure.

With this coefficient function at hand, the microstructural deformation field $\boldsymbol{u}$ can be reconstructed by the following gradient $L^2$ projection problem
\begin{equation}
	J(\boldsymbol{u}) = \frac{1}{2} \lVert \nabla \boldsymbol{u} - \boldsymbol{F}^{\pm}\rVert \rightarrow \min,
\end{equation}
subject to periodic boundary conditions. This projection solves for $\boldsymbol{u}$ which is the closest possible function whose gradient matches the oscillating lamination gradients $\boldsymbol{F}^{\pm}$. Taking the first variation and setting it to zero leads to
\begin{equation}
    DJ(\boldsymbol{u})\nabla \delta \boldsymbol{u} = (\nabla \boldsymbol{u} - \boldsymbol{F}^{\pm},\nabla \delta \boldsymbol{u}) = 0
\end{equation}
for all $\delta \boldsymbol{u} \in V$, where $V$ is a suitable test space. Note that the relation $\nabla \delta \boldsymbol{u} = \delta \nabla \boldsymbol{u}$ was used in order to express everything with respect to the desired unknown $\boldsymbol{u}$.
Using linearity of the inner product results in
\begin{equation}
	\label{eq:microstructure_displacement}
	\int_{\Omega} \nabla \boldsymbol{u} : \nabla \delta \boldsymbol{u}\, \text{d}V = \int_{\Omega} \boldsymbol{F}^{\pm} : \nabla \delta \boldsymbol{u} \, \text{d}V.
\end{equation}
Equation \eqref{eq:microstructure_displacement} can now be discretised by finite elements and solved for an approximation $\boldsymbol{u}_h$ in a discrete space $V_h$.
This procedure will be used for a visualization for a two-dimensional example in Section~\ref{sec:examples} (also see Figure~\ref{fig:microstructure} below).

%% file: figures/tikz/h_search_begin.tex
\tikzsetfigurename{h_search_begin}
\begin{tikzpicture}
    \pgfmathsetmacro\dirdeg{0.125} 
    \pgfmathsetmacro\firstlinelength{1.75} 
    \pgfmathsetmacro\firstdirdeg{0.25} 
    \pgfmathsetmacro\secondlinelength{1.5} 
    \pgfmathsetmacro\firstfirstdirdeg{0.75} 
    \pgfmathsetmacro\firstseconddirdeg{0.875} 
    \pgfmathsetmacro\thirdlinelength{1.} 
    \node (F) at (0, 0) {};
    \node (F1) at ({0.9 * \firstlinelength * cos(deg(\firstdirdeg * pi))}, {0.9 * \firstlinelength * sin(deg(\firstdirdeg * pi))}) {};
    \node (F2) at (-{0.7 *\firstlinelength * cos(deg(\firstdirdeg * pi))}, -{0.7 * \firstlinelength * sin(deg(\firstdirdeg * pi))}) {};
    \node (F11) at ({0.6 * \secondlinelength * cos(deg(\firstfirstdirdeg * pi)) + 0.9 * \firstlinelength * cos(deg(\firstdirdeg * pi))}, {0.6 * \secondlinelength * sin(deg(\firstfirstdirdeg * pi)) + 0.9 * \firstlinelength * sin(deg(\firstdirdeg * pi))}) {};
    \node (F12) at ({-0.8* \secondlinelength * cos(deg(\firstfirstdirdeg * pi)) + 0.9 * \firstlinelength * cos(deg(\firstdirdeg * pi))}, {- 0.8 * \secondlinelength * sin(deg(\firstfirstdirdeg * pi)) + 0.9 * \firstlinelength * sin(deg(\firstdirdeg * pi))}) {};
    \node (F21) at ({0.7 * \secondlinelength * cos(deg(\firstseconddirdeg * pi)) - 0.7 *\firstlinelength * cos(deg(\firstdirdeg * pi))}, {0.7 * \secondlinelength * sin(deg(\firstseconddirdeg * pi)) - 0.7 * \firstlinelength * sin(deg(\firstdirdeg * pi)) }) {};
    \node (F22) at ({-0.9 * \secondlinelength * cos(deg(\firstseconddirdeg * pi)) - 0.7 *\firstlinelength * cos(deg(\firstdirdeg * pi))}, {- 0.9 * \secondlinelength * sin(deg(\firstseconddirdeg * pi))  - 0.7 * \firstlinelength * sin(deg(\firstdirdeg * pi))}) {};

    \filldraw[black] (F) circle (2pt) node[anchor=east]{\small $\boldsymbol{F}$};

    \foreach \n in {0, \dirdeg,...,2} {
        \draw[thin, dashed,gray] (F) -- ++({\firstlinelength * cos(deg(\n * pi))}, {\firstlinelength * sin(deg(\n * pi))}) {};
    }

	\phantom{
        \foreach \n in {0, \dirdeg,...,2} {
            \draw[thin, dashed,gray] (F1) -- ++({\secondlinelength * cos(deg(\n * pi))}, {\secondlinelength * sin(deg(\n * pi))}) {};
            \draw[thin, dashed,gray] (F2) -- ++({\secondlinelength * cos(deg(\n * pi))}, {\secondlinelength * sin(deg(\n * pi))}) {};
        }
	}



        \draw[thick,black] (-{\firstlinelength * cos(deg(\firstdirdeg * pi))}, -{\firstlinelength * sin(deg(\firstdirdeg * pi))}) -- ({\firstlinelength * cos(deg(\firstdirdeg * pi))}, {\firstlinelength * sin(deg(\firstdirdeg * pi))}) {};
        \filldraw[black] (F1) circle (2pt) node[anchor=east]{\small $\boldsymbol{F}^{+}$};
        \filldraw[black] (F2) circle (2pt) node[anchor=east]{\small $\boldsymbol{F}^{-}$};

	\end{tikzpicture}

%% file: figures/tikz/h_tree_begin.tex
\tikzsetfigurename{h_tree_begin}
\begin{tikzpicture}
    \pgfmathsetmacro\verticalsize{1.5} %
    \pgfmathsetmacro\firstsize{2.} %
    \pgfmathsetmacro\secondsize{1.} %
    \pgfmathsetmacro\thirdsize{1.0} %
    \pgfmathsetmacro\fourthsize{1.0} %
    \node (F) at (0,0) {\small $\boldsymbol{F}$};
    \node (F1) at (-\firstsize,-\verticalsize) {\small $\boldsymbol{F}^{+}$};
    \node (F2) at (\firstsize,-\verticalsize) {\small $\boldsymbol{F}^{-}$};
    \draw (F) -- (F1) node[midway,left,yshift=1em] {}; 
    \draw (F) -- (F2) node[midway,right,yshift=1em] {}; 

    \phantom{
    \node (F11) at (-\firstsize-\secondsize,-2*\verticalsize) {\small $\boldsymbol{F}_{11}$};
    \node (F12) at (-\firstsize+\secondsize,-2*\verticalsize) {\small $\boldsymbol{F}_{12}$};
    \draw (F1) -- (F11) node[midway,left] {}; 
    \draw (F1) -- (F12) node[midway,right] {}; 
    \node (F21) at (\firstsize-\secondsize,-2*\verticalsize) {\small $\boldsymbol{F}_{21}$};
    \node (F22) at (\firstsize+\secondsize,-2*\verticalsize) {\small $\boldsymbol{F}_{22}$};
    \draw (F2) -- (F21) node[midway,left] {}; 
    \draw (F2) -- (F22) node[midway,right] {}; 
	}
    %
    %
    %
\end{tikzpicture}

%% file: figures/tikz/h_search_end.tex
\tikzsetfigurename{h_search_end}
\begin{tikzpicture}
    \pgfmathsetmacro\dirdeg{0.125} 
    \pgfmathsetmacro\firstlinelength{1.75} 
    \pgfmathsetmacro\firstdirdeg{0.25} 
    \pgfmathsetmacro\secondlinelength{1.5} 
    \pgfmathsetmacro\firstfirstdirdeg{0.75} 
    \pgfmathsetmacro\firstseconddirdeg{0.875} 
    \pgfmathsetmacro\thirdlinelength{1.} 
    \node (F) at (0, 0) {};
    \node (F1) at ({0.9 * \firstlinelength * cos(deg(\firstdirdeg * pi))}, {0.9 * \firstlinelength * sin(deg(\firstdirdeg * pi))}) {};
    \node (F2) at (-{0.7 *\firstlinelength * cos(deg(\firstdirdeg * pi))}, -{0.7 * \firstlinelength * sin(deg(\firstdirdeg * pi))}) {};
    \node (F11) at ({0.6 * \secondlinelength * cos(deg(\firstfirstdirdeg * pi)) + 0.9 * \firstlinelength * cos(deg(\firstdirdeg * pi))}, {0.6 * \secondlinelength * sin(deg(\firstfirstdirdeg * pi)) + 0.9 * \firstlinelength * sin(deg(\firstdirdeg * pi))}) {};
    \node (F12) at ({-0.8* \secondlinelength * cos(deg(\firstfirstdirdeg * pi)) + 0.9 * \firstlinelength * cos(deg(\firstdirdeg * pi))}, {- 0.8 * \secondlinelength * sin(deg(\firstfirstdirdeg * pi)) + 0.9 * \firstlinelength * sin(deg(\firstdirdeg * pi))}) {};
    \node (F21) at ({0.7 * \secondlinelength * cos(deg(\firstseconddirdeg * pi)) - 0.7 *\firstlinelength * cos(deg(\firstdirdeg * pi))}, {0.7 * \secondlinelength * sin(deg(\firstseconddirdeg * pi)) - 0.7 * \firstlinelength * sin(deg(\firstdirdeg * pi)) }) {};
    \node (F22) at ({-0.9 * \secondlinelength * cos(deg(\firstseconddirdeg * pi)) - 0.7 *\firstlinelength * cos(deg(\firstdirdeg * pi))}, {- 0.9 * \secondlinelength * sin(deg(\firstseconddirdeg * pi))  - 0.7 * \firstlinelength * sin(deg(\firstdirdeg * pi))}) {};

    \filldraw[black] (F) circle (2pt) node[anchor=east]{\small $\boldsymbol{F}$};


        \foreach \n in {0, \dirdeg,...,2} {
            \draw[thin, dashed,gray] (F1) -- ++({\secondlinelength * cos(deg(\n * pi))}, {\secondlinelength * sin(deg(\n * pi))}) {};
            \draw[thin, dashed,gray] (F2) -- ++({\secondlinelength * cos(deg(\n * pi))}, {\secondlinelength * sin(deg(\n * pi))}) {};
        }

        \draw[thick, black] ({-\secondlinelength * cos(deg(\firstfirstdirdeg * pi)) + 0.9 * \firstlinelength * cos(deg(\firstdirdeg * pi))}, {-\secondlinelength * sin(deg(\firstfirstdirdeg * pi)) + 0.9 * \firstlinelength * sin(deg(\firstdirdeg * pi))}) --
        ({\secondlinelength * cos(deg(\firstfirstdirdeg * pi)) + 0.9 * \firstlinelength * cos(deg(\firstdirdeg * pi))}, {\secondlinelength * sin(deg(\firstfirstdirdeg * pi))+ 0.9 * \firstlinelength * sin(deg(\firstdirdeg * pi))}) {};

        \draw[thick, black] ({-\secondlinelength * cos(deg(\firstseconddirdeg * pi)) - 0.7 *\firstlinelength * cos(deg(\firstdirdeg * pi))}, {-\secondlinelength * sin(deg(\firstseconddirdeg * pi))  - 0.7 * \firstlinelength * sin(deg(\firstdirdeg * pi))}) -- ({\secondlinelength * cos(deg(\firstseconddirdeg * pi)) - 0.7 *\firstlinelength * cos(deg(\firstdirdeg * pi))}, {\secondlinelength * sin(deg(\firstseconddirdeg * pi)) - 0.7 * \firstlinelength * sin(deg(\firstdirdeg * pi)) }) {};
        \filldraw[black] (F11) circle (2pt) node[anchor=east]{\small $\boldsymbol{F}^{++}$};
        \filldraw[black] (F12) circle (2pt) node[anchor=east]{\small $\boldsymbol{F}^{+-}$};
        \filldraw[black] (F21) circle (2pt) node[anchor=east]{\small $\boldsymbol{F}^{-+}$};
        \filldraw[black] (F22) circle (2pt) node[anchor=east]{\small $\boldsymbol{F}^{--}$};

        \draw[thick, black] (-{\firstlinelength * cos(deg(\firstdirdeg * pi))}, -{\firstlinelength * sin(deg(\firstdirdeg * pi))}) -- ({\firstlinelength * cos(deg(\firstdirdeg * pi))}, {\firstlinelength * sin(deg(\firstdirdeg * pi))}) {};
        \filldraw[black] (F1) circle (2pt) node[anchor=east]{\small $\boldsymbol{F}^{+}$};
        \filldraw[black] (F2) circle (2pt) node[anchor=east]{\small $\boldsymbol{F}^{-}$};

	\end{tikzpicture}

%% file: figures/tikz/h_tree_end.tex
\tikzsetfigurename{h_tree_end}
\begin{tikzpicture}
    \pgfmathsetmacro\verticalsize{1.5} %
    \pgfmathsetmacro\firstsize{2.} %
    \pgfmathsetmacro\secondsize{1.} %
    \pgfmathsetmacro\thirdsize{1.0} %
    \pgfmathsetmacro\fourthsize{1.0} %
    \node (F) at (0,0) {\small $\boldsymbol{F}$};
    \node (F1) at (-\firstsize,-\verticalsize) {\small $\boldsymbol{F}^{+}$};
	\node (F2) at (\firstsize,-\verticalsize) {\small $\boldsymbol{F}^{-}$};
    \draw (F) -- (F1) node[midway,left,yshift=1em] {}; 
    \draw (F) -- (F2) node[midway,right,yshift=1em] {}; 

    \node (F11) at (-\firstsize-\secondsize,-2*\verticalsize) {\small $\boldsymbol{F}^{++}$};
    \node (F12) at (-\firstsize+\secondsize,-2*\verticalsize) {\small $\boldsymbol{F}^{+-}$};
    \draw (F1) -- (F11) node[midway,left] {}; 
    \draw (F1) -- (F12) node[midway,right] {}; 
    \node (F21) at (\firstsize-\secondsize,-2*\verticalsize) {\small $\boldsymbol{F}^{-+}$};
    \node (F22) at (\firstsize+\secondsize,-2*\verticalsize) {\small $\boldsymbol{F}^{--}$};
    \draw (F2) -- (F21) node[midway,left] {}; 
    \draw (F2) -- (F22) node[midway,right] {}; 
    %
    %
    %
\end{tikzpicture}

%% file: sections/sec4.tex
%

\section{Benchmark Problems} \label{sec:benchmark}
We begin the application of the algorithm by studying a variety of theoretical benchmark problems to demonstrate the approximation quality of the algorithm and its limitations.
The Julia implementation of the algorithm and the used benchmark functions can be found in the repository \texttt{https://github.com/koehlerson/NumericalRelaxation.jl}.
In all computational experiments, the set 
\begin{equation}\label{eq:discreteR1dir}
	\mathcal{R} = \{\boldsymbol{a} \otimes \boldsymbol{b} \mid \boldsymbol{a}, \boldsymbol{b} \in \Z^d, |\boldsymbol{a} |_\infty, |\boldsymbol{b}|_\infty \leq 1\}
\end{equation}
is used as discretisation for the rank-one directions, as proposed in \cite[Section 5.1]{DolWal:2000:ena}.
In general, this set of rank-one directions can be extended, but due to the insights gained in \cite{BalKohNeuPetPet:2023:mrc}, our numerical experiments focus on the set \eqref{eq:discreteR1dir}.
The appropriate scaling of the directions by the factor \(\delta\) is included in Algorithm~\ref{alg:hroc-kernel}.
The maximum laminate depth $k_{\max}$ is set to $10$.

\subsection{Kohn--Strang--Dolzmann Example}
\begin{figure}[h]
    \begin{subfigure}{0.49\textwidth}
		\ifthenelse{\boolean{professormode}}
		{\includegraphics{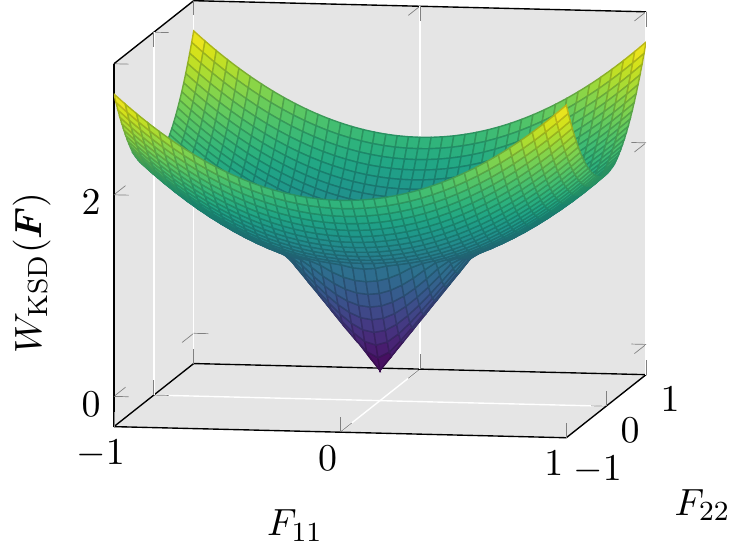}}
        {\input{figures/tikz/KSD.tex}}
        \caption{}
        \label{fig:KSD_W}
    \end{subfigure}
    \begin{subfigure}{0.49\textwidth}
		\ifthenelse{\boolean{professormode}}
		{\includegraphics{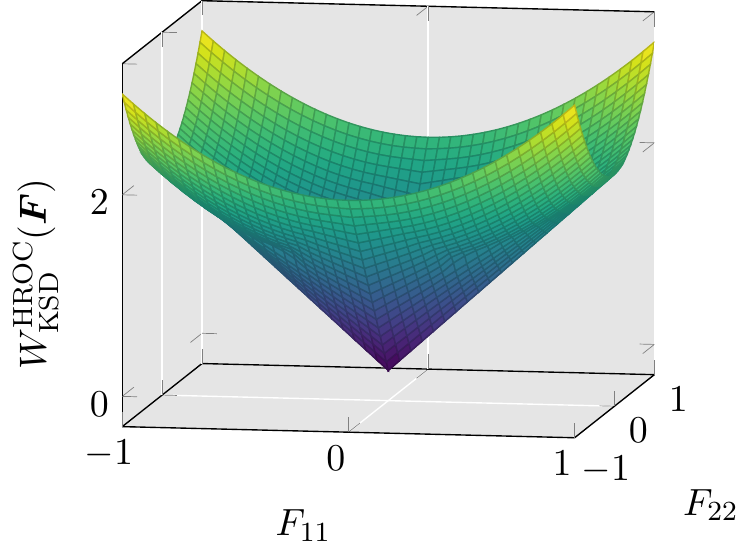}}
        {\input{figures/tikz/KSD_hroc.tex}}
        \caption{}
        \label{fig:KSD_hroc}
    \end{subfigure}
    \caption{(\subref{fig:KSD_W}) Kohn-Strang-Dolzmann energy density and (\subref{fig:KSD_hroc}) approximation of the rank-one convex envelope by HROC algorithm in the $F_{11}$--$F_{22}$ plane for $F_{12} = F_{21} = 0$.
             The relative error is $\max_{\boldsymbol{F} \in \mathcal{N}_{\delta, r}} \| \frac{W_{\text{KSD}}^{\text{HROC}}({\boldsymbol{F}}) - W_{\text{KSD}}^{\text{rc}}({\boldsymbol{F}})}{W_{\text{KSD}}^{\text{rc}}({\boldsymbol{F}})}\| = 0.0386$ and the absolute error $0.0472$ with convexification parameter $N = 5000$ on the grid in the $F_{11}$--$F_{22}$ plane represented by the parameters $\delta \approx 1 / 20, r = 1$.}
    \label{fig:KSD-energies}
\end{figure}

In \cite{KohStr:1986:odr:a, KohStr:1986:odr}, an example which can be relaxed analytically was proposed.
\cite{Dol:1999:ncr, DolWal:2000:ena} modified the example such that the function's behaviour at zero is continuous.
The modified example reads
\begin{equation}
    W_{\text{KSD}}(\boldsymbol{F}) = 
    \begin{cases*}
        1 + \vert \boldsymbol{F} \vert ^2 & \text{if } $\vert \boldsymbol{F} \vert \geq \sqrt{2} - 1$, \\
        2 \sqrt{2} \vert\boldsymbol{F}\vert & \text{if } $\vert \boldsymbol{F}\vert \leq \sqrt{2} - 1$.
    \end{cases*}
\end{equation}
Its rank-one convex envelope, which equals its quasiconvex and polyconvex envelope but differs from its convex hull, is given in \cite[Section~5.1]{Dol:1999:ncr} and reads
\begin{equation}
    W_{\text{KSD}}^{\text{rc}}(\boldsymbol{F}) = \begin{cases*}
        1 + \vert \boldsymbol{F} \vert ^2 & \text{if } $\rho(\boldsymbol{F}) \geq 1$, \\
        2\left(\rho(\boldsymbol{F}) - \vert \det \boldsymbol{F} \vert\right) & \text{if } $\rho(\boldsymbol{F}) \leq 1$,
    \end{cases*}
\end{equation}
with $\rho(\boldsymbol{F}) = \sqrt{\vert \boldsymbol{F} \vert ^2 + 2 \vert \det \boldsymbol{F} \vert}$.
The function $W_{\text{KSD}}$ is shown in Figure~\ref{fig:KSD_W} and the approximation of the rank-one convex envelope by the HROC algorithm in Figure~\ref{fig:KSD_hroc}.
In order to construct an energetic landscape, the HROC algorithm was evaluated at different points in the diagonal matrix plane and linearly interpolated in between.
The error on the box $\mathcal{N}_{\delta, r}$ shows that the HROC algorithm indeed approximates the rank-one convex envelope.

\begin{figure}[h]
	\ifthenelse{\boolean{professormode}}
	{\includegraphics{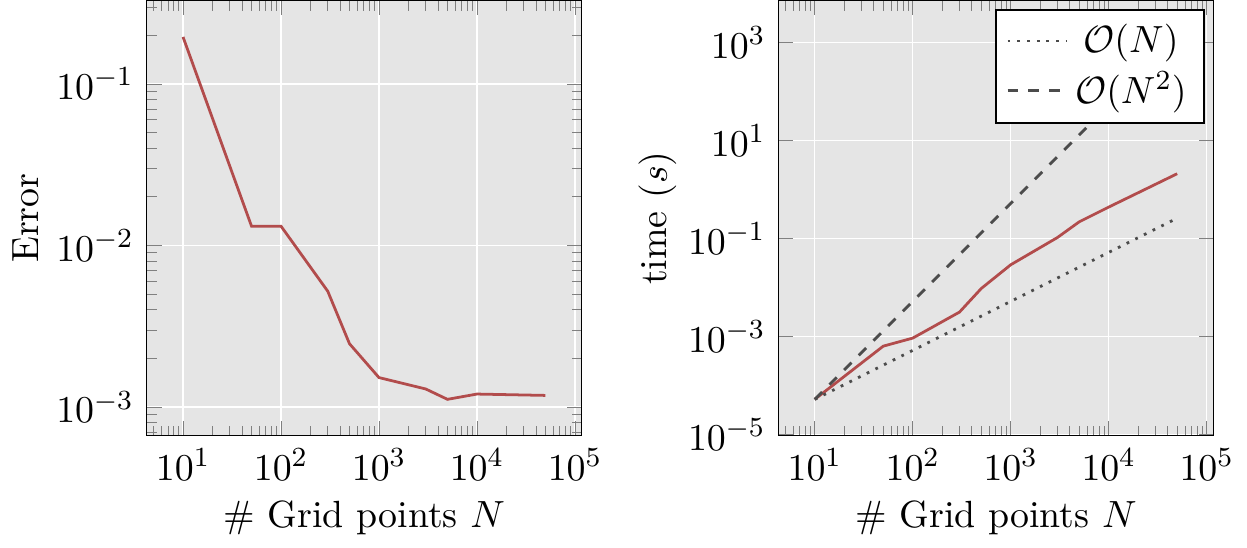}}
	{\input{figures/tikz/KSD_convergence.tex}} 
    \caption{On the left-hand side, the pointwise approximation error of the HROC algorithm in $\hat{\boldsymbol{F}}$ versus the one-dimensional convexification discretisation parameter $N$ is shown for the Kohn--Strang--Dolzmann example.
    A plateau for an error of $10^{-3}$ is reached for $10^3$ or more points along the discretised rank-one lines.
    The right-hand side shows how the one-dimensional discretisation points scale in terms of computational time.
    A discretisation with $N \approx 10^3$ points leads to acceptable computational times of milliseconds which is suited for the use in constitutive models while at the same time preserving possible accuracy of the algorithm, which, however, is already at the limit when it comes to feasibility for realistic boundary value problems.}
    \label{fig:KSD}
\end{figure}

The approximation quality and computational performance of a pointwise evaluation of the HROC algorithm is visualized in Figure~\ref{fig:KSD}.
The convergence of the pointwise error ${W_{\text{KSD}}^{\text{HROC}}(\hat{\boldsymbol{F}}) - W_{\text{KSD}}^{\text{rc}}(\hat{\boldsymbol{F}})}$ in the evaluation point
\begin{equation}
	\hat{\boldsymbol{F}} = \begin{bmatrix} 0.2 & 0.1 \\ 0.1 & 0.3 \end{bmatrix},
\end{equation}
which is not part of the $F_{11}$--$F_{22}$ plane, is plotted on the left-hand side of Figure~\ref{fig:KSD}.
The point is purposely chosen to match the numerical studies of \cite[Section~4.1]{NeuPetPetWie:2023:cpi}.
On the right-hand side of this figure, the scaling of the computational time with respect to the discretisation parameter $N$ for the one-dimensional convexifications is visualized.
Since the tree depth $k_{\max}$ and set of discretised rank-one directions $\mathcal{R}$ is kept fixed one can observe linear complexity in the one-dimensional convexification discretisation parameter $N$ in the asymptotic regime. 
In the practically relevant regime for $N \approx 10^{-3}$, which allows the incorporation into a nonlinear finite element simulation, the scaling is of order $3N$.
The prefactor $3$ is due to the second order laminates present in the problem.
In general, a factor of $2^t-1$ can be expected, where $t \leq k_{\max}$ is the problem dependent maximum laminate depth.
The observed plateau in the convergence plot in Figure~\ref{fig:KSD} starting at $N \approx 10^3$ indicates that the approximation accuracy is limited, i.e.~at some point a finer one-dimensional convexification does not necessarily lead to higher accuracy.
In order to achieve a higher approximation quality, one could enrich the rank-one discretisation set and increase the maximal limit for the tree depth.

\subsection{Multiwell Example}
\begin{figure}[h]
    \begin{subfigure}{0.49\textwidth}
		\ifthenelse{\boolean{professormode}}
		{\includegraphics{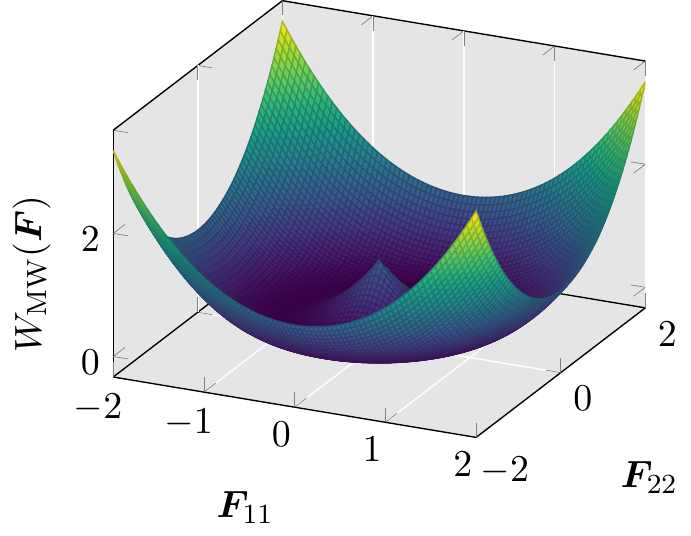}}
        {\input{figures/tikz/Multi.tex}}
        \caption{}
        \label{fig:multi}
    \end{subfigure}
    \begin{subfigure}{0.49\textwidth}
		\ifthenelse{\boolean{professormode}}
		{\includegraphics{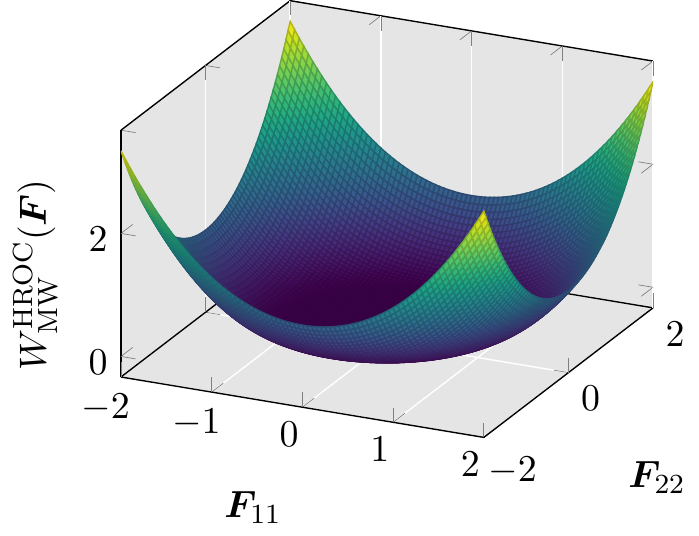}}
        {\input{figures/tikz/Multi_hroc.tex}} 
        \caption{}
        \label{fig:multi_hroc}
    \end{subfigure}
    \caption{$F_{11}$--$F_{22}$ plane of the multiwell energy density benchmark problem in Figure~\ref{fig:multi}.
    Figure~\ref{fig:multi_hroc} shows the approximated rank-one convex envelope using the HROC algorithm. Comparing the two, a negligible difference is obtained.
    Here, the HROC algorithm was used for the pointwise evaluation in $6561$ grid points in the $F_{11}$--$F_{22}$ plane.}
    \label{fig:multiwell-energies}
\end{figure}

Next, we analyse a benchmark problem in three spatial dimensions, namely the following multiwell function
\begin{equation}
    W_{\text{MW}}(\boldsymbol{F}) = \left(\vert \boldsymbol{F} \vert ^2 - 1\right)^2.
\end{equation}
Its analytic rank-one convex envelope, which equals its convex envelope, see e.g.~\cite[Section~9.1.1]{Bar:2015:nmn}, is given by
\begin{equation}
    W_{\text{MW}}^{\text{rc}}(\boldsymbol{F}) =
    \begin{cases*}
    	\left(\vert \boldsymbol{F} \vert^2 - 1 \right)^2 & \text{if } $\vert \boldsymbol{F} \vert \geq 1$, \\
        0 & \text{else}.
    \end{cases*}
\end{equation}
The benchmark problem is visualized in the $F_{11}$--$F_{22}$ plane (all other components of $\boldsymbol{F}$ are set to zero, including $F_{33}$) in Figure~\ref{fig:multi}.
Figure~\ref{fig:multi_hroc} shows the HROC algorithm response sampled at points in this plane and linearly interpolated in between.

\begin{figure}[h]
	\ifthenelse{\boolean{professormode}}
	{\includegraphics{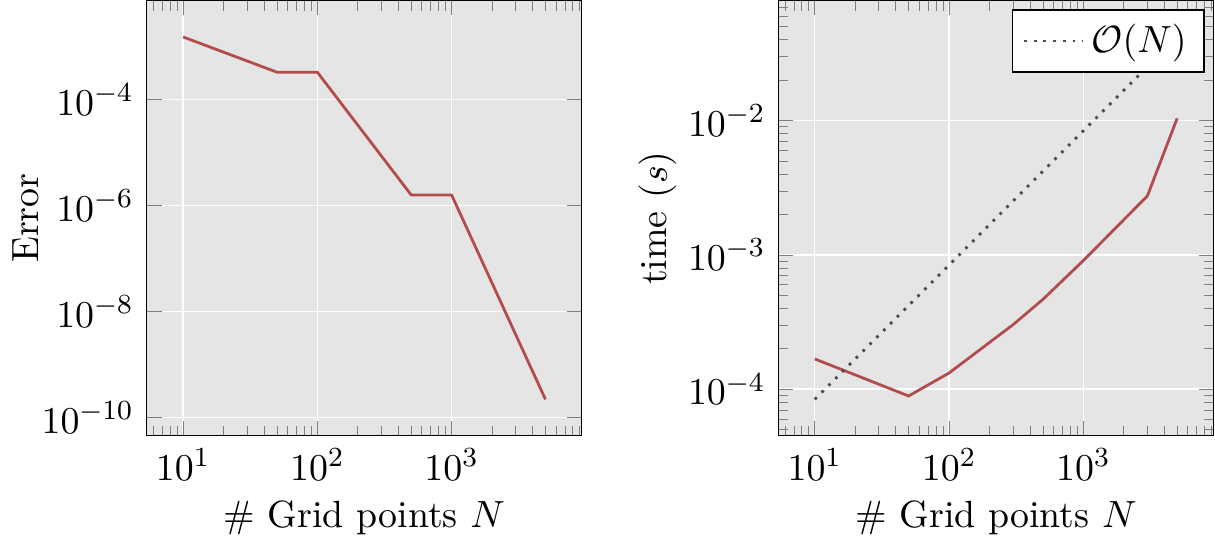}}
    {\input{figures/tikz/Multi_convergence.tex}} 
    \caption{Error development (left) and computational times (right) versus one-dimensional convexification discretisation parameter $N$ for the multiwell benchmark problem in three spatial dimensions. The use within constitutive models lies in an acceptable range for $N \leq 10^4$ points.}
    \label{fig:Multi}
\end{figure}

The convergence and performance of the algorithm for this three-dimensional example is illustrated in Figure~\ref{fig:Multi}.
Again, the pointwise error $W^{\text{HROC}}_{\text{MW}}(\hat{\boldsymbol{F}})-W^{\text{rc}}_{\text{MW}}(\hat{\boldsymbol{F}})$ at $\hat{\boldsymbol{F}} = \boldsymbol{0}$ is plotted on the left-hand side in Figure~\ref{fig:Multi}. 
Here, it is noteworthy, that the error approaches zero rather quickly.
The computational times are shown on the right-hand side of Figure~\ref{fig:Multi}. 
The performance in the nine-dimensional $\R^{3\times 3}$-space is excellent; for $N \approx 10^4$ discretisation points, the application of the algorithm is still feasible in constitutive models within finite element applications.
The linear complexity in three spatial dimensions of the HROC algorithm can be observed by the parallel line to the perfect linear scaling.
For a comparison to state-of-the-art polyconvexification methods, see \cite[Figure~4.4]{NeuPetPetWie:2023:cpi}.

\subsection{Failure of Approximation} \label{sec:benchmark:counter_example}
In what follows, we illustrate the limitations of the HROC algorithm when it comes to the approximation of the rank-one convex envelope for arbitrary functions, i.e.~in general it does not deliver the correct point evaluation of the rank-one convex envelope but only an upper bound.
The HROC algorithm computes a valid hierarchical rank-one sequence, which, however, might not be optimal in the sense of the rank-one convex envelope.
The computed convexified function value is smaller or equal to the original function value and an upper bound for the rank-one convex envelope evaluation.
This is due to the fact that the algorithm only minimises over a subset of hierarchical rank-one connected sets (laminates), namely only the local optimal ones.

In order to illustrate this, we discuss a counter example for which the algorithm fails due to the nonlocality of the convexification problem.
Consider the function
\begin{equation} \label{eq:Wcounter}
	W_{\text{fail}}(\boldsymbol{F}) = \left((\nu_1 - 3)^2 (\nu_1 + 3)^2 + (\nu_2 - 3)^2 (\nu_2 + 3)^2\right) \left(\left(\sqrt{\nu_1^2 + \nu_2^2} - 1\right)^2 + 1\right),
\end{equation}
where $\nu_1$ and $\nu_2$ describe the signed singular values of the $2\times 2$ matrix $\boldsymbol{F}$, so that, $\nu_1 \nu_2 = \det F$.

\begin{figure}[h]
    \begin{subfigure}{0.49\textwidth}
        \ifthenelse{\boolean{professormode}}
        {\includegraphics{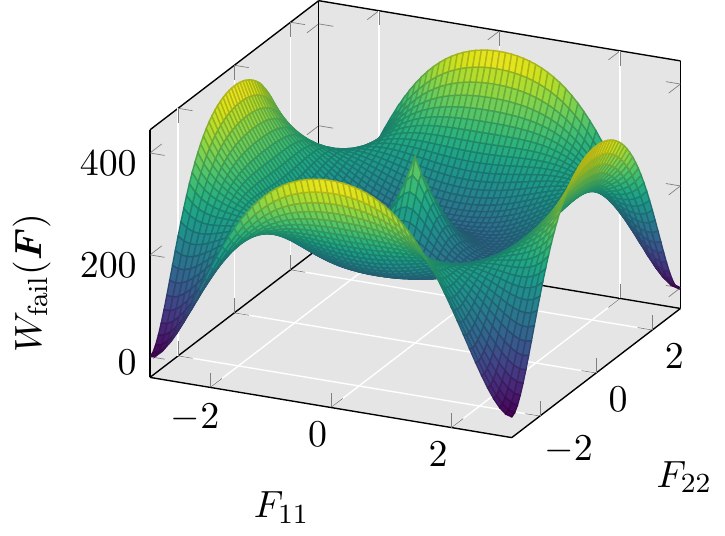}}
        {\input{figures/tikz/counter_example.tex}} 
        \caption{}
        \label{fig:counter_example}
    \end{subfigure}
    \begin{subfigure}{0.49\textwidth}
        \ifthenelse{\boolean{professormode}}
        {\includegraphics{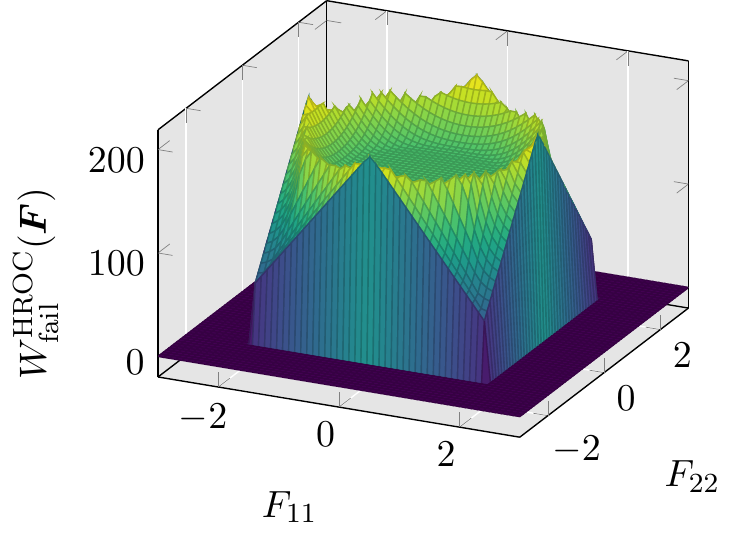}}
        {\input{figures/tikz/counter_example_hroc.tex}} 
        \caption{}
        \label{fig:counter_example_hroc}
    \end{subfigure}
    \caption{Figure~\ref{fig:counter_example} illustrates $W_{\text{fail}}$ in the $F_{11}$--$F_{22}$ plane while Figure~\ref{fig:counter_example_hroc} shows the application of the HROC algorithm. 
    	The inability to approximate the rank-one convex envelope is noticeable due to the missing zero plateau in the computed hull within the inner radius $r = 1$.
    	This is a result of the nested minima, where the outer zero-valued global minima cannot be reached by HROC-selected rank-one connected sets.
    	However, they could be reached by taking into account non-optimal intermediate laminates.}
    \label{fig:counter_example_energies}
\end{figure}

Figure~\ref{fig:counter_example} shows a surface plot of the function $W_{\text{fail}}$ in the $F_{11}$--$F_{22}$ plane, while Figure~\ref{fig:counter_example_hroc} shows the results of the HROC algorithm within the inner radius of the local minima.
Within this radius, the function value is lowered to approximately $200$.
However, the exact function value of the rank-one convex envelope is $0$ in the ball of radius $3$.

\begin{figure}[h]
    \ifthenelse{\boolean{professormode}}
    {\includegraphics{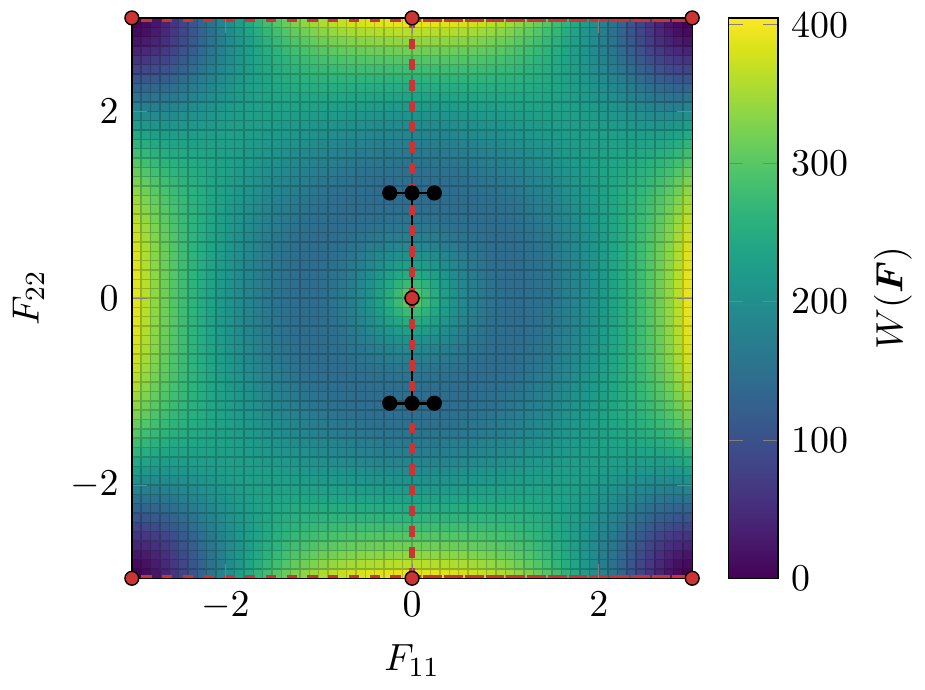}}
    {\input{figures/tikz/counter_example_contour.tex}} 
    \caption{Failure of the HROC algorithm for $W_{\text{fail}}$ as in \eqref{eq:Wcounter}.
    	$W_{\text{fail}}$ is plotted in the diagonal $F_{11}$--$F_{22}$ plane.
    	Evaluation of the HROC algorithm at $\hat{\boldsymbol{F}} = \boldsymbol{0}$. 
    	The computed hierarchical rank-one connected set of matrices is visualized by the black dots and connecting black lines.
		The HROC algorithm constructs a non-optimal $\mathcal{H}$-sequence.
		The optimal $\mathcal{H}$-sequence can be seen by the red dots with their associated rank-one connections depicted as red dashed lines.
    }
    \label{fig:counter_example_contour}
\end{figure}

The cause of this discrepancy is illustrated in Figure~\ref{fig:counter_example_contour}.
There, the computed hierarchical sequence is visualized by the directions as solid lines and matrices as dots.
Note that rank-one directions which point out of the $F_{11}$--$F_{22}$ plane are included in the discretisation set. 
However, the example is designed in such a way that the matrices involved in the hierarchical rank-one connected set are still diagonal matrices and hence lie in the illustrated $F_{11}$--$F_{22}$ plane.
Within the $F_{11}$--$F_{22}$ plane, there is no chance for the HROC algorithm to notice that a possible suboptimal laminate will yield the zero-valued envelope target value.
For this, the algorithm would need to extend the first hierarchical points to the very border (radius $3$) of the figure and thus, to choose a non-optimal intermediate laminate.
Instead, the algorithm chooses the level-wise optimal laminate, which is similar to the one computed in the Multiwell example of the previous section.
An optimal $\mathcal{H}$-sequence is shown in red, making it comprehensible that, for the rank-one convex envelope of $W_{\text{fail}}$, everything in the box of radius $3$ needs to be set to zero.

This illustrates that the algorithm has a local character and is not always able to incorporate the global behaviour of the function to be rank-one convexified, which has to be kept in mind when applying it.
A possible remedy for this issue could be to enforce more globality of the algorithm with the drawback of higher computational complexity, or allowing an extended set of directions to be checked for convexification.
The discrepancy could be omitted by introducing rank-two connections, which, however, would contradict the rank-one (only) connectivity and would lead to incompatible microstructures.

%% file: figures/tikz/KSD_convergence.tex
\tikzsetfigurename{KSD_convergence}
\begin{tikzpicture}
\begin{groupplot}[group style={group size={2 by 1}, horizontal sep={2cm}}, height={6cm}, width={6cm}]
    \nextgroupplot[axis background/.style={{fill={white!89.803921568!black}}}, x grid style={{white}}, y grid style={{white}}, xmajorgrids, ymajorgrids, ylabel={Error}, xlabel={\# Grid points $N$}, xmode={log}, ymode={log}]
    \addplot[mark={none}, thick, red!40!gray]
        table[row sep={\\}]
        {
            \\
            10.0  0.19544511501033235  \\
            50.0  0.01316501093370348  \\
            100.0  0.01316501093370348  \\
            300.0  0.005206886657101895  \\
            500.0  0.0024635218859133667  \\
            1000.0  0.001520443555545925  \\
            3000.0  0.0012923336402422825  \\
            5000.0  0.0011148003928612704  \\
            10000.0  0.0012020251919390157  \\
            50000.0  0.0011790213332263377  \\
        }
        ;
    \nextgroupplot[axis background/.style={{fill={white!89.803921568!black}}}, x grid style={{white}}, y grid style={{white}}, xmajorgrids, ymajorgrids, ylabel={time $(s)$}, xlabel={\# Grid points $N$}, xmode={log}, ymode={log}]
    \addplot[mark={none}, thick, red!40!gray,forget plot]
        table[row sep={\\}]
        {
            \\
            10.0  5.2490266666666675e-5  \\
            50.0  0.0006392171666666667  \\
            100.0  0.0009335751666666667  \\
            300.0  0.0031677162333333337  \\
            500.0  0.0095953706  \\
            1000.0  0.029288178166666668  \\
            3000.0  0.10561996876666667  \\
            5000.0  0.2204355212  \\
            10000.0  0.4407850014  \\
            50000.0  2.116016829033333  \\
        }
        ;
    \addplot[mark={none}, dotted, thick, black!40!gray]
        table[row sep={\\}]
        {
            \\
            10.0 5.2490266666e-5  \\
            50000  0.262451\\
        }
        ;
    \addlegendentry{$\mathcal{O}(N)$};
    \addplot[mark={none}, dashed, thick, black!40!gray]
        table[row sep={\\}]
        {
            \\
            10.0 5.2490266666e-5  \\
            50000  1312.256666\\
        }
        ;
    \addlegendentry{$\mathcal{O}(N^2)$};
\end{groupplot}
\end{tikzpicture}

%% file: figures/tikz/Multi_convergence.tex
\tikzsetfigurename{Multi_convergence}
\begin{tikzpicture}
\begin{groupplot}[group style={group size={2 by 1}, horizontal sep={2cm}}, height={6cm}, width={6cm}]
    \nextgroupplot[axis background/.style={{fill={white!89.803921568!black}}},
		x grid style={{white}},
		y grid style={{white}},
		xmajorgrids, ymajorgrids,
		ylabel={Error},
		xlabel={\# Grid points $N$},
		xmode={log}, ymode={log},]
    \addplot[mark={none}, thick, red!40!gray]
        table[row sep={\\}]
		{
			\\
			10.0  0.001539030917347257  \\
			50.0  0.0003324701827431277  \\
			100.0  0.0003324701827431277  \\
			300.0  0.0  \\
			500.0  1.5956796974109105e-6  \\
			1000.0  1.5956796974109105e-6  \\
			3000.0  0.0  \\
			5000.0  2.1903675826697366e-10  \\
		}
		;
    \nextgroupplot[axis background/.style={{fill={white!89.803921568!black}}}, 
    	x grid style={{white}}, 
    	y grid style={{white}}, 
    	xmajorgrids, ymajorgrids,     
    	ylabel={time $(s)$}, 
    	xlabel={\# Grid points $N$}, 
    	xmode={log}, ymode={log}]
    \addplot[mark={none}, thick, red!40!gray, forget plot]
        table[row sep={\\}]
		{
			\\
			10.0  0.00016805295000000002  \\
			50.0  8.900605000000001e-5  \\
			100.0  0.0001320029  \\
			300.0  0.000302451  \\
			500.0  0.00046628145  \\
			1000.0  0.0009068584000000001  \\
			3000.0  0.00274963005  \\
			5000.0  0.0104040059  \\
		}
		;
    \addplot[mark={none}, dotted, thick, black!40!gray]
        table[row sep={\\}]
        {
            \\
            10.0 0.000084026475  \\
            5000 0.0420132375 \\
        }
        ;
    \addlegendentry{$\mathcal{O}(N)$};    
\end{groupplot}
\end{tikzpicture}

%% file: sections/sec5.tex
%

\section{Application to Continuum Damage Mechanics}\label{sec:examples}
While benchmark problems are useful to test the performance of a single algorithm call, they often neglect real-world applications.
We test the novel algorithm for the finite-strain phenomenological continuum damage model also considered in \cite{BalOrt:2012:riv}.
As in all dissipative formulations, the incremental stress potential changes within each time increment, so that an efficient online convexification is crucial to the overall feasibility of a numerical simulation.

First, we briefly recall the damage model of \cite{BalOrt:2012:riv}. The foundation is the strain-energy density
\begin{equation}
    \psi(\boldsymbol{F},\alpha) = (1-D(\alpha))\,\psi^0(\boldsymbol{F}),
\end{equation}
where the internal variable is denoted by $\alpha$ and the non-decreasing damage function $D$ maps $\alpha$ to values in the range $[0,1)$. 
In this setting, $D$ taking the value $0$ corresponds to the intact state of the material while $D$ approaching $1$ the completely damaged state, where the value $1$ is usually omitted for numerical reasons. 
In the following experiments, the damage function
\begin{equation}
    D(\alpha) = D_\infty \left(1-\exp\left(-\frac{\alpha}{D_0}\right)\right)
\end{equation}
is used.
The effective strain energy density $\psi^0$ models the virtually undamaged response of the underlying material.
The thermodynamic force associated to $\alpha$ is $\beta = \psi^0(\boldsymbol{F})$ in case of damage evolution.
According to \cite{BalOrt:2012:riv}, the incremental stress potential reads
\begin{equation}
    W(\boldsymbol{F}) = \psi(\boldsymbol{F},\alpha) - \psi(\boldsymbol{F}_k,\alpha_k) + \alpha D - \alpha_k D_k - \overline{D} + \overline{D}_k.
\end{equation}
Note that the index of the current time step $(\cdot)_{k+1}$ is omitted if there is no potential of confusion.

\subsection{Material Point Experiments}

\begin{figure}[h]
    \begin{subfigure}{0.49\textwidth}
        \ifthenelse{\boolean{professormode}}
        {\includegraphics{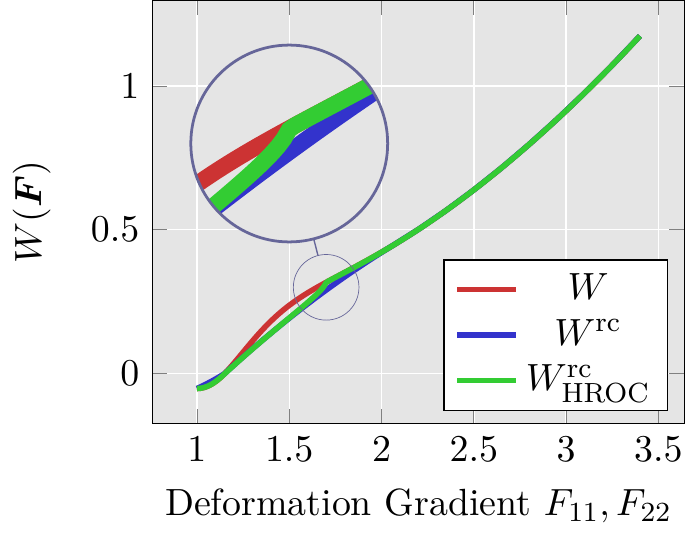}}
        {\input{figures/tikz/neohooke-energy-balt.tex}} 
        \caption{}
        \label{fig:energy}
    \end{subfigure}
    \begin{subfigure}{0.49\textwidth}
        \ifthenelse{\boolean{professormode}}
        {\includegraphics{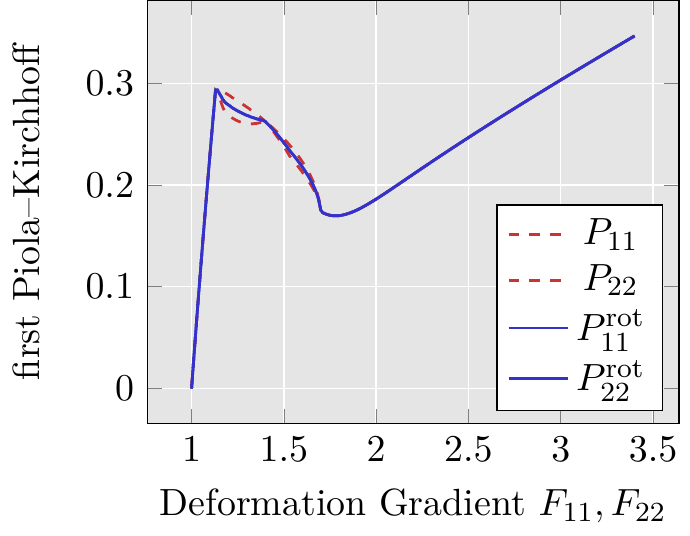}}
        {\input{figures/tikz/rotationaverage.tex}} 
        \caption{}
        \label{fig:energyderivative}
    \end{subfigure}
    \caption{Comparison of HROC algorithm with results from \cite{BalKohNeuPetPet:2023:mrc}.
             Figure~\ref{fig:energy} shows the incremental stress potential for a single increment along the biaxial path $F_{11}=F_{22}$ for $F_{12} = F_{21} = 0$.
             The novel algorithm matches the predicted rank-one convex hull in the non-convex regime except at $F_{11}=F_{22}\approx 1.75$.
             There, a small dip upwards can be seen.
             Note, however, that this relaxed regime is non-convex also for the results in \cite{BalKohNeuPetPet:2023:mrc} due to followed deformation path of $\text{rank}=2$.
             Figure~\ref{fig:energyderivative} shows the comparison of the rotationally averaged first derivative (Piola--Kirchhoff stresses) and the non-averaged ones.
             The rotational averaged stresses correspond to the finite difference derivative of the incremental stress potential curve and are therefore considered exact.}
    \label{fig:energycomparison}
\end{figure}

For a single time increment, i.e.~a fixed incremental stress potential, we compare the convexification via the HROC approach to the results of \cite{BalKohNeuPetPet:2023:mrc}, where an expensive algorithm for the full rank-one convexification is proposed.
The results of this algorithm are obtained with significantly increased computational effort compared to the here proposed HROC approach, however, they serve as reference here.
Figure~\ref{fig:energy} shows the computed hulls of $W$ along the diagonal line $F_{11} = F_{22}$ with $F_{12} = F_{21} = 0$, where we used the compressible Neo-Hookean effective strain energy density
\begin{equation}
    \psi^0_{\text{NH}1}(\boldsymbol{F}) = \frac{\mu}{2} (I_1 - 3) - \mu \ln(J) + \frac{\lambda}{2}\ln(J)^2,
\end{equation}
with the first invariant $I_1 = \text{tr}(\boldsymbol{F}^T\boldsymbol{F})$ and Jacobian $J = \det \boldsymbol{F}$.
The material parameters are set to $D_0 = 0.3,D_\infty = 0.9,\lambda=0.5,\mu=1.0$ and the internal variable value of the previous increment $\alpha_k$ is kept constant at $0.0625$.
Notably, the novel HROC algorithm approximates the rank-one convex envelope very well until the deformation gradient $F_{11}=F_{22}\approx 1.75$ is reached.
There, a small deviation can be noticed between the HROC algorithm and the result from \cite[Figure~11]{BalKohNeuPetPet:2023:mrc} plotted as $W^\text{rc}$ in Figure~\ref{fig:energycomparison}.

Figure~\ref{fig:energyderivative} shows the derivative obtained via \eqref{eq:deriv}.
At the beginning of the algorithm development, we observed that the derivative was oscillating and jumped between the values of the red curves $P_{11}$ and $P_{22}$ of Figure~\ref{fig:energyderivative}.
This behaviour can be omitted by enforcing laminate direction continuity over successive HROC calls.
Further, we noticed that, for this isotropic deformation behaviour and a given isotropic incremental stress potential, the obtained stress response was not isotropic.
This behaviour is a result of the loss of rotational invariance, since we are operating directly in the space of deformation gradients.

\begin{figure}[h]
    \begin{subfigure}{0.47\textwidth}
        \begin{center}
        	\includegraphics[width=0.7\textwidth]{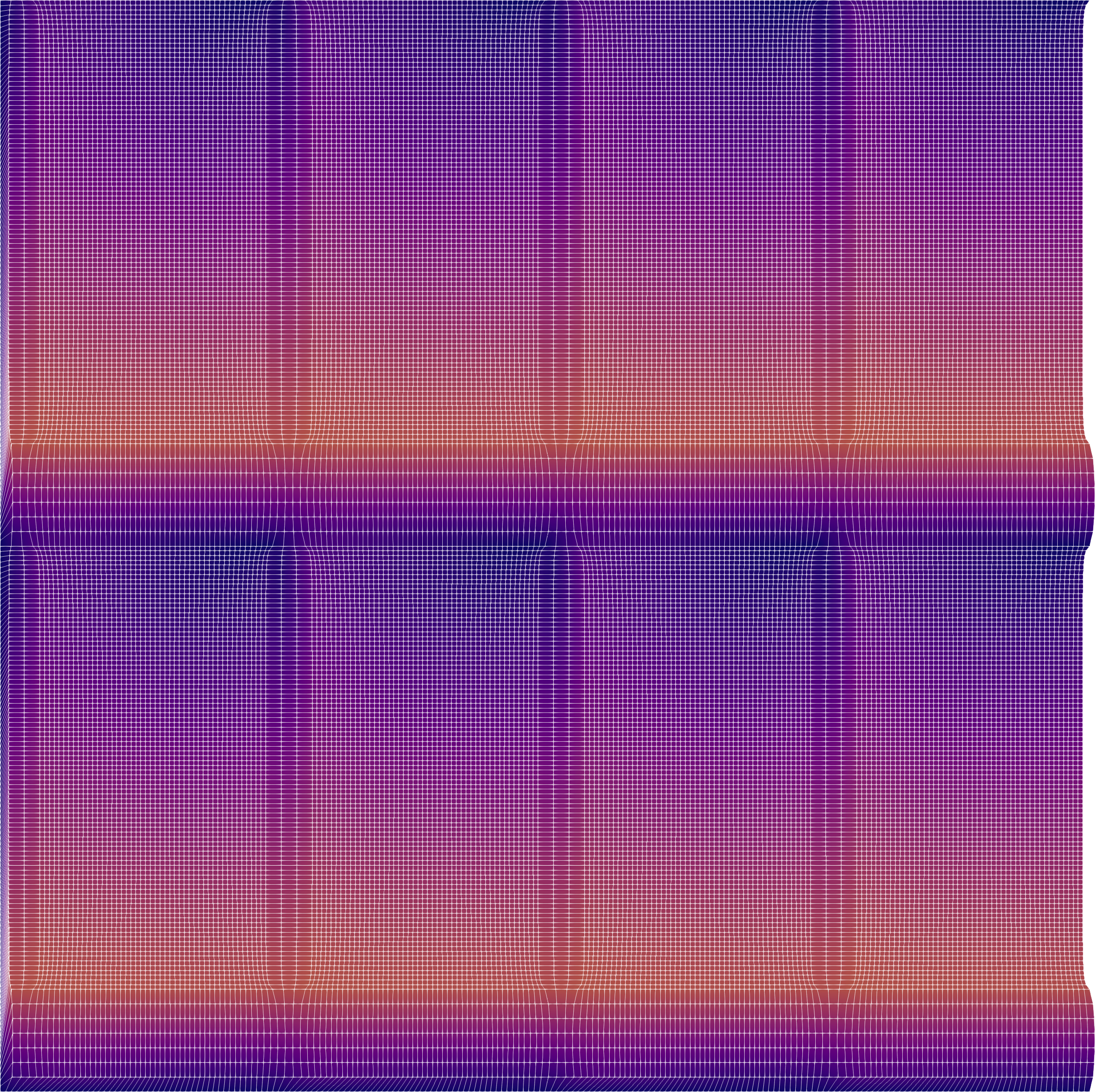}
        \end{center}
        \caption{Reconstructed microstructure at ${F_{11}=F_{22}=1.24}$.}
        \label{fig:micro1.24}
    \end{subfigure}
	\hfill
    \begin{subfigure}{0.47\textwidth}
    	\begin{center}
    		\includegraphics[width=0.7\textwidth]{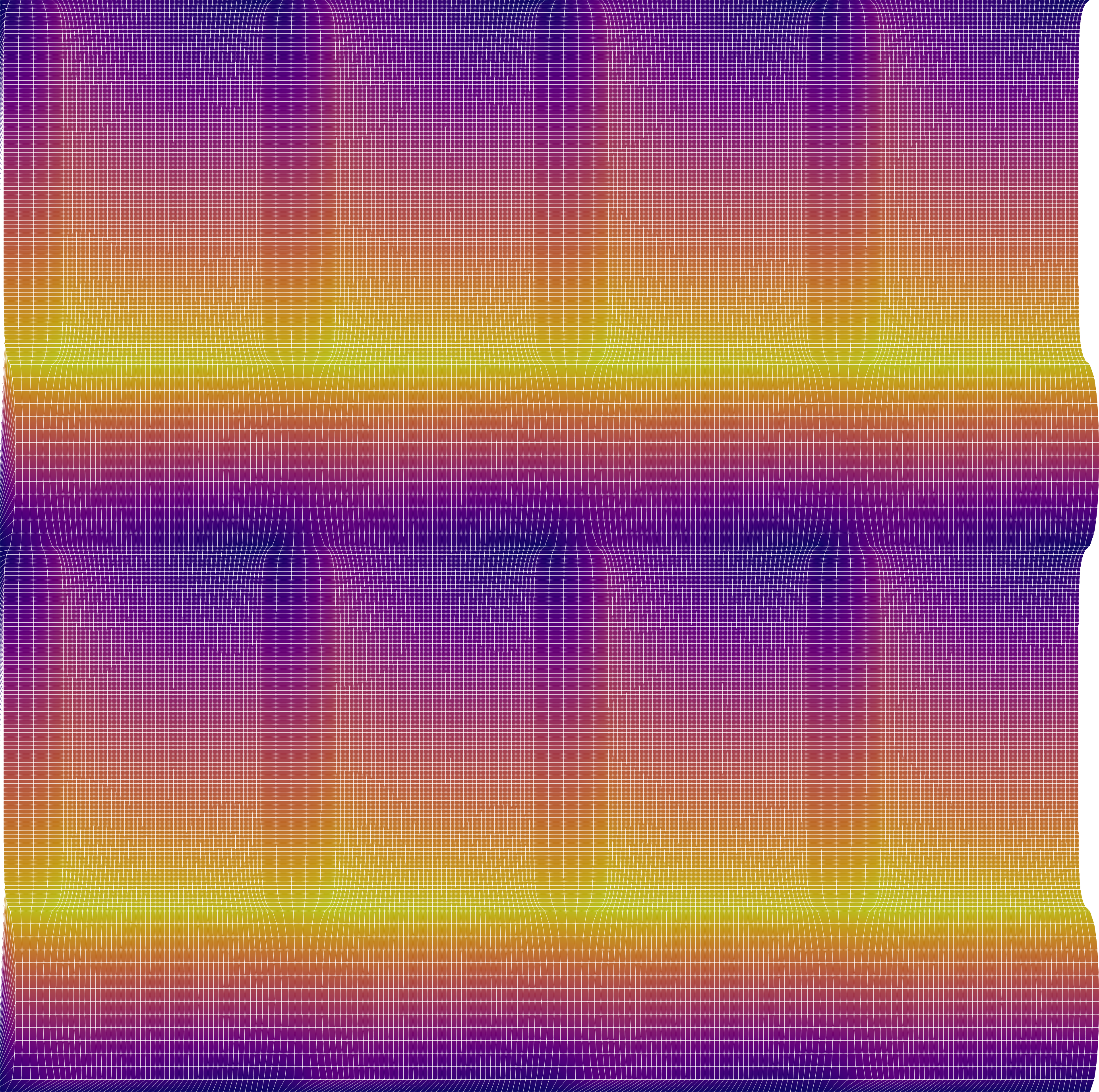}
    	\end{center}
        \caption{Reconstructed microstructure at ${F_{11}=F_{22}=1.34}$.}
        \label{fig:micro1.34}
    \end{subfigure}
    \caption{Continuous development of the described laminate visualized by a reconstructed displaced microstructure.
    		The microstructures correspond to positions on the macroscopic biaxial deformation path which is illustrated in Figure~\ref{fig:energycomparison}.}
    \label{fig:microstructure}
\end{figure}

From an energetic point of view, the resulting microstructure is equivalent to a microstructure that is rotated by $90^{\circ}$.
Due to isotropy, a laminate can overshoot in, e.g.~$x_2$-direction, while it undershoots the deformation in the $x_1$-direction, as visible for the prescribed macroscopic homogeneous biaxial deformation of $F_{11}=F_{22}=1.24$ in Figure~\ref{fig:micro1.24}.
Thus, an average over all possible rotations over the microstructure should result in the desired isotropic stress response.
In Figure~\ref{fig:energyderivative}, this procedure is applied for the curves $P_{11}^{\text{rot}}$ and $P_{22}^{\text{rot}}$ and it can be seen that this indeed restores the desired rotational invariance since the curves for $P_{11}^{\text{rot}}$ and $P_{22}^{\text{rot}}$ coincide.
While, at first sight, this seems like a disadvantage of isotropic materials, it is at the same time an opportunity due to the ambiguity of the constructed laminates and the implied possible $\mathcal{H}$-sequences that describe them.
It is notable that the obtained rotational average matches the finite differences of the green curve in Figure~\ref{fig:energy} exactly, thus validating the obtained procedure to take derivatives with the obtained $\mathcal{H}$-sequences.
Further, it becomes evident that the energy density feature of strain softening is not lost due to rank-one convexification.
On the contrary, it is preserved in a realistic manner, but only for deformation paths which differ by rank greater than one.
This is a distinct feature of semi-convexity.
If the direction set is enriched by directions with rank greater than one, this feature is lost and the classical relaxation plateau is obtained again.

\subsection{Two-dimensional Plate with a Hole}
\begin{figure}[h]
    \centering
    \sbox{\measurebox}{%
      \begin{minipage}[b]{.55\textwidth}
      \vspace{3em}
      \subfloat
        []{
            \ifthenelse{\boolean{professormode}}
            {\includegraphics[width=\textwidth]{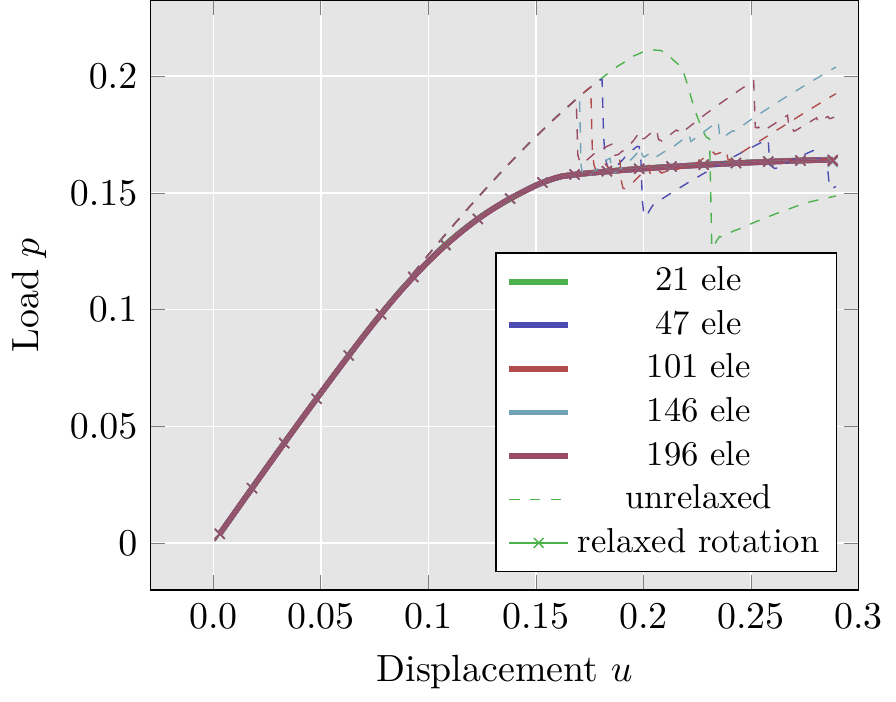}}
            {\input{figures/tikz/load-displacement_quarterplatehole.tex}} 
            \label{fig:load-displacement_quarterplatehole}
        }
      \end{minipage}}
    \usebox{\measurebox}\qquad
    \begin{minipage}[b][\ht\measurebox][s]{.3\textwidth}
    \subfloat
      []{
        \includegraphics[height=4.0cm]{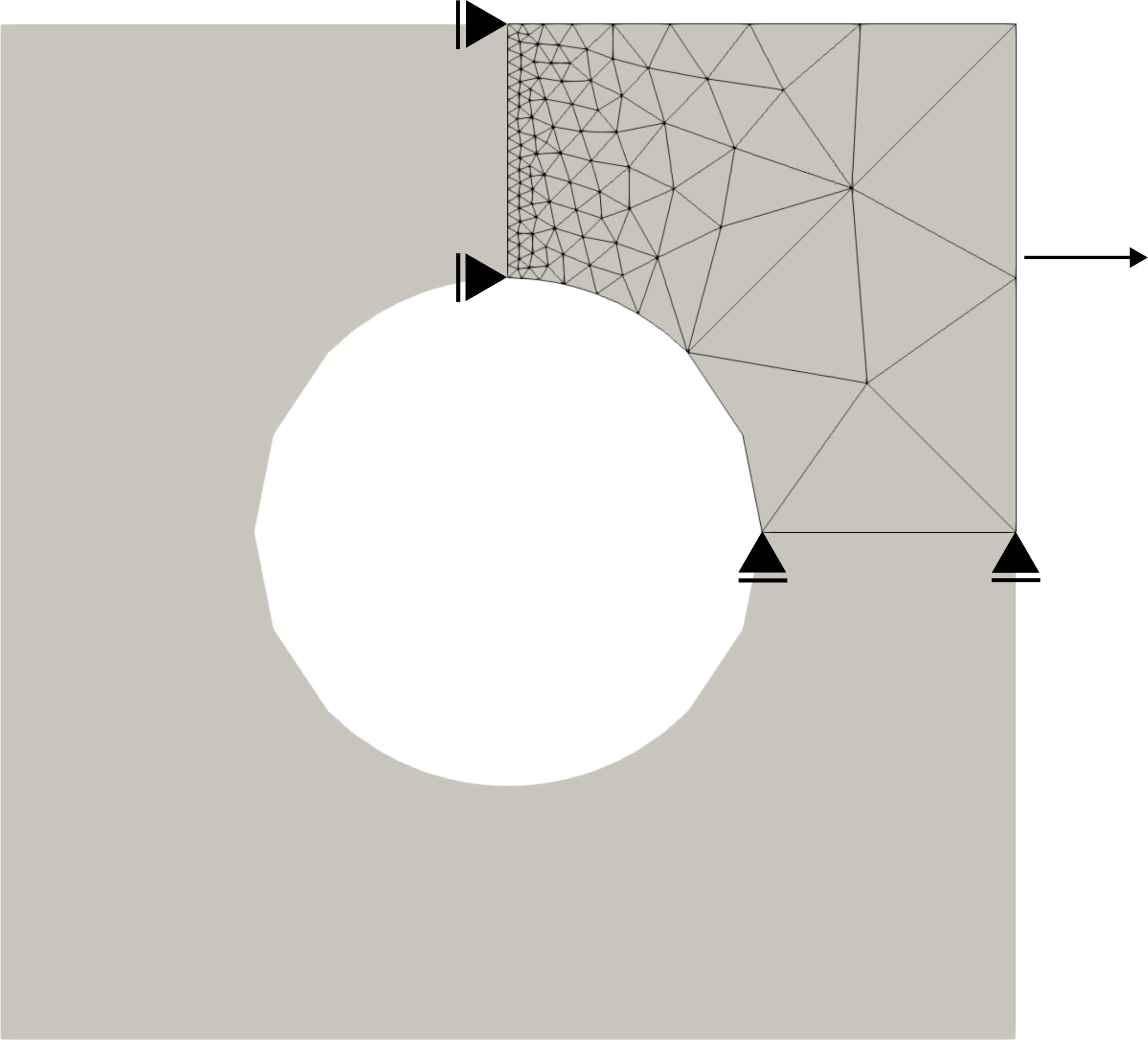}
        \label{fig:platehole-micro-bvp}
      }

    \subfloat
      []{
        \includegraphics[height=4.0cm]{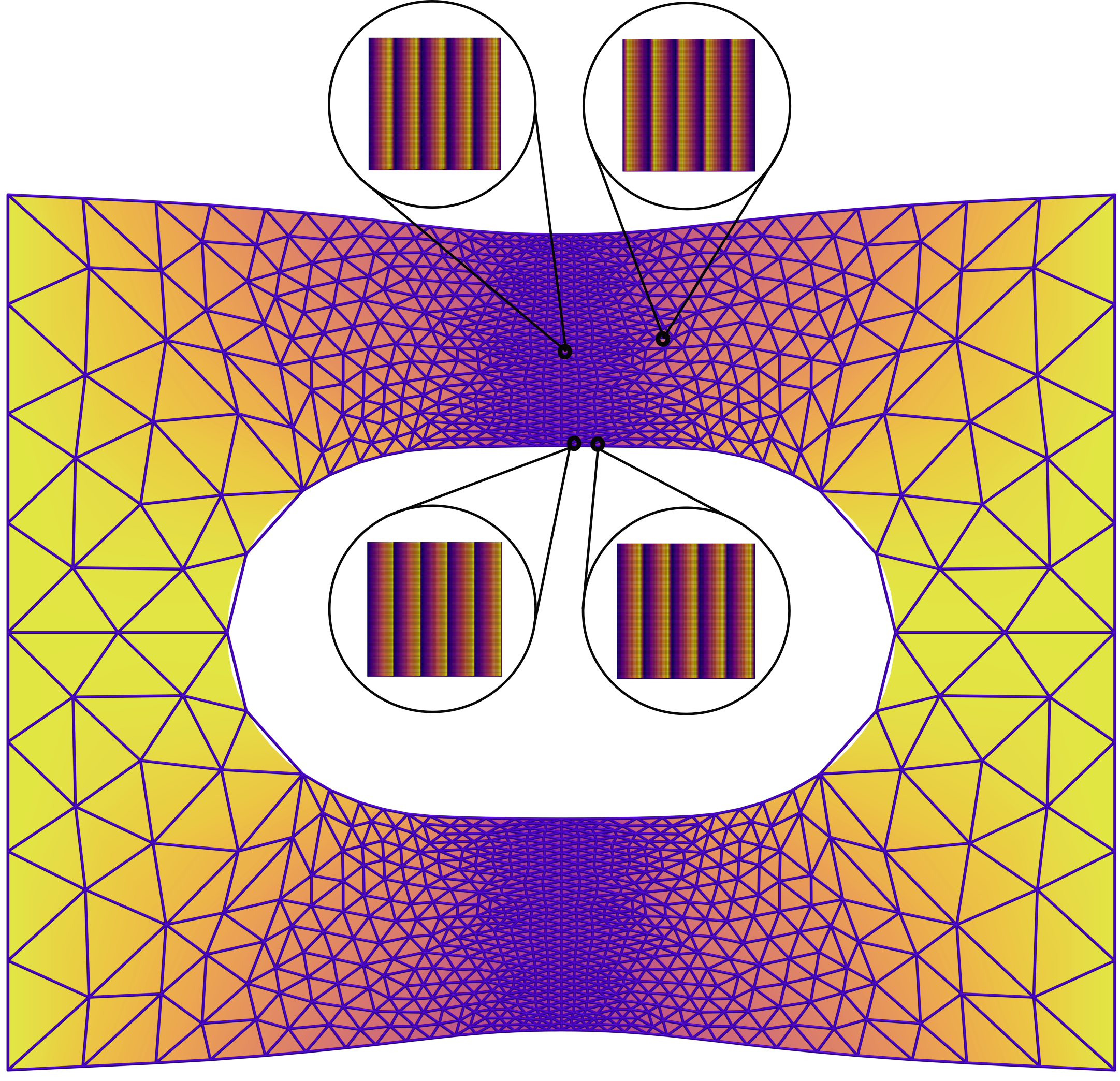}
        \label{fig:platehole-micro}
      }
    \end{minipage}
    \vspace{2em}
    \caption{Plate with a hole two-dimensional finite element problem with the isotropic continuum damage mechanics model.
    Left-hand side (A) shows the load--displacement diagram for the unrelaxed, relaxed and rotational averaged relaxed response.
    For the discretisation, quadratic triangles were used and the relaxed response converges quickly (all solid lines visually coincide) in contrast to the non-converging unrelaxed response.
    The boundary value problem and the associated finest mesh can be seen at the top right in subfigure (B).
    The bottom of the right-hand side (C) shows the obtained minimiser. The deformed mesh is coloured in the Euclidean norm of the deformation.
    At four Gauss points the laminate microstructures are illustrated.
    It can be seen that the laminates are continuous in terms of their direction and volume fraction of the phases.
    }
    \label{fig:platehole}
\end{figure}

The HROC algorithm is now tested for the two-dimensional plate with a hole benchmark problem.
Here, the $\psi^0_{\text{NH1}}$ effective strain energy density is used with the material parameters $D_\infty=0.9,D_0=0.3, {\mu = 0.9}, {\lambda = 0.4}$.
For the convexification, we used $N = 8000$ points along each rank-one convexified line as the discretisation parameter.
The symmetry of the problem is exploited and only a quarter is discretised.
Quadratic basis functions on a tetrahedral mesh are used for the discretisation of the domain.
In order to enforce symmetry, homogeneous Dirichlet boundary conditions are considered for the displacement component perpendicular to the symmetry line.
On the right-hand side of the domain, a heterogeneous Dirichlet boundary condition is used that ramps up linearly in the $x_1$-component to a maximum pull of 0.3 while the $x_2$-component is set to zero.
In Figure~\ref{fig:platehole-micro}, the deformed domain with the displacement norm as contour colour can be seen.
Further, the microstructures at four distinct Gauss points are shown as well.
Here, it can be seen that the laminate propagates continuously through the domain.
Figure~\ref{fig:load-displacement_quarterplatehole} shows the force--displacement diagram.
A strong mesh-dependent response of the unrelaxed formulation can be observed while the relaxed response is clearly mesh-insensitive.
Noteably, according to the force--displacement diagram, it does not matter if the rotational average is computed or not.
Both curves match almost identically.

\subsection{Three-dimensional Cube with a Hole}
We discuss a similar boundary value problem in three spatial dimensions.
Here, a standard algorithm rank-one convexification algorithm would be too computationally demanding due to the requirement to discretise the $d\times d$-space.
To illustrate this, consider a grid of deformation gradients in the nine-dimensional space.
Discretising each axis only by 10 points would already lead to a memory demand of 72 GBs if 64 bit floats are used.
Therefore, a pointwise algorithm in the sense of obtaining a single point of the approximated rank-one convex envelope is required.

The domain of the problem is the unitbox with a sphere cut out in the middle of the box.
Due to symmetry, we only discretise one eighth of the domain.
On each side along the $x_1$-axis, Dirichlet boundary conditions are applied to pull the box.
For the simulation, the following effective strain energy density is used
\begin{equation}
    \psi_{\text{NH2}}^0(\boldsymbol{F}) = C_1 (\overline{I} - 3) + ((C_1/6) + (D_1/4))(J^2 + (1/J^2) - 2),
\end{equation}
where ${C_1=\mu/2}, {D_1=\lambda/2}$, and $\overline{I}=J^{-2/3} I$.
The material parameters ${\mu=0.4}, {\lambda=0.1}$, ${D_{\infty} = 0.95},$ and ${D_0 = 0.1}$ are used.
For the convexification, we used $N=9000$ convexification points along each convexified rank-one line.
The domain is discretised by a tetrahedral mesh and quadratic basis functions are used.

\begin{figure}[h]
    \centering
    \sbox{\measurebox}{%
      \begin{minipage}[b]{.55\textwidth}
      \vspace{3em}
      \subfloat
        []{
        \ifthenelse{\boolean{professormode}}
        {\includegraphics{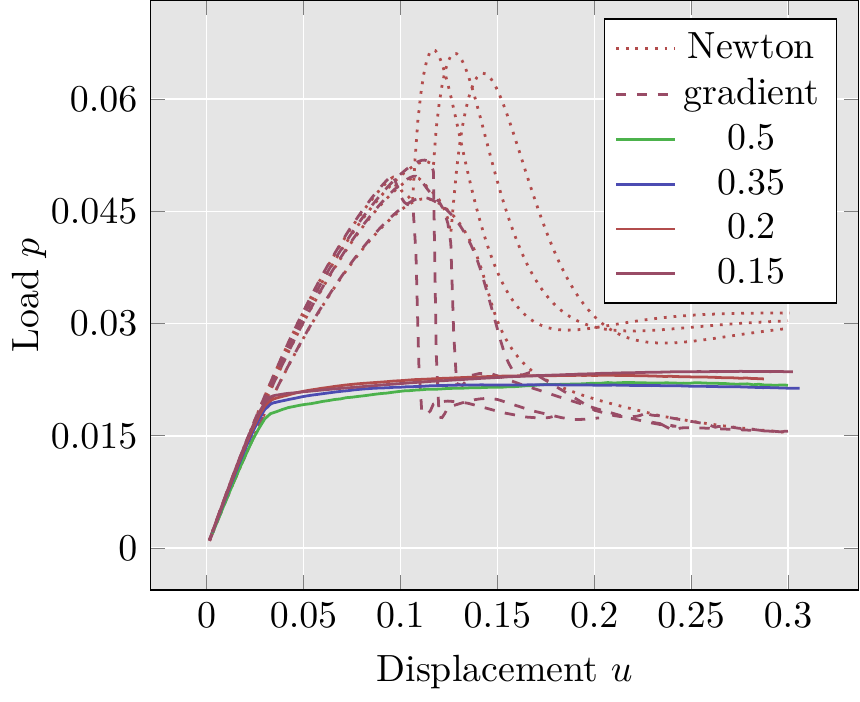}}
        {\input{figures/tikz/boxhole3d-load-displacement.tex}} 
        \label{fig:load-displacement_boxhole}
        }
      \end{minipage}}
    \usebox{\measurebox}\qquad
    \begin{minipage}[b][\ht\measurebox][s]{.3\textwidth}
    \subfloat
      []{
        \includegraphics[height=4.0cm]{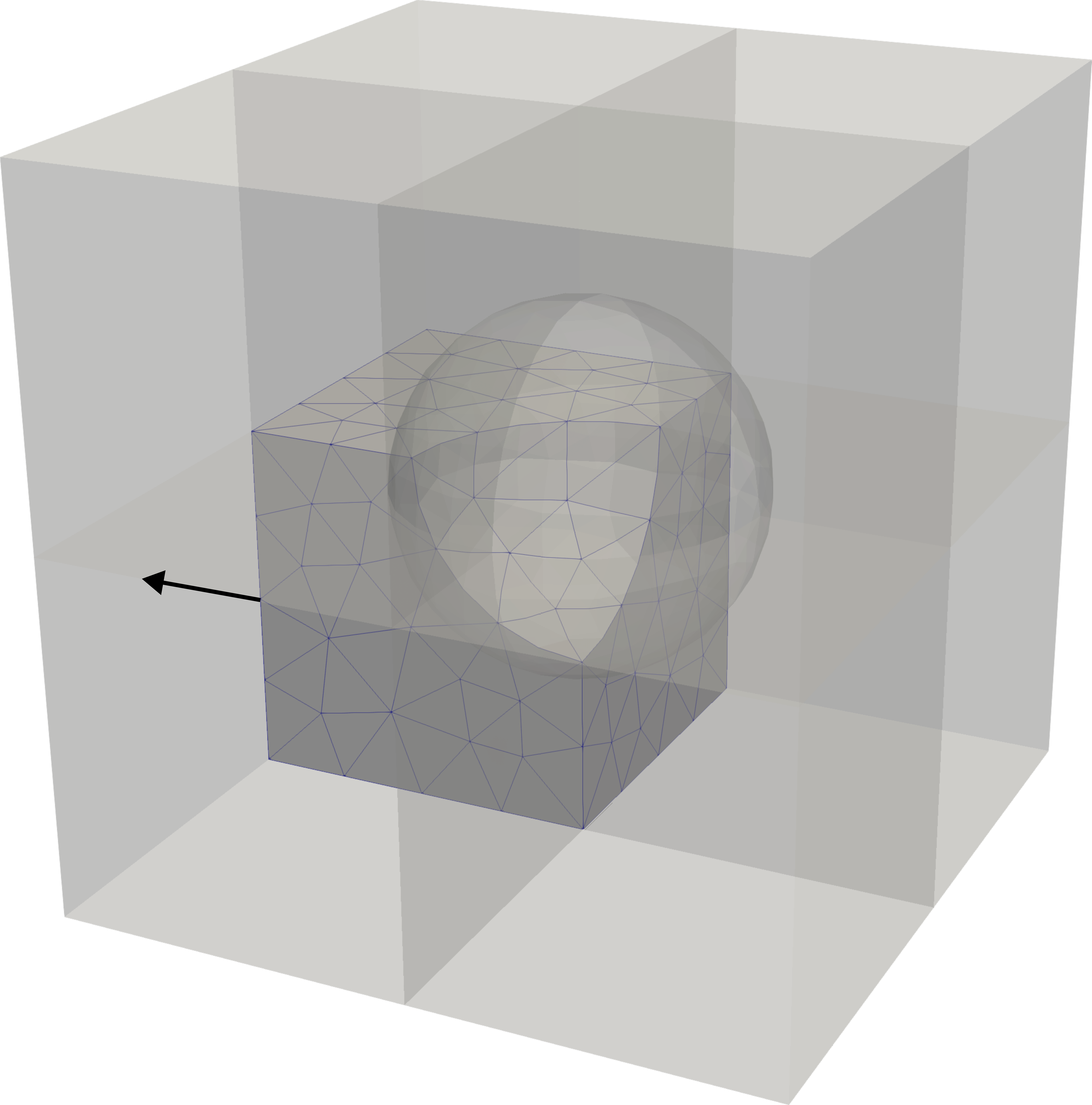}
        \label{fig:boxhole-micro}
      }

    \subfloat
      []{
        \includegraphics[width=0.9\textwidth]{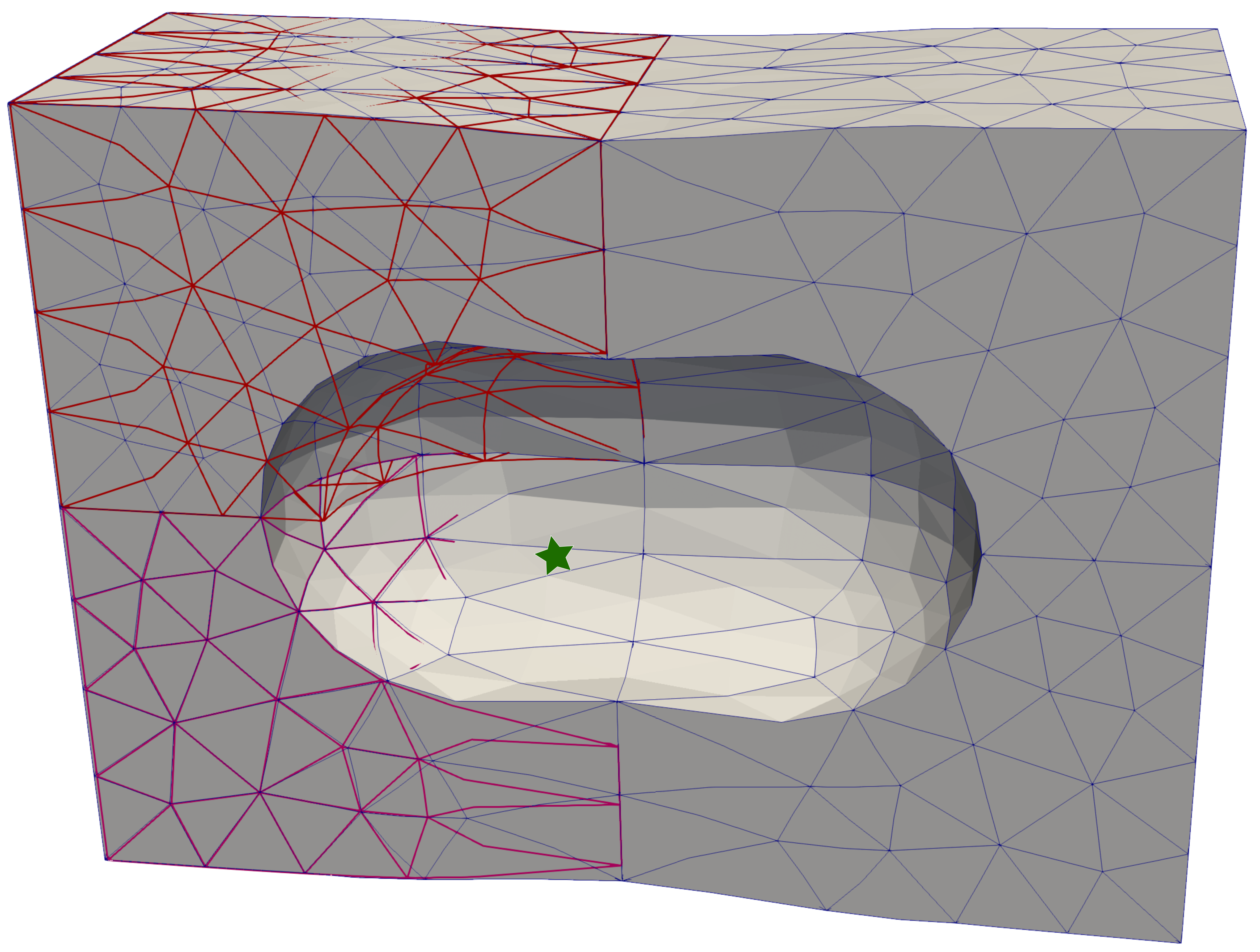}
        \label{fig:boxhole-minimizers}
      }
    \end{minipage}
    \caption{Three-dimensional finite element problem considering a pulled cube with a hole.
    The force--displacement diagram is shown on the left-hand side (A).
    The boundary value problem and its associated finest discretization is shown in (B).
    There, the discretized one eighth of the domain with quadratic tetrahedrons is visualized.
    At the bottom of the right-hand side subfigure (C) shows the three different obtained minimisers.
    Different minimisers for Newton and gradient descent optimization schemes are obtained for the unrelaxed response.
    Four different meshes are used.
    The numbers in the legend refer to the maximum edge length of the used quadratic tetrahedral elements.
    Subfigure (C) clearly shows the different responses obtained from a Newton (red wireframe) and gradient (purple wireframe) based finite element solver for the unrelaxed response.
    Both lead to a mesh-dependent, non-converging behaviour.
    Considering the relaxed response (blue wireframe), obtained by the HROC algorithm, a mostly converging response can be observed, although there is a slight deviation at the end of the numerical experiment.
    It is emphasized that, to the best of the authors' knowledge, thanks to the proposed HROC algorithm, this is the first time where the calculation of a relevant three-dimensional boundary value problem is shown for a fully three-dimensional, relaxed material model only using standard single workstation performance.}
    \label{fig:boxhole}
\end{figure}

In Figure~\ref{fig:load-displacement_boxhole}, the results in terms of the load--displacement diagram can be seen.
The unrelaxed response is calculated using a Newton and a steepest-descent scheme.
For the relaxed response, the Newton method was used.
From the load--displacement diagram, it is clear that the unrelaxed response is mesh-dependent.
On the contrary, the relaxed response shows the typical relaxation plateau.
Note that this full three-dimensional relaxed simulation is only possible due to the performance of the HROC algorithm.
Although, the curves at the end of the numerical experiment are not fully aligned, which may be the result of too coarse convexification and therefore a too poor rank-one envelope approximation, or a coarse finite element discretisation.
The discretisations used are the finest possible, assembled by multithreading assembly.
If a more refined discretisation is to be considered, distributed computing is required.
As the variation at the end is rather small and the curves at the beginning are almost identical and overlapping, this mainly seems to be a result of a too coarse convexification.

Figure~\ref{fig:boxhole-minimizers} shows the different minimisers obtained by the unrelaxed response and Newton's method (red wireframe), unrelaxed response and steepest descent method (purple wireframe), and relaxed response with Newton's method (blue wireframe).
The variation in the Newton's method and the steepest descent are likely due to the local character of Newton's method.
At the peak of the load--displacement curve, the Newton diverges and needs to be reset, thereby, finding a minimum where the elements at the pulled boundary localize further.
In contrast, the steepest descent method is more robust in terms of convergence and therefore stays in the local minima of localizing the elements that are near the cut-out sphere, due to the weakened cross section.

\begin{figure}[h]
    \begin{subfigure}{0.45\textwidth}
        \includegraphics[width=0.9\textwidth]{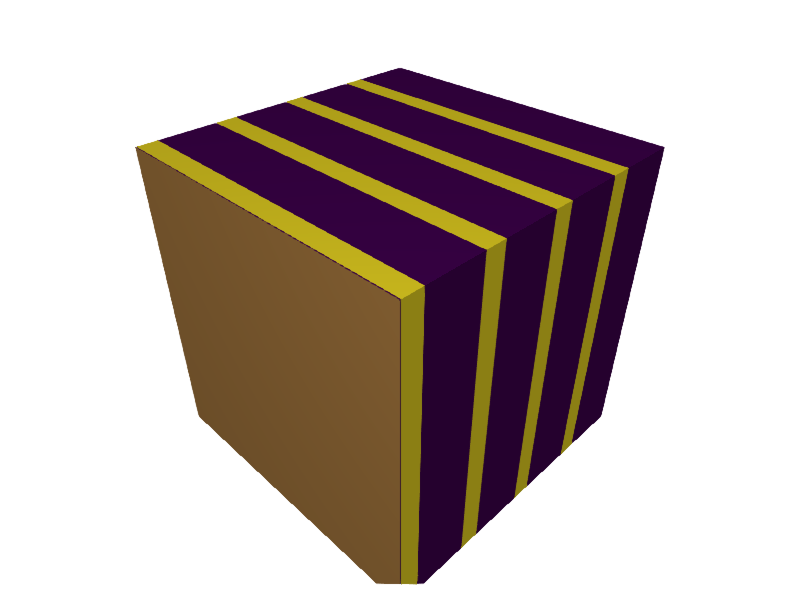}
        \caption{}
        \label{fig:t=50_ms}
    \end{subfigure}
    \begin{subfigure}{0.45\textwidth}
        \includegraphics[width=0.9\textwidth]{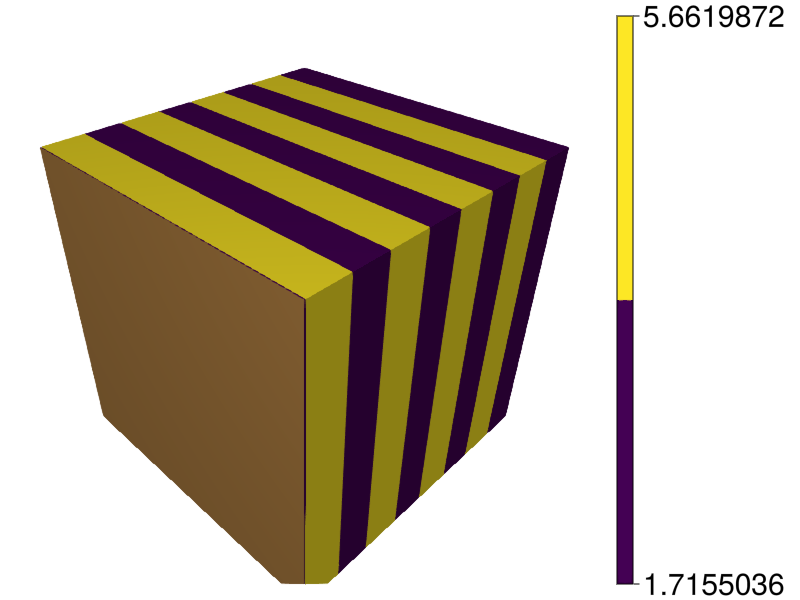}
        \caption{}
        \label{fig:t=100_ms}
    \end{subfigure}
    \caption{Reconstructed microstructure of the cube with a hole example at $t=50$ and $t=100$ at the 213th element marked with a green star in Figure~\ref{fig:boxhole-minimizers}.
             In both Figures~\ref{fig:t=50_ms} and \ref{fig:t=100_ms}, the Frobenius norm of the laminate's gradient is shown in colour.
             Notably, the norm of the phase gradient stays almost identical over the simulation, despite the changing components that relate to the shearing of the element.
             A continuous development of the laminate can be observed.}
    \label{fig:boxhole-ms}
\end{figure}

Figure~\ref{fig:boxhole-ms} shows the continuous evolution of the microstructure over time for an element marked in Figure~\ref{fig:boxhole-minimizers}. 
For the illustrated laminate of order one, the two phases are coloured according to the norms of the associated deformation gradients.
It can be observed that only the volume fractions of the two phases change over time while the phases remain unchanged.

%% file: figures/tikz/neohooke-energy-balt.tex
\tikzsetfigurename{neohooke-energy-balt}
\begin{tikzpicture}[spy using outlines=
	{circle, magnification=3, connect spies}]
\begin{axis}[axis background/.style={{fill={white!89.803921568!black}}},
x grid style={{white}},
y grid style={{white}},
xtick={1,1.5,2,2.5,3,3.5},
xmajorgrids,
ymajorgrids,
ylabel={$W(\boldsymbol{F})$},
xlabel={Deformation Gradient $F_{11}, F_{22}$},
legend pos=south east,
width=0.95\textwidth,
height=0.8\textwidth,
]
    \addplot[mark={none}, ultra thick, red!60!gray]
        table[row sep={\\}]
        {
            \\
            1.0  -0.0519271582295569  \\
            1.01  -0.051679314466603066  \\
            1.02  -0.050941182917584926  \\
            1.03  -0.049720662027090246  \\
            1.04  -0.0480253910311052  \\
            1.05  -0.04586276074388651  \\
            1.06  -0.04323992380095504  \\
            1.07  -0.04016380439037848  \\
            1.08  -0.036641107502330406  \\
            1.09  -0.032678327724901535  \\
            1.1  -0.02828175761227547  \\
            1.11  -0.023457495649661486  \\
            1.12  -0.018211453837780944  \\
            1.13  -0.012549364918229278  \\
            1.14  -0.006476789259662719  \\
            1.15  8.785751429152278e-7  \\
            1.16  0.0067956706404064815  \\
            1.17  0.013803915922745491  \\
            1.18  0.020997137486839457  \\
            1.19  0.02834753379767363  \\
            1.2  0.035828120857679946  \\
            1.21  0.043412859084377836  \\
            1.22  0.05107676476129769  \\
            1.23  0.05879600607489313  \\
            1.24  0.06654798391541217  \\
            1.25  0.07431139776938114  \\
            1.26  0.08206629716480332  \\
            1.27  0.08979411924707098  \\
            1.28  0.0974777131638663  \\
            1.29  0.10510135202113483  \\
            1.3  0.11265073323999095  \\
            1.31  0.12011296819669609  \\
            1.32  0.12747656206540567  \\
            1.33  0.1347313848070838  \\
            1.34  0.14186863425881135  \\
            1.35  0.14888079227673948  \\
            1.36  0.15576157487426728  \\
            1.37  0.16250587727584576  \\
            1.38  0.16910971477726072  \\
            1.39  0.1755701602665064  \\
            1.4  0.18188527921659914  \\
            1.41  0.18805406291395782  \\
            1.42  0.1940763606344097  \\
            1.43  0.1999528114244269  \\
            1.44  0.20568477608884228  \\
            1.45  0.2112742699288503  \\
            1.46  0.21672389671641945  \\
            1.47  0.22203678433397356  \\
            1.48  0.22721652245202573  \\
            1.49  0.23226710256288652  \\
            1.5  0.2371928606360949  \\
            1.51  0.24199842261123916  \\
            1.52  0.24668865289664688  \\
            1.53  0.2512686059982969  \\
            1.54  0.25574348136240244  \\
            1.55  0.26011858147758327  \\
            1.56  0.2643992732484216  \\
            1.57  0.26859095262153176  \\
            1.58  0.27269901241800976  \\
            1.59  0.2767288133022221  \\
            1.6  0.280685657796236  \\
            1.61  0.28457476723164465  \\
            1.62  0.28840126151598544  \\
            1.63  0.292170141579161  \\
            1.64  0.29588627435613046  \\
            1.65  0.2995543801553937  \\
            1.66  0.30317902225827975  \\
            1.67  0.3067645985915445  \\
            1.68  0.31031533531509914  \\
            1.69  0.31383528216760836  \\
            1.7  0.3173283094150309  \\
            1.71  0.3207981062507427  \\
            1.72  0.32424818050047455  \\
            1.73  0.32768185949078277  \\
            1.74  0.33110229194594554  \\
            1.75  0.33451245078493386  \\
            1.76  0.3379151366972658  \\
            1.77  0.34131298238402874  \\
            1.78  0.3447084573579938  \\
            1.79  0.3481038732044893  \\
            1.8  0.35150138921241203  \\
            1.81  0.3549030182923988  \\
            1.82  0.35831063310666156  \\
            1.83  0.36172597234224846  \\
            1.84  0.36515064706651945  \\
            1.85  0.36858614711030857  \\
            1.86  0.37203384743065315  \\
            1.87  0.37549501441097  \\
            1.88  0.37897081206223227  \\
            1.89  0.38246230809397563  \\
            1.9  0.3859704798288536  \\
            1.91  0.38949621993897654  \\
            1.92  0.3930403419864078  \\
            1.93  0.3966035857539435  \\
            1.94  0.4001866223557159  \\
            1.95  0.4037900591202118  \\
            1.96  0.4074144442410309  \\
            1.97  0.4110602711931156  \\
            1.98  0.4147279829143077  \\
            1.99  0.41841797575392403  \\
            2.0  0.4221306031916311  \\
            2.01  0.42586617933123894  \\
            2.02  0.4296249821751609  \\
            2.03  0.43340725668620805  \\
            2.04  0.4372132176441478  \\
            2.05  0.44104305230499896  \\
            2.06  0.4448969228715035  \\
            2.07  0.4487749687835011  \\
            2.08  0.4526773088371029  \\
            2.09  0.45660404314167946  \\
            2.1  0.4605552549236237  \\
            2.11  0.4645310121858318  \\
            2.12  0.4685313692316341  \\
            2.13  0.4725563680617614  \\
            2.14  0.4766060396526627  \\
            2.15  0.4806804051242203  \\
            2.16  0.48477947680460864  \\
            2.17  0.4889032591996866  \\
            2.18  0.49305174987402844  \\
            2.19  0.4972249402502683  \\
            2.2  0.5014228163331572  \\
            2.21  0.5056453593643114  \\
            2.22  0.5098925464133027  \\
            2.23  0.5141643509103958  \\
            2.24  0.5184607431258764  \\
            2.25  0.5227816906005981  \\
            2.26  0.527127158532044  \\
            2.27  0.5314971101198996  \\
            2.28  0.5358915068748232  \\
            2.29  0.5403103088938382  \\
            2.3  0.5447534751054873  \\
            2.31  0.549220963487646  \\
            2.32  0.553712731260659  \\
            2.33  0.5582287350582184  \\
            2.34  0.5627689310782354  \\
            2.35  0.5673332752157084  \\
            2.36  0.5719217231794657  \\
            2.37  0.5765342305944462  \\
            2.38  0.5811707530910585  \\
            2.39  0.5858312463829946  \\
            2.4  0.5905156663347609  \\
            2.41  0.5952239690200453  \\
            2.42  0.599956110771949  \\
            2.43  0.6047120482260054  \\
            2.44  0.6094917383567989  \\
            2.45  0.614295138508928  \\
            2.46  0.6191222064229928  \\
            2.47  0.6239729002571742  \\
            2.48  0.6288471786049524  \\
            2.49  0.633745000509437  \\
            2.5  0.6386663254747237  \\
            2.51  0.643611113474653  \\
            2.52  0.6485793249593116  \\
            2.53  0.6535709208595603  \\
            2.54  0.6585858625898581  \\
            2.55  0.6636241120496086  \\
            2.56  0.6686856316232367  \\
            2.57  0.6737703841791619  \\
            2.58  0.6788783330678564  \\
            2.59  0.6840094421190862  \\
            2.6  0.6891636756384875  \\
            2.61  0.6943409984035774  \\
            2.62  0.6995413756592871  \\
            2.63  0.7047647731130876  \\
            2.64  0.7100111569298265  \\
            2.65  0.7152804937262646  \\
            2.66  0.7205727505654277  \\
            2.67  0.7258878949507949  \\
            2.68  0.7312258948203526  \\
            2.69  0.736586718540574  \\
            2.7  0.7419703349003391  \\
            2.71  0.747376713104804  \\
            2.72  0.7528058227692916  \\
            2.73  0.7582576339131512  \\
            2.74  0.7637321169536756  \\
            2.75  0.7692292427000215  \\
            2.76  0.7747489823472135  \\
            2.77  0.7802913074701573  \\
            2.78  0.7858561900177617  \\
            2.79  0.791443602307098  \\
            2.8  0.7970535170176513  \\
            2.81  0.8026859071856366  \\
            2.82  0.8083407461984059  \\
            2.83  0.8140180077889422  \\
            2.84  0.8197176660304212  \\
            2.85  0.8254396953308752  \\
            2.86  0.8311840704279468  \\
            2.87  0.8369507663837069  \\
            2.88  0.8427397585795905  \\
            2.89  0.8485510227113972  \\
            2.9  0.8543845347843882  \\
            2.91  0.8602402711084695  \\
            2.92  0.8661182082934452  \\
            2.93  0.8720183232443878  \\
            2.94  0.8779405931570489  \\
            2.95  0.8838849955133886  \\
            2.96  0.8898515080771613  \\
            2.97  0.895840108889597  \\
            2.98  0.9018507762651419  \\
            2.99  0.9078834887872972  \\
            3.0  0.9139382253045149  \\
            3.01  0.9200149649261785  \\
            3.02  0.9261136870186485  \\
            3.03  0.9322343712013849  \\
            3.04  0.9383769973431381  \\
            3.05  0.9445415455582035  \\
            3.06  0.9507279962027425  \\
            3.07  0.9569363298711847  \\
            3.08  0.9631665273926678  \\
            3.09  0.9694185698275714  \\
            3.1  0.9756924384640774  \\
            3.11  0.9819881148148232  \\
            3.12  0.9883055806135962  \\
            3.13  0.9946448178120922  \\
            3.14  1.0010058085767244  \\
            3.15  1.007388535285497  \\
            3.16  1.0137929805249315  \\
            3.17  1.0202191270870355  \\
            3.18  1.026666957966341  \\
            3.19  1.0331364563569845  \\
            3.2  1.0396276056498324  \\
            3.21  1.0461403894296688  \\
            3.22  1.0526747914724246  \\
            3.23  1.059230795742447  \\
            3.24  1.0658083863898322  \\
            3.25  1.0724075477477846  \\
            3.26  1.0790282643300375  \\
            3.27  1.0856705208283093  \\
            3.28  1.0923343021097947  \\
            3.29  1.0990195932147189  \\
            3.3  1.1057263793539103  \\
            3.31  1.1124546459064315  \\
            3.32  1.1192043784172383  \\
            3.33  1.125975562594879  \\
            3.34  1.1327681843092399  \\
            3.35  1.1395822295893185  \\
            3.36  1.1464176846210417  \\
            3.37  1.1532745357451097  \\
            3.38  1.1601527694548852  \\
            3.39  1.167052372394307  \\
            3.4  1.1739733313558616  \\
        }
        ;
    \addlegendentry {$W$}
    \addplot[mark={none}, ultra thick, blue!60!gray]
        table[row sep={\\}]
        {
            \\
            1.0  -0.0519271582295569  \\
            1.01  -0.04897012374858168  \\
            1.02  -0.04594097003341536  \\
            1.03  -0.04283969708405787  \\
            1.04  -0.03966630490050919  \\
            1.05  -0.03642079348276941  \\
            1.06  -0.033103162830838466  \\
            1.07  -0.029713412944716324  \\
            1.08  -0.026251543824403083  \\
            1.09  -0.022717555469898686  \\
            1.1  -0.01911144788120314  \\
            1.11  -0.015433221058316435  \\
            1.12  -0.011682875001238532  \\
            1.13  -0.007860409709969627  \\
            1.14  -0.003965825184509419  \\
            1.15  8.785751418904551e-7  \\
            1.16  0.005865513747558595  \\
            1.17  0.011708274852214415  \\
            1.18  0.01752916188910941  \\
            1.19  0.023328174858243567  \\
            1.2  0.029105313759616898  \\
            1.21  0.0348605785932294  \\
            1.22  0.04059396935908107  \\
            1.23  0.046305486057172066  \\
            1.24  0.05199512868750207  \\
            1.25  0.05766289725007125  \\
            1.26  0.06330879174487976  \\
            1.27  0.06893281217192727  \\
            1.28  0.07453495853121397  \\
            1.29  0.08011523082273983  \\
            1.3  0.08567362904650486  \\
            1.31  0.09121028005637369  \\
            1.32  0.09672531070621099  \\
            1.33  0.10221872099601688  \\
            1.34  0.1076905109257911  \\
            1.35  0.11314068049553377  \\
            1.36  0.11856922970524487  \\
            1.37  0.12397615855492447  \\
            1.38  0.1293614670445725  \\
            1.39  0.13472515517418898  \\
            1.4  0.14006722294377394  \\
            1.41  0.14538767035332734  \\
            1.42  0.1506864974028492  \\
            1.43  0.15596370409233953  \\
            1.44  0.16121929042179847  \\
            1.45  0.16645325639122568  \\
            1.46  0.17166560200062136  \\
            1.47  0.17685632724998546  \\
            1.48  0.18202543213931804  \\
            1.49  0.1871729166686191  \\
            1.5  0.1922987808378887  \\
            1.51  0.1974030246471266  \\
            1.52  0.20248564809633302  \\
            1.53  0.20754665118550786  \\
            1.54  0.21258603391465114  \\
            1.55  0.21760379628376295  \\
            1.56  0.22259993829284325  \\
            1.57  0.22757445994189193  \\
            1.58  0.23252736123090906  \\
            1.59  0.23745864215989465  \\
            1.6  0.24236830272884866  \\
            1.61  0.24725638498334246  \\
            1.62  0.25212293096894745  \\
            1.63  0.25696794068566325  \\
            1.64  0.2617914141334903  \\
            1.65  0.2665933513124284  \\
            1.66  0.2713737522224776  \\
            1.67  0.27613261686363805  \\
            1.68  0.2808699452359094  \\
            1.69  0.2855857373392919  \\
            1.7  0.2902799931737854  \\
            1.71  0.29495271273939005  \\
            1.72  0.2996038960361057  \\
            1.73  0.3042335430639327  \\
            1.74  0.30884165382287054  \\
            1.75  0.3134282283129195  \\
            1.76  0.3179932985970365  \\
            1.77  0.3225368967381788  \\
            1.78  0.3270590227363463  \\
            1.79  0.3315596765915392  \\
            1.8  0.33603885830375707  \\
            1.81  0.3404965678730003  \\
            1.82  0.3449328052992687  \\
            1.83  0.3493475705825623  \\
            1.84  0.35374086372288105  \\
            1.85  0.3581126847202252  \\
            1.86  0.3624630335745945  \\
            1.87  0.366791910285989  \\
            1.88  0.37109931485440856  \\
            1.89  0.37538524727985345  \\
            1.9  0.37964970756232375  \\
            1.91  0.3838927094692083  \\
            1.92  0.38811426676789673  \\
            1.93  0.3923143794583891  \\
            1.94  0.3964930475406852  \\
            1.95  0.40065027101478523  \\
            1.96  0.4047860498806891  \\
            1.97  0.40890038413839686  \\
            1.98  0.4129932737879084  \\
            1.99  0.4170647188292238  \\
            2.0  0.4211147192623431  \\
            2.01  0.4251432750872662  \\
            2.02  0.42915038630399327  \\
            2.03  0.43313605291252394  \\
            2.04  0.43710027491285874  \\
            2.05  0.4410430523049972  \\
            2.06  0.4450561827438591  \\
            2.07  0.44907105420600885  \\
            2.08  0.4530876666914466  \\
            2.09  0.4571060202001721  \\
            2.1  0.4611261147321858  \\
            2.11  0.4651479502874871  \\
            2.12  0.46917152686607655  \\
            2.13  0.4731968444679539  \\
            2.14  0.47722390309311924  \\
            2.15  0.4812527027415723  \\
            2.16  0.4852832434133135  \\
            2.17  0.4893155251083424  \\
            2.18  0.49334954782665946  \\
            2.19  0.49738531156826443  \\
            2.2  0.5014228163331574  \\
            2.21  0.5058038070468778  \\
            2.22  0.5101866606003769  \\
            2.23  0.5145713769936543  \\
            2.24  0.5189579562267104  \\
            2.25  0.523346398299545  \\
            2.26  0.5277367032121579  \\
            2.27  0.5321288709645495  \\
            2.28  0.5365229015567194  \\
            2.29  0.5409187949886682  \\
            2.3  0.5453165512603952  \\
            2.31  0.549716170371901  \\
            2.32  0.554117652323185  \\
            2.33  0.5585209971142476  \\
            2.34  0.5629262047450886  \\
            2.35  0.5673332752157084  \\
            2.36  0.5720769607914304  \\
            2.37  0.5768223393826113  \\
            2.38  0.5815694109892512  \\
            2.39  0.5863181756113498  \\
            2.4  0.5910686332489075  \\
            2.41  0.5958207839019241  \\
            2.42  0.6005746275703996  \\
            2.43  0.6053301642543341  \\
            2.44  0.6100873939537274  \\
            2.45  0.6148463166685794  \\
            2.46  0.6196069323988904  \\
            2.47  0.6243692411446604  \\
            2.48  0.6291332429058893  \\
            2.49  0.6338989376825773  \\
            2.5  0.6386663254747237  \\
            2.51  0.6437633796314342  \\
            2.52  0.6488619419395817  \\
            2.53  0.6539620123991661  \\
            2.54  0.6590635910101875  \\
            2.55  0.6641666777726459  \\
            2.56  0.6692712726865415  \\
            2.57  0.6743773757518736  \\
            2.58  0.6794849869686433  \\
            2.59  0.6845941063368494  \\
            2.6  0.689704733856493  \\
            2.61  0.6948168695275733  \\
            2.62  0.6999305133500908  \\
            2.63  0.705045665324045  \\
            2.64  0.7101623254494362  \\
            2.65  0.7152804937262646  \\
            2.66  0.7207225975293886  \\
            2.67  0.7261660486300794  \\
            2.68  0.7316108470283376  \\
            2.69  0.7370569927241621  \\
            2.7  0.742504485717554  \\
            2.71  0.7479533260085123  \\
            2.72  0.753403513597038  \\
            2.73  0.7588550484831305  \\
            2.74  0.7643079306667899  \\
            2.75  0.7697621601480162  \\
            2.76  0.775217736926809  \\
            2.77  0.7806746610031694  \\
            2.78  0.7861329323770964  \\
            2.79  0.7915925510485904  \\
            2.8  0.7970535170176513  \\
            2.81  0.8028338096317982  \\
            2.82  0.8086153126200738  \\
            2.83  0.8143980259824785  \\
            2.84  0.820181949719012  \\
            2.85  0.8259670838296749  \\
            2.86  0.8317534283144659  \\
            2.87  0.8375409831733864  \\
            2.88  0.8433297484064353  \\
            2.89  0.8491197240136135  \\
            2.9  0.8549109099949206  \\
            2.91  0.8607033063503564  \\
            2.92  0.866496913079921  \\
            2.93  0.8722917301836147  \\
            2.94  0.8780877576614372  \\
            2.95  0.8838849955133886  \\
            2.96  0.8899978388249895  \\
            2.97  0.8961117754058924  \\
            2.98  0.9022268052560962  \\
            2.99  0.9083429283756025  \\
            3.0  0.9144601447644098  \\
            3.01  0.920578454422519  \\
            3.02  0.92669785734993  \\
            3.03  0.9328183535466422  \\
            3.04  0.9389399430126568  \\
            3.05  0.9450626257479724  \\
            3.06  0.9511864017525904  \\
            3.07  0.9573112710265095  \\
            3.08  0.9634372335697304  \\
            3.09  0.969564289382253  \\
            3.1  0.9756924384640774  \\
            3.11  0.9821331658182499  \\
            3.12  0.9885748855435763  \\
            3.13  0.9950175976400559  \\
            3.14  1.0014613021076897  \\
            3.15  1.0079059989464758  \\
            3.16  1.0143516881564163  \\
            3.17  1.0207983697375098  \\
            3.18  1.027246043689757  \\
            3.19  1.0336947100131577  \\
            3.2  1.0401443687077119  \\
            3.21  1.0465950197734193  \\
            3.22  1.0530466632102804  \\
            3.23  1.0594992990182952  \\
            3.24  1.0659529271974633  \\
            3.25  1.0724075477477846  \\
            3.26  1.0791722662213228  \\
            3.27  1.0859378895187186  \\
            3.28  1.0927044176399714  \\
            3.29  1.0994718505850811  \\
            3.3  1.1062401883540482  \\
            3.31  1.1130094309468728  \\
            3.32  1.1197795783635538  \\
            3.33  1.1265506306040927  \\
            3.34  1.1333225876684878  \\
            3.35  1.1400954495567408  \\
            3.36  1.1468692162688507  \\
            3.37  1.1536438878048176  \\
            3.38  1.160419464164642  \\
            3.39  1.1671959453483232  \\
            3.4  1.1739733313558616  \\
        }
        ;
    \addlegendentry {$W^{\text{rc}}$}
    \addplot[mark={none}, ultra thick, green!60!gray]
        table[row sep={\\}]
        {
            \\
            1.0  -0.0519271582295569  \\
            1.01  -0.051679314466603066  \\
            1.02  -0.050941182917584926  \\
            1.03  -0.049720662027090246  \\
            1.04  -0.0480253910311052  \\
            1.05  -0.04586276074388651  \\
            1.06  -0.04323992380095504  \\
            1.07  -0.04016380439037848  \\
            1.08  -0.036641107502330406  \\
            1.09  -0.032678327724901535  \\
            1.1  -0.02828175761227547  \\
            1.11  -0.023457495649661486  \\
            1.12  -0.018211453837780944  \\
            1.13  -0.012549364918229278  \\
            1.14  -0.006633774068413792  \\
            1.15  -0.000725897920736596  \\
            1.16  0.005170147233835376  \\
            1.17  0.011054086251970676  \\
            1.18  0.0169158648867486  \\
            1.19  0.022730045589203184  \\
            1.2  0.028519302264971902  \\
            1.21  0.03428694004793856  \\
            1.22  0.04003309681354423  \\
            1.23  0.04575445752449507  \\
            1.24  0.05145119916220815  \\
            1.25  0.05712704823956101  \\
            1.26  0.06277886074856988  \\
            1.27  0.0684103659594863  \\
            1.28  0.07401572552763748  \\
            1.29  0.07960489170454267  \\
            1.3  0.08516934523481828  \\
            1.31  0.09071841791449534  \\
            1.32  0.09624466679940984  \\
            1.33  0.10175206236399799  \\
            1.34  0.10724348767219072  \\
            1.35  0.11271810045985386  \\
            1.36  0.11817832551339286  \\
            1.37  0.1236229064673009  \\
            1.38  0.12905548408643638  \\
            1.39  0.13447683384340642  \\
            1.4  0.13988688433933041  \\
            1.41  0.14528613899811477  \\
            1.42  0.15067391894731635  \\
            1.43  0.15605080684580316  \\
            1.44  0.16141803125968068  \\
            1.45  0.16677737542950816  \\
            1.46  0.17213036248539007  \\
            1.47  0.17747701368084678  \\
            1.48  0.18281985210679333  \\
            1.49  0.188160882566355  \\
            1.5  0.1935018113591483  \\
            1.51  0.1988448353153981  \\
            1.52  0.20419235594830148  \\
            1.53  0.2095470103175865  \\
            1.54  0.21490186670266315  \\
            1.55  0.22026891175246271  \\
            1.56  0.22565247304401104  \\
            1.57  0.23105740754827292  \\
            1.58  0.23649159212977558  \\
            1.59  0.2419629259702331  \\
            1.6  0.2474802640428524  \\
            1.61  0.25306485038115295  \\
            1.62  0.2587171041090727  \\
            1.63  0.2644731493927916  \\
            1.64  0.27036692053653755  \\
            1.65  0.276443392915827  \\
            1.66  0.28277561842280274  \\
            1.67  0.2894882703615139  \\
            1.68  0.2968044502544996  \\
            1.69  0.3052313795644684  \\
            1.7  0.3161462429663042  \\
            1.71  0.3207981062507427  \\
            1.72  0.32424818050047455  \\
            1.73  0.32768185949078277  \\
            1.74  0.33110229194594554  \\
            1.75  0.33451245078493386  \\
            1.76  0.3379151366972658  \\
            1.77  0.34131298238402874  \\
            1.78  0.3447084573579938  \\
            1.79  0.3481038732044893  \\
            1.8  0.35150138921241203  \\
            1.81  0.3549030182923988  \\
            1.82  0.35831063310666156  \\
            1.83  0.36172597234224846  \\
            1.84  0.36515064706651945  \\
            1.85  0.36858614711030857  \\
            1.86  0.37203384743065315  \\
            1.87  0.37549501441097  \\
            1.88  0.37897081206223227  \\
            1.89  0.38246230809397563  \\
            1.9  0.3859704798288536  \\
            1.91  0.38949621993897654  \\
            1.92  0.3930403419864078  \\
            1.93  0.3966035857539435  \\
            1.94  0.4001866223557159  \\
            1.95  0.4037900591202118  \\
            1.96  0.4074144442410309  \\
            1.97  0.4110602711931156  \\
            1.98  0.4147279829143077  \\
            1.99  0.41841797575392403  \\
            2.0  0.4221306031916311  \\
            2.01  0.42586617933123894  \\
            2.02  0.4296249821751609  \\
            2.03  0.43340725668620805  \\
            2.04  0.4372132176441478  \\
            2.05  0.44104305230499896  \\
            2.06  0.4448969228715035  \\
            2.07  0.4487749687835011  \\
            2.08  0.4526773088371029  \\
            2.09  0.45660404314167946  \\
            2.1  0.4605552549236237  \\
            2.11  0.4645310121858318  \\
            2.12  0.4685313692316341  \\
            2.13  0.4725563680617614  \\
            2.14  0.4766060396526627  \\
            2.15  0.4806804051242203  \\
            2.16  0.48477947680460864  \\
            2.17  0.4889032591996866  \\
            2.18  0.49305174987402844  \\
            2.19  0.4972249402502683  \\
            2.2  0.5014228163331572  \\
            2.21  0.5056453593643114  \\
            2.22  0.5098925464133027  \\
            2.23  0.5141643509103958  \\
            2.24  0.5184607431258764  \\
            2.25  0.5227816906005981  \\
            2.26  0.527127158532044  \\
            2.27  0.5314971101198996  \\
            2.28  0.5358915068748232  \\
            2.29  0.5403103088938382  \\
            2.3  0.5447534751054873  \\
            2.31  0.549220963487646  \\
            2.32  0.553712731260659  \\
            2.33  0.5582287350582184  \\
            2.34  0.5627689310782354  \\
            2.35  0.5673332752157084  \\
            2.36  0.5719217231794657  \\
            2.37  0.5765342305944462  \\
            2.38  0.5811707530910585  \\
            2.39  0.5858312463829946  \\
            2.4  0.5905156663347609  \\
            2.41  0.5952239690200453  \\
            2.42  0.599956110771949  \\
            2.43  0.6047120482260054  \\
            2.44  0.6094917383567989  \\
            2.45  0.614295138508928  \\
            2.46  0.6191222064229928  \\
            2.47  0.6239729002571742  \\
            2.48  0.6288471786049524  \\
            2.49  0.633745000509437  \\
            2.5  0.6386663254747237  \\
            2.51  0.643611113474653  \\
            2.52  0.6485793249593116  \\
            2.53  0.6535709208595603  \\
            2.54  0.6585858625898581  \\
            2.55  0.6636241120496086  \\
            2.56  0.6686856316232367  \\
            2.57  0.6737703841791619  \\
            2.58  0.6788783330678564  \\
            2.59  0.6840094421190862  \\
            2.6  0.6891636756384875  \\
            2.61  0.6943409984035774  \\
            2.62  0.6995413756592871  \\
            2.63  0.7047647731130876  \\
            2.64  0.7100111569298265  \\
            2.65  0.7152804937262646  \\
            2.66  0.7205727505654277  \\
            2.67  0.7258878949507949  \\
            2.68  0.7312258948203526  \\
            2.69  0.736586718540574  \\
            2.7  0.7419703349003391  \\
            2.71  0.747376713104804  \\
            2.72  0.7528058227692916  \\
            2.73  0.7582576339131512  \\
            2.74  0.7637321169536756  \\
            2.75  0.7692292427000215  \\
            2.76  0.7747489823472135  \\
            2.77  0.7802913074701573  \\
            2.78  0.7858561900177617  \\
            2.79  0.791443602307098  \\
            2.8  0.7970535170176513  \\
            2.81  0.8026859071856366  \\
            2.82  0.8083407461984059  \\
            2.83  0.8140180077889422  \\
            2.84  0.8197176660304212  \\
            2.85  0.8254396953308752  \\
            2.86  0.8311840704279468  \\
            2.87  0.8369507663837069  \\
            2.88  0.8427397585795905  \\
            2.89  0.8485510227113972  \\
            2.9  0.8543845347843882  \\
            2.91  0.8602402711084695  \\
            2.92  0.8661182082934452  \\
            2.93  0.8720183232443878  \\
            2.94  0.8779405931570489  \\
            2.95  0.8838849955133886  \\
            2.96  0.8898515080771613  \\
            2.97  0.895840108889597  \\
            2.98  0.9018507762651419  \\
            2.99  0.9078834887872972  \\
            3.0  0.9139382253045149  \\
            3.01  0.9200149649261785  \\
            3.02  0.9261136870186485  \\
            3.03  0.9322343712013849  \\
            3.04  0.9383769973431381  \\
            3.05  0.9445415455582035  \\
            3.06  0.9507279962027425  \\
            3.07  0.9569363298711847  \\
            3.08  0.9631665273926678  \\
            3.09  0.9694185698275714  \\
            3.1  0.9756924384640774  \\
            3.11  0.9819881148148232  \\
            3.12  0.9883055806135962  \\
            3.13  0.9946448178120922  \\
            3.14  1.0010058085767244  \\
            3.15  1.007388535285497  \\
            3.16  1.0137929805249315  \\
            3.17  1.0202191270870355  \\
            3.18  1.026666957966341  \\
            3.19  1.0331364563569845  \\
            3.2  1.0396276056498324  \\
            3.21  1.0461403894296688  \\
            3.22  1.0526747914724246  \\
            3.23  1.059230795742447  \\
            3.24  1.0658083863898322  \\
            3.25  1.0724075477477846  \\
            3.26  1.0790282643300375  \\
            3.27  1.0856705208283093  \\
            3.28  1.0923343021097947  \\
            3.29  1.0990195932147189  \\
            3.3  1.1057263793539103  \\
            3.31  1.1124546459064315  \\
            3.32  1.1192043784172383  \\
            3.33  1.125975562594879  \\
            3.34  1.1327681843092399  \\
            3.35  1.1395822295893185  \\
            3.36  1.1464176846210417  \\
            3.37  1.1532745357451097  \\
            3.38  1.1601527694548852  \\
            3.39  1.167052372394307  \\
            3.4  1.1739733313558616  \\
        }
        ;
    \addlegendentry {$W^{\text{rc}}_{\text{HROC}}$}
  \coordinate (spypoint) at (axis cs:1.7,0.3);
  \coordinate (magnifyglass) at (axis cs:1.5,0.8);
\end{axis}
\spy [blue!20!gray, size=2.0cm] on (spypoint)
   in node[fill=white] at (magnifyglass);
\end{tikzpicture}

%% file: figures/tikz/rotationaverage.tex
\tikzsetfigurename{rotationaverage}
\begin{tikzpicture}
\begin{axis}[axis background/.style={{fill={white!89.803921568!black}}},
x grid style={{white}},
y grid style={{white}},
xtick={1,1.5,2,2.5,3,3.5},
ytick={0,0.1,0.2,0.3},
xmajorgrids,
ymajorgrids,
ylabel={first Piola--Kirchhoff},
xlabel={Deformation Gradient $F_{11},F_{22}$},
legend pos=south east,
width=0.95\textwidth,
height=0.8\textwidth,
]
    \addplot[mark={none}, thick, red!60!gray, dashed]
        table[row sep={\\}]
        {
            \\
            1.0  0.0  \\
            1.01  0.024716311120714  \\
            1.02  0.04903103287619028  \\
            1.03  0.0729574048583441  \\
            1.04  0.09650811315335567  \\
            1.05  0.11969531837834253  \\
            1.06  0.14253068205221184  \\
            1.07  0.1650253914140702  \\
            1.08  0.18719018279389232  \\
            1.09  0.2090353636321955  \\
            1.1  0.23057083323820005  \\
            1.11  0.2518061023692831  \\
            1.12  0.27275031170841013  \\
            1.13  0.293412249310631  \\
            1.14  0.29531608414373633  \\
            1.15  0.29404013594594297  \\
            1.16  0.29314951651584525  \\
            1.17  0.291870230096465  \\
            1.18  0.2908740072370407  \\
            1.19  0.28969217608717335  \\
            1.2  0.28860014643773574  \\
            1.21  0.28749747495197325  \\
            1.22  0.2860268783304284  \\
            1.23  0.2849952097633391  \\
            1.24  0.2839501380598803  \\
            1.25  0.28262980605812277  \\
            1.26  0.2816406871413596  \\
            1.27  0.28037983582425186  \\
            1.28  0.27910212597131523  \\
            1.29  0.2778087466159  \\
            1.3  0.27657887888604027  \\
            1.31  0.27540761160899657  \\
            1.32  0.273894824224668  \\
            1.33  0.272677480687091  \\
            1.34  0.2712857233342147  \\
            1.35  0.26978672782821944  \\
            1.36  0.2686356044022036  \\
            1.37  0.2672308517218501  \\
            1.38  0.2655837264445078  \\
            1.39  0.2641901441045979  \\
            1.4  0.26234444238521815  \\
            1.41  0.26012831267246245  \\
            1.42  0.2579690700946554  \\
            1.43  0.2557220836298238  \\
            1.44  0.25346812979088973  \\
            1.45  0.2509470804515139  \\
            1.46  0.24848184122961783  \\
            1.47  0.24594066972112197  \\
            1.48  0.24322034987319846  \\
            1.49  0.2406660535840144  \\
            1.5  0.2377060128647094  \\
            1.51  0.23497217581861962  \\
            1.52  0.2320200743153526  \\
            1.53  0.22886312196927822  \\
            1.54  0.2266711346397931  \\
            1.55  0.2244389510295011  \\
            1.56  0.2222820876679295  \\
            1.57  0.21993869020409623  \\
            1.58  0.21751786283468397  \\
            1.59  0.2151913276993702  \\
            1.6  0.21272462198615957  \\
            1.61  0.21019942629614013  \\
            1.62  0.20754161396944024  \\
            1.63  0.20483787304687484  \\
            1.64  0.2020781325916567  \\
            1.65  0.19916249667366065  \\
            1.66  0.19601337948942407  \\
            1.67  0.19258832451635144  \\
            1.68  0.1885790936038626  \\
            1.69  0.18345297580433642  \\
            1.7  0.1752477678876005  \\
            1.71  0.17296829968645033  \\
            1.72  0.17206685141200295  \\
            1.73  0.1713272846425571  \\
            1.74  0.1707407322050607  \\
            1.75  0.17029848722626761  \\
            1.76  0.16999203917085337  \\
            1.77  0.1698131043836777  \\
            1.78  0.16975365152152344  \\
            1.79  0.16980592226138502  \\
            1.8  0.16996244766988022  \\
            1.81  0.17021606061224048  \\
            1.82  0.17055990457004344  \\
            1.83  0.17098743922496817  \\
            1.84  0.17149244315177925  \\
            1.85  0.17206901394796104  \\
            1.86  0.17271156611028826  \\
            1.87  0.17341482695053645  \\
            1.88  0.1741738308237799  \\
            1.89  0.1749839119236436  \\
            1.9  0.17584069587967432  \\
            1.91  0.1767400903729277  \\
            1.92  0.17767827496712862  \\
            1.93  0.17865169033448436  \\
            1.94  0.17965702703757258  \\
            1.95  0.18069121401180843  \\
            1.96  0.18175140687686753  \\
            1.97  0.18283497619020062  \\
            1.98  0.18393949574145596  \\
            1.99  0.18506273097324932  \\
            2.0  0.18620262760132067  \\
            2.01  0.18735730049566612  \\
            2.02  0.18852502287374695  \\
            2.03  0.18970421584732344  \\
            2.04  0.19089343835580141  \\
            2.05  0.19209137751121563  \\
            2.06  0.19329683937300324  \\
            2.07  0.19450874016458286  \\
            2.08  0.1957260979383133  \\
            2.09  0.19694802469069983  \\
            2.1  0.1981737189256152  \\
            2.11  0.1994024586598293  \\
            2.12  0.20063359486219934  \\
            2.13  0.2018665453154116  \\
            2.14  0.20310078888720584  \\
            2.15  0.20433586019640138  \\
            2.16  0.20557134465783644  \\
            2.17  0.20680687388943286  \\
            2.18  0.20804212146398937  \\
            2.19  0.20927679898793805  \\
            2.2  0.21051065248916112  \\
            2.21  0.21174345909600278  \\
            2.22  0.21297502398980384  \\
            2.23  0.21420517761363045  \\
            2.24  0.2154337731203  \\
            2.25  0.21666068404333383  \\
            2.26  0.21788580217507336  \\
            2.27  0.21910903563684792  \\
            2.28  0.22033030712676083  \\
            2.29  0.22154955233139662  \\
            2.3  0.22276671848846918  \\
            2.31  0.2239817630881702  \\
            2.32  0.225194652701719  \\
            2.33  0.2264053619263393  \\
            2.34  0.22761387243657216  \\
            2.35  0.22882017213257466  \\
            2.36  0.23002425437667107  \\
            2.37  0.23122611731008913  \\
            2.38  0.23242576324242004  \\
            2.39  0.23362319810692433  \\
            2.4  0.23481843097535116  \\
            2.41  0.2360114736264638  \\
            2.42  0.2372023401629515  \\
            2.43  0.2383910466718686  \\
            2.44  0.2395776109241483  \\
            2.45  0.2407620521091829  \\
            2.46  0.24194439060076922  \\
            2.47  0.2431246477511122  \\
            2.48  0.24430284570986593  \\
            2.49  0.24547900726548777  \\
            2.5  0.24665315570645507  \\
            2.51  0.24782531470012348  \\
            2.52  0.24899550818724547  \\
            2.53  0.25016376029035836  \\
            2.54  0.2513300952344375  \\
            2.55  0.2524945372783969  \\
            2.56  0.25365711065613716  \\
            2.57  0.25481783952601145  \\
            2.58  0.255976747927698  \\
            2.59  0.25713385974555214  \\
            2.6  0.25828919867766365  \\
            2.61  0.25944278820988353  \\
            2.62  0.26059465159419926  \\
            2.63  0.2617448118309018  \\
            2.64  0.2628932916540476  \\
            2.65  0.2640401135197865  \\
            2.66  0.26518529959716153  \\
            2.67  0.26632887176106435  \\
            2.68  0.2674708515870301  \\
            2.69  0.2686112603476283  \\
            2.7  0.2697501190102125  \\
            2.71  0.27088744823584143  \\
            2.72  0.27202326837918855  \\
            2.73  0.2731575994892963  \\
            2.74  0.27429046131104196  \\
            2.75  0.27542187328720136  \\
            2.76  0.2765518545610104  \\
            2.77  0.2776804239791417  \\
            2.78  0.2788076000950235  \\
            2.79  0.27993340117243015  \\
            2.8  0.2810578451892969  \\
            2.81  0.2821809498417136  \\
            2.82  0.28330273254804306  \\
            2.83  0.28442321045314944  \\
            2.84  0.2855424004326905  \\
            2.85  0.28666031909745815  \\
            2.86  0.2877769827977432  \\
            2.87  0.28889240762770907  \\
            2.88  0.29000660942974976  \\
            2.89  0.29111960379883817  \\
            2.9  0.2922314060868332  \\
            2.91  0.29334203140675186  \\
            2.92  0.2944514946369914  \\
            2.93  0.2955598104255093  \\
            2.94  0.296666993193925  \\
            2.95  0.29777305714158286  \\
            2.96  0.2988780162495388  \\
            2.97  0.299981884284485  \\
            2.98  0.3010846748026057  \\
            2.99  0.30218640115336903  \\
            3.0  0.3032870764832407  \\
            3.01  0.3043867137393437  \\
            3.02  0.305485325673033  \\
            3.03  0.30658292484341176  \\
            3.04  0.30767952362077544  \\
            3.05  0.3087751341899909  \\
            3.06  0.30986976855380455  \\
            3.07  0.3109634385360902  \\
            3.08  0.3120561557850259  \\
            3.09  0.3131479317762147  \\
            3.1  0.3142387778157336  \\
            3.11  0.3153287050431288  \\
            3.12  0.3164177244343467  \\
            3.13  0.3175058468046057  \\
            3.14  0.3185930828112107  \\
            3.15  0.3196794429563128  \\
            3.16  0.3207649375896078  \\
            3.17  0.32184957691098814  \\
            3.18  0.32293337097313346  \\
            3.19  0.3240163296840537  \\
            3.2  0.3250984628095832  \\
            3.21  0.32617977997581704  \\
            3.22  0.32726029067150764  \\
            3.23  0.32834000425040655  \\
            3.24  0.3294189299335657  \\
            3.25  0.330497076811588  \\
            3.26  0.3315744538468359  \\
            3.27  0.33265106987559356  \\
            3.28  0.3337269336101949  \\
            3.29  0.33480205364109583  \\
            3.3  0.3358764384389188  \\
            3.31  0.33695009635645023  \\
            3.32  0.3380230356306049  \\
            3.33  0.3390952643843397  \\
            3.34  0.34016679062854926  \\
            3.35  0.3412376222639084  \\
            3.36  0.3423077670826875  \\
            3.37  0.34337723277052845  \\
            3.38  0.3444460269081939  \\
            3.39  0.34551415697327403  \\
            3.4  0.3465816303418654  \\
        }
        ;
    \addlegendentry {$P_{11}$}
    \addplot[mark={none}, thick, red!60!gray, dashed]
        table[row sep={\\}]
        {
            \\
            1.0  0.0  \\
            1.01  0.024716311120714  \\
            1.02  0.04903103287619028  \\
            1.03  0.0729574048583441  \\
            1.04  0.09650811315335567  \\
            1.05  0.11969531837834253  \\
            1.06  0.14253068205221184  \\
            1.07  0.1650253914140702  \\
            1.08  0.18719018279389232  \\
            1.09  0.2090353636321955  \\
            1.1  0.23057083323820005  \\
            1.11  0.2518061023692831  \\
            1.12  0.27275031170841013  \\
            1.13  0.293412249310631  \\
            1.14  0.29178837012663184  \\
            1.15  0.2864709858837749  \\
            1.16  0.28133192671305074  \\
            1.17  0.2761761820672176  \\
            1.18  0.2722335108388601  \\
            1.19  0.2704055925540909  \\
            1.2  0.2689621455839007  \\
            1.21  0.26752094692248685  \\
            1.22  0.2659939191858812  \\
            1.23  0.2649149457465928  \\
            1.24  0.26383000628858544  \\
            1.25  0.2629988589555392  \\
            1.26  0.2622380636355927  \\
            1.27  0.2617156477307592  \\
            1.28  0.26116628197364694  \\
            1.29  0.2605907997142425  \\
            1.3  0.2603137067729125  \\
            1.31  0.26032018815148444  \\
            1.32  0.26019792888544185  \\
            1.33  0.26010763097759104  \\
            1.34  0.26051056947655266  \\
            1.35  0.26053467345342085  \\
            1.36  0.2608691736280857  \\
            1.37  0.2613505849443159  \\
            1.38  0.2619683571339419  \\
            1.39  0.2625582398868926  \\
            1.4  0.2624246086905432  \\
            1.41  0.2608434896655936  \\
            1.42  0.2595002369915835  \\
            1.43  0.25784703743358156  \\
            1.44  0.2561678354049542  \\
            1.45  0.25439242554749375  \\
            1.46  0.2528283718611441  \\
            1.47  0.2509729169213375  \\
            1.48  0.24926266966895735  \\
            1.49  0.2473257902105618  \\
            1.5  0.2454577629987389  \\
            1.51  0.2435934342395479  \\
            1.52  0.24161912653297568  \\
            1.53  0.23953075022913828  \\
            1.54  0.2374323734294847  \\
            1.55  0.23527131790881728  \\
            1.56  0.23288431042263638  \\
            1.57  0.23057415373977397  \\
            1.58  0.22815811213418175  \\
            1.59  0.22567972387357688  \\
            1.6  0.2229171113362677  \\
            1.61  0.22018514150033017  \\
            1.62  0.21727785531771637  \\
            1.63  0.21407428902792575  \\
            1.64  0.2106820285163382  \\
            1.65  0.2070017268379127  \\
            1.66  0.20295178210234327  \\
            1.67  0.19828201152793468  \\
            1.68  0.19274894732900283  \\
            1.69  0.18573347409434643  \\
            1.7  0.17548279189606586  \\
            1.71  0.17296829968645033  \\
            1.72  0.17206685141200295  \\
            1.73  0.1713272846425571  \\
            1.74  0.1707407322050607  \\
            1.75  0.17029848722626761  \\
            1.76  0.16999203917085337  \\
            1.77  0.1698131043836777  \\
            1.78  0.16975365152152344  \\
            1.79  0.16980592226138502  \\
            1.8  0.16996244766988022  \\
            1.81  0.17021606061224048  \\
            1.82  0.17055990457004344  \\
            1.83  0.17098743922496817  \\
            1.84  0.17149244315177925  \\
            1.85  0.17206901394796104  \\
            1.86  0.17271156611028826  \\
            1.87  0.17341482695053645  \\
            1.88  0.1741738308237799  \\
            1.89  0.1749839119236436  \\
            1.9  0.17584069587967432  \\
            1.91  0.1767400903729277  \\
            1.92  0.17767827496712862  \\
            1.93  0.17865169033448436  \\
            1.94  0.17965702703757258  \\
            1.95  0.18069121401180843  \\
            1.96  0.18175140687686753  \\
            1.97  0.18283497619020062  \\
            1.98  0.18393949574145596  \\
            1.99  0.18506273097324932  \\
            2.0  0.18620262760132067  \\
            2.01  0.18735730049566612  \\
            2.02  0.18852502287374695  \\
            2.03  0.18970421584732344  \\
            2.04  0.19089343835580141  \\
            2.05  0.19209137751121563  \\
            2.06  0.19329683937300324  \\
            2.07  0.19450874016458286  \\
            2.08  0.1957260979383133  \\
            2.09  0.19694802469069983  \\
            2.1  0.1981737189256152  \\
            2.11  0.1994024586598293  \\
            2.12  0.20063359486219934  \\
            2.13  0.2018665453154116  \\
            2.14  0.20310078888720584  \\
            2.15  0.20433586019640138  \\
            2.16  0.20557134465783644  \\
            2.17  0.20680687388943286  \\
            2.18  0.20804212146398937  \\
            2.19  0.20927679898793805  \\
            2.2  0.21051065248916112  \\
            2.21  0.21174345909600278  \\
            2.22  0.21297502398980384  \\
            2.23  0.21420517761363045  \\
            2.24  0.2154337731203  \\
            2.25  0.21666068404333383  \\
            2.26  0.21788580217507336  \\
            2.27  0.21910903563684792  \\
            2.28  0.22033030712676083  \\
            2.29  0.22154955233139662  \\
            2.3  0.22276671848846918  \\
            2.31  0.2239817630881702  \\
            2.32  0.225194652701719  \\
            2.33  0.2264053619263393  \\
            2.34  0.22761387243657216  \\
            2.35  0.22882017213257466  \\
            2.36  0.23002425437667107  \\
            2.37  0.23122611731008913  \\
            2.38  0.23242576324242004  \\
            2.39  0.23362319810692433  \\
            2.4  0.23481843097535116  \\
            2.41  0.2360114736264638  \\
            2.42  0.2372023401629515  \\
            2.43  0.2383910466718686  \\
            2.44  0.2395776109241483  \\
            2.45  0.2407620521091829  \\
            2.46  0.24194439060076922  \\
            2.47  0.2431246477511122  \\
            2.48  0.24430284570986593  \\
            2.49  0.24547900726548777  \\
            2.5  0.24665315570645507  \\
            2.51  0.24782531470012348  \\
            2.52  0.24899550818724547  \\
            2.53  0.25016376029035836  \\
            2.54  0.2513300952344375  \\
            2.55  0.2524945372783969  \\
            2.56  0.25365711065613716  \\
            2.57  0.25481783952601145  \\
            2.58  0.255976747927698  \\
            2.59  0.25713385974555214  \\
            2.6  0.25828919867766365  \\
            2.61  0.25944278820988353  \\
            2.62  0.26059465159419926  \\
            2.63  0.2617448118309018  \\
            2.64  0.2628932916540476  \\
            2.65  0.2640401135197865  \\
            2.66  0.26518529959716153  \\
            2.67  0.26632887176106435  \\
            2.68  0.2674708515870301  \\
            2.69  0.2686112603476283  \\
            2.7  0.2697501190102125  \\
            2.71  0.27088744823584143  \\
            2.72  0.27202326837918855  \\
            2.73  0.2731575994892963  \\
            2.74  0.27429046131104196  \\
            2.75  0.27542187328720136  \\
            2.76  0.2765518545610104  \\
            2.77  0.2776804239791417  \\
            2.78  0.2788076000950235  \\
            2.79  0.27993340117243015  \\
            2.8  0.2810578451892969  \\
            2.81  0.2821809498417136  \\
            2.82  0.28330273254804306  \\
            2.83  0.28442321045314944  \\
            2.84  0.2855424004326905  \\
            2.85  0.28666031909745815  \\
            2.86  0.2877769827977432  \\
            2.87  0.28889240762770907  \\
            2.88  0.29000660942974976  \\
            2.89  0.29111960379883817  \\
            2.9  0.2922314060868332  \\
            2.91  0.29334203140675186  \\
            2.92  0.2944514946369914  \\
            2.93  0.2955598104255093  \\
            2.94  0.296666993193925  \\
            2.95  0.29777305714158286  \\
            2.96  0.2988780162495388  \\
            2.97  0.299981884284485  \\
            2.98  0.3010846748026057  \\
            2.99  0.30218640115336903  \\
            3.0  0.3032870764832407  \\
            3.01  0.3043867137393437  \\
            3.02  0.305485325673033  \\
            3.03  0.30658292484341176  \\
            3.04  0.30767952362077544  \\
            3.05  0.3087751341899909  \\
            3.06  0.30986976855380455  \\
            3.07  0.3109634385360902  \\
            3.08  0.3120561557850259  \\
            3.09  0.3131479317762147  \\
            3.1  0.3142387778157336  \\
            3.11  0.3153287050431288  \\
            3.12  0.3164177244343467  \\
            3.13  0.3175058468046057  \\
            3.14  0.3185930828112107  \\
            3.15  0.3196794429563128  \\
            3.16  0.3207649375896078  \\
            3.17  0.32184957691098814  \\
            3.18  0.32293337097313346  \\
            3.19  0.3240163296840537  \\
            3.2  0.3250984628095832  \\
            3.21  0.32617977997581704  \\
            3.22  0.32726029067150764  \\
            3.23  0.32834000425040655  \\
            3.24  0.3294189299335657  \\
            3.25  0.330497076811588  \\
            3.26  0.3315744538468359  \\
            3.27  0.33265106987559356  \\
            3.28  0.3337269336101949  \\
            3.29  0.33480205364109583  \\
            3.3  0.3358764384389188  \\
            3.31  0.33695009635645023  \\
            3.32  0.3380230356306049  \\
            3.33  0.3390952643843397  \\
            3.34  0.34016679062854926  \\
            3.35  0.3412376222639084  \\
            3.36  0.3423077670826875  \\
            3.37  0.34337723277052845  \\
            3.38  0.3444460269081939  \\
            3.39  0.34551415697327403  \\
            3.4  0.3465816303418654  \\
        }
        ;
    \addlegendentry {$P_{22}$}
    \addplot[mark={none}, thick, blue!60!gray]
        table[row sep={\\}]
        {
            \\
            1.0  0.0  \\
            1.01  0.024716311120714  \\
            1.02  0.04903103287619028  \\
            1.03  0.0729574048583441  \\
            1.04  0.09650811315335567  \\
            1.05  0.11969531837834253  \\
            1.06  0.14253068205221184  \\
            1.07  0.1650253914140702  \\
            1.08  0.18719018279389232  \\
            1.09  0.2090353636321955  \\
            1.1  0.23057083323820005  \\
            1.11  0.2518061023692831  \\
            1.12  0.27275031170841013  \\
            1.13  0.293412249310631  \\
            1.14  0.2935522271351841  \\
            1.15  0.29025556091485893  \\
            1.16  0.287240721614448  \\
            1.17  0.2840232060818413  \\
            1.18  0.28155375903795044  \\
            1.19  0.28004888432063213  \\
            1.2  0.2787811460108182  \\
            1.21  0.27750921093723  \\
            1.22  0.2760103987581548  \\
            1.23  0.27495507775496597  \\
            1.24  0.2738900721742329  \\
            1.25  0.27281433250683096  \\
            1.26  0.2719393753884761  \\
            1.27  0.27104774177750557  \\
            1.28  0.2701342039724811  \\
            1.29  0.26919977316507127  \\
            1.3  0.2684462928294764  \\
            1.31  0.2678638998802405  \\
            1.32  0.2670463765550549  \\
            1.33  0.266392555832341  \\
            1.34  0.2658981464053837  \\
            1.35  0.26516070064082015  \\
            1.36  0.26475238901514464  \\
            1.37  0.26429071833308304  \\
            1.38  0.26377604178922487  \\
            1.39  0.26337419199574524  \\
            1.4  0.26238452553788066  \\
            1.41  0.26048590116902803  \\
            1.42  0.2587346535431194  \\
            1.43  0.2567845605317027  \\
            1.44  0.254817982597922  \\
            1.45  0.25266975299950384  \\
            1.46  0.250655106545381  \\
            1.47  0.24845679332122972  \\
            1.48  0.2462415097710779  \\
            1.49  0.2439959218972881  \\
            1.5  0.24158188793172414  \\
            1.51  0.23928280502908378  \\
            1.52  0.23681960042416414  \\
            1.53  0.23419693609920825  \\
            1.54  0.2320517540346389  \\
            1.55  0.22985513446915917  \\
            1.56  0.22758319904528296  \\
            1.57  0.2252564219719351  \\
            1.58  0.22283798748443284  \\
            1.59  0.22043552578647352  \\
            1.6  0.21782086666121364  \\
            1.61  0.21519228389823514  \\
            1.62  0.21240973464357832  \\
            1.63  0.2094560810374003  \\
            1.64  0.20638008055399742  \\
            1.65  0.20308211175578667  \\
            1.66  0.19948258079588366  \\
            1.67  0.19543516802214306  \\
            1.68  0.19066402046643272  \\
            1.69  0.18459322494934144  \\
            1.7  0.17536527989183318  \\
            1.71  0.17296829968645033  \\
            1.72  0.17206685141200295  \\
            1.73  0.1713272846425571  \\
            1.74  0.1707407322050607  \\
            1.75  0.17029848722626761  \\
            1.76  0.16999203917085337  \\
            1.77  0.1698131043836777  \\
            1.78  0.16975365152152344  \\
            1.79  0.16980592226138502  \\
            1.8  0.16996244766988022  \\
            1.81  0.17021606061224048  \\
            1.82  0.17055990457004344  \\
            1.83  0.17098743922496817  \\
            1.84  0.17149244315177925  \\
            1.85  0.17206901394796104  \\
            1.86  0.17271156611028826  \\
            1.87  0.17341482695053645  \\
            1.88  0.1741738308237799  \\
            1.89  0.1749839119236436  \\
            1.9  0.17584069587967432  \\
            1.91  0.1767400903729277  \\
            1.92  0.17767827496712862  \\
            1.93  0.17865169033448436  \\
            1.94  0.17965702703757258  \\
            1.95  0.18069121401180843  \\
            1.96  0.18175140687686753  \\
            1.97  0.18283497619020062  \\
            1.98  0.18393949574145596  \\
            1.99  0.18506273097324932  \\
            2.0  0.18620262760132067  \\
            2.01  0.18735730049566612  \\
            2.02  0.18852502287374695  \\
            2.03  0.18970421584732344  \\
            2.04  0.19089343835580141  \\
            2.05  0.19209137751121563  \\
            2.06  0.19329683937300324  \\
            2.07  0.19450874016458286  \\
            2.08  0.1957260979383133  \\
            2.09  0.19694802469069983  \\
            2.1  0.1981737189256152  \\
            2.11  0.1994024586598293  \\
            2.12  0.20063359486219934  \\
            2.13  0.2018665453154116  \\
            2.14  0.20310078888720584  \\
            2.15  0.20433586019640138  \\
            2.16  0.20557134465783644  \\
            2.17  0.20680687388943286  \\
            2.18  0.20804212146398937  \\
            2.19  0.20927679898793805  \\
            2.2  0.21051065248916112  \\
            2.21  0.21174345909600278  \\
            2.22  0.21297502398980384  \\
            2.23  0.21420517761363045  \\
            2.24  0.2154337731203  \\
            2.25  0.21666068404333383  \\
            2.26  0.21788580217507336  \\
            2.27  0.21910903563684792  \\
            2.28  0.22033030712676083  \\
            2.29  0.22154955233139662  \\
            2.3  0.22276671848846918  \\
            2.31  0.2239817630881702  \\
            2.32  0.225194652701719  \\
            2.33  0.2264053619263393  \\
            2.34  0.22761387243657216  \\
            2.35  0.22882017213257466  \\
            2.36  0.23002425437667107  \\
            2.37  0.23122611731008913  \\
            2.38  0.23242576324242004  \\
            2.39  0.23362319810692433  \\
            2.4  0.23481843097535116  \\
            2.41  0.2360114736264638  \\
            2.42  0.2372023401629515  \\
            2.43  0.2383910466718686  \\
            2.44  0.2395776109241483  \\
            2.45  0.2407620521091829  \\
            2.46  0.24194439060076922  \\
            2.47  0.2431246477511122  \\
            2.48  0.24430284570986593  \\
            2.49  0.24547900726548777  \\
            2.5  0.24665315570645507  \\
            2.51  0.24782531470012348  \\
            2.52  0.24899550818724547  \\
            2.53  0.25016376029035836  \\
            2.54  0.2513300952344375  \\
            2.55  0.2524945372783969  \\
            2.56  0.25365711065613716  \\
            2.57  0.25481783952601145  \\
            2.58  0.255976747927698  \\
            2.59  0.25713385974555214  \\
            2.6  0.25828919867766365  \\
            2.61  0.25944278820988353  \\
            2.62  0.26059465159419926  \\
            2.63  0.2617448118309018  \\
            2.64  0.2628932916540476  \\
            2.65  0.2640401135197865  \\
            2.66  0.26518529959716153  \\
            2.67  0.26632887176106435  \\
            2.68  0.2674708515870301  \\
            2.69  0.2686112603476283  \\
            2.7  0.2697501190102125  \\
            2.71  0.27088744823584143  \\
            2.72  0.27202326837918855  \\
            2.73  0.2731575994892963  \\
            2.74  0.27429046131104196  \\
            2.75  0.27542187328720136  \\
            2.76  0.2765518545610104  \\
            2.77  0.2776804239791417  \\
            2.78  0.2788076000950235  \\
            2.79  0.27993340117243015  \\
            2.8  0.2810578451892969  \\
            2.81  0.2821809498417136  \\
            2.82  0.28330273254804306  \\
            2.83  0.28442321045314944  \\
            2.84  0.2855424004326905  \\
            2.85  0.28666031909745815  \\
            2.86  0.2877769827977432  \\
            2.87  0.28889240762770907  \\
            2.88  0.29000660942974976  \\
            2.89  0.29111960379883817  \\
            2.9  0.2922314060868332  \\
            2.91  0.29334203140675186  \\
            2.92  0.2944514946369914  \\
            2.93  0.2955598104255093  \\
            2.94  0.296666993193925  \\
            2.95  0.29777305714158286  \\
            2.96  0.2988780162495388  \\
            2.97  0.299981884284485  \\
            2.98  0.3010846748026057  \\
            2.99  0.30218640115336903  \\
            3.0  0.3032870764832407  \\
            3.01  0.3043867137393437  \\
            3.02  0.305485325673033  \\
            3.03  0.30658292484341176  \\
            3.04  0.30767952362077544  \\
            3.05  0.3087751341899909  \\
            3.06  0.30986976855380455  \\
            3.07  0.3109634385360902  \\
            3.08  0.3120561557850259  \\
            3.09  0.3131479317762147  \\
            3.1  0.3142387778157336  \\
            3.11  0.3153287050431288  \\
            3.12  0.3164177244343467  \\
            3.13  0.3175058468046057  \\
            3.14  0.3185930828112107  \\
            3.15  0.3196794429563128  \\
            3.16  0.3207649375896078  \\
            3.17  0.32184957691098814  \\
            3.18  0.32293337097313346  \\
            3.19  0.3240163296840537  \\
            3.2  0.3250984628095832  \\
            3.21  0.32617977997581704  \\
            3.22  0.32726029067150764  \\
            3.23  0.32834000425040655  \\
            3.24  0.3294189299335657  \\
            3.25  0.330497076811588  \\
            3.26  0.3315744538468359  \\
            3.27  0.33265106987559356  \\
            3.28  0.3337269336101949  \\
            3.29  0.33480205364109583  \\
            3.3  0.3358764384389188  \\
            3.31  0.33695009635645023  \\
            3.32  0.3380230356306049  \\
            3.33  0.3390952643843397  \\
            3.34  0.34016679062854926  \\
            3.35  0.3412376222639084  \\
            3.36  0.3423077670826875  \\
            3.37  0.34337723277052845  \\
            3.38  0.3444460269081939  \\
            3.39  0.34551415697327403  \\
            3.4  0.3465816303418654  \\
        }
        ;
    \addlegendentry {$P_{11}^{\text{rot}}$}
    \addplot[mark={none}, thick, blue!60!gray]
        table[row sep={\\}]
        {
            \\
            1.0  0.0  \\
            1.01  0.024716311120714  \\
            1.02  0.04903103287619028  \\
            1.03  0.0729574048583441  \\
            1.04  0.09650811315335567  \\
            1.05  0.11969531837834253  \\
            1.06  0.14253068205221184  \\
            1.07  0.1650253914140702  \\
            1.08  0.18719018279389232  \\
            1.09  0.2090353636321955  \\
            1.1  0.23057083323820005  \\
            1.11  0.2518061023692831  \\
            1.12  0.27275031170841013  \\
            1.13  0.293412249310631  \\
            1.14  0.2935522271351841  \\
            1.15  0.29025556091485893  \\
            1.16  0.287240721614448  \\
            1.17  0.2840232060818413  \\
            1.18  0.28155375903795044  \\
            1.19  0.28004888432063213  \\
            1.2  0.2787811460108182  \\
            1.21  0.27750921093723  \\
            1.22  0.2760103987581548  \\
            1.23  0.27495507775496597  \\
            1.24  0.2738900721742329  \\
            1.25  0.27281433250683096  \\
            1.26  0.2719393753884761  \\
            1.27  0.27104774177750557  \\
            1.28  0.2701342039724811  \\
            1.29  0.26919977316507127  \\
            1.3  0.2684462928294764  \\
            1.31  0.2678638998802405  \\
            1.32  0.2670463765550549  \\
            1.33  0.266392555832341  \\
            1.34  0.2658981464053837  \\
            1.35  0.26516070064082015  \\
            1.36  0.26475238901514464  \\
            1.37  0.26429071833308304  \\
            1.38  0.26377604178922487  \\
            1.39  0.26337419199574524  \\
            1.4  0.26238452553788066  \\
            1.41  0.26048590116902803  \\
            1.42  0.2587346535431194  \\
            1.43  0.2567845605317027  \\
            1.44  0.254817982597922  \\
            1.45  0.25266975299950384  \\
            1.46  0.250655106545381  \\
            1.47  0.24845679332122972  \\
            1.48  0.2462415097710779  \\
            1.49  0.2439959218972881  \\
            1.5  0.24158188793172414  \\
            1.51  0.23928280502908378  \\
            1.52  0.23681960042416414  \\
            1.53  0.23419693609920825  \\
            1.54  0.2320517540346389  \\
            1.55  0.22985513446915917  \\
            1.56  0.22758319904528296  \\
            1.57  0.2252564219719351  \\
            1.58  0.22283798748443284  \\
            1.59  0.22043552578647352  \\
            1.6  0.21782086666121364  \\
            1.61  0.21519228389823514  \\
            1.62  0.21240973464357832  \\
            1.63  0.2094560810374003  \\
            1.64  0.20638008055399742  \\
            1.65  0.20308211175578667  \\
            1.66  0.19948258079588366  \\
            1.67  0.19543516802214306  \\
            1.68  0.19066402046643272  \\
            1.69  0.18459322494934144  \\
            1.7  0.17536527989183318  \\
            1.71  0.17296829968645033  \\
            1.72  0.17206685141200295  \\
            1.73  0.1713272846425571  \\
            1.74  0.1707407322050607  \\
            1.75  0.17029848722626761  \\
            1.76  0.16999203917085337  \\
            1.77  0.1698131043836777  \\
            1.78  0.16975365152152344  \\
            1.79  0.16980592226138502  \\
            1.8  0.16996244766988022  \\
            1.81  0.17021606061224048  \\
            1.82  0.17055990457004344  \\
            1.83  0.17098743922496817  \\
            1.84  0.17149244315177925  \\
            1.85  0.17206901394796104  \\
            1.86  0.17271156611028826  \\
            1.87  0.17341482695053645  \\
            1.88  0.1741738308237799  \\
            1.89  0.1749839119236436  \\
            1.9  0.17584069587967432  \\
            1.91  0.1767400903729277  \\
            1.92  0.17767827496712862  \\
            1.93  0.17865169033448436  \\
            1.94  0.17965702703757258  \\
            1.95  0.18069121401180843  \\
            1.96  0.18175140687686753  \\
            1.97  0.18283497619020062  \\
            1.98  0.18393949574145596  \\
            1.99  0.18506273097324932  \\
            2.0  0.18620262760132067  \\
            2.01  0.18735730049566612  \\
            2.02  0.18852502287374695  \\
            2.03  0.18970421584732344  \\
            2.04  0.19089343835580141  \\
            2.05  0.19209137751121563  \\
            2.06  0.19329683937300324  \\
            2.07  0.19450874016458286  \\
            2.08  0.1957260979383133  \\
            2.09  0.19694802469069983  \\
            2.1  0.1981737189256152  \\
            2.11  0.1994024586598293  \\
            2.12  0.20063359486219934  \\
            2.13  0.2018665453154116  \\
            2.14  0.20310078888720584  \\
            2.15  0.20433586019640138  \\
            2.16  0.20557134465783644  \\
            2.17  0.20680687388943286  \\
            2.18  0.20804212146398937  \\
            2.19  0.20927679898793805  \\
            2.2  0.21051065248916112  \\
            2.21  0.21174345909600278  \\
            2.22  0.21297502398980384  \\
            2.23  0.21420517761363045  \\
            2.24  0.2154337731203  \\
            2.25  0.21666068404333383  \\
            2.26  0.21788580217507336  \\
            2.27  0.21910903563684792  \\
            2.28  0.22033030712676083  \\
            2.29  0.22154955233139662  \\
            2.3  0.22276671848846918  \\
            2.31  0.2239817630881702  \\
            2.32  0.225194652701719  \\
            2.33  0.2264053619263393  \\
            2.34  0.22761387243657216  \\
            2.35  0.22882017213257466  \\
            2.36  0.23002425437667107  \\
            2.37  0.23122611731008913  \\
            2.38  0.23242576324242004  \\
            2.39  0.23362319810692433  \\
            2.4  0.23481843097535116  \\
            2.41  0.2360114736264638  \\
            2.42  0.2372023401629515  \\
            2.43  0.2383910466718686  \\
            2.44  0.2395776109241483  \\
            2.45  0.2407620521091829  \\
            2.46  0.24194439060076922  \\
            2.47  0.2431246477511122  \\
            2.48  0.24430284570986593  \\
            2.49  0.24547900726548777  \\
            2.5  0.24665315570645507  \\
            2.51  0.24782531470012348  \\
            2.52  0.24899550818724547  \\
            2.53  0.25016376029035836  \\
            2.54  0.2513300952344375  \\
            2.55  0.2524945372783969  \\
            2.56  0.25365711065613716  \\
            2.57  0.25481783952601145  \\
            2.58  0.255976747927698  \\
            2.59  0.25713385974555214  \\
            2.6  0.25828919867766365  \\
            2.61  0.25944278820988353  \\
            2.62  0.26059465159419926  \\
            2.63  0.2617448118309018  \\
            2.64  0.2628932916540476  \\
            2.65  0.2640401135197865  \\
            2.66  0.26518529959716153  \\
            2.67  0.26632887176106435  \\
            2.68  0.2674708515870301  \\
            2.69  0.2686112603476283  \\
            2.7  0.2697501190102125  \\
            2.71  0.27088744823584143  \\
            2.72  0.27202326837918855  \\
            2.73  0.2731575994892963  \\
            2.74  0.27429046131104196  \\
            2.75  0.27542187328720136  \\
            2.76  0.2765518545610104  \\
            2.77  0.2776804239791417  \\
            2.78  0.2788076000950235  \\
            2.79  0.27993340117243015  \\
            2.8  0.2810578451892969  \\
            2.81  0.2821809498417136  \\
            2.82  0.28330273254804306  \\
            2.83  0.28442321045314944  \\
            2.84  0.2855424004326905  \\
            2.85  0.28666031909745815  \\
            2.86  0.2877769827977432  \\
            2.87  0.28889240762770907  \\
            2.88  0.29000660942974976  \\
            2.89  0.29111960379883817  \\
            2.9  0.2922314060868332  \\
            2.91  0.29334203140675186  \\
            2.92  0.2944514946369914  \\
            2.93  0.2955598104255093  \\
            2.94  0.296666993193925  \\
            2.95  0.29777305714158286  \\
            2.96  0.2988780162495388  \\
            2.97  0.299981884284485  \\
            2.98  0.3010846748026057  \\
            2.99  0.30218640115336903  \\
            3.0  0.3032870764832407  \\
            3.01  0.3043867137393437  \\
            3.02  0.305485325673033  \\
            3.03  0.30658292484341176  \\
            3.04  0.30767952362077544  \\
            3.05  0.3087751341899909  \\
            3.06  0.30986976855380455  \\
            3.07  0.3109634385360902  \\
            3.08  0.3120561557850259  \\
            3.09  0.3131479317762147  \\
            3.1  0.3142387778157336  \\
            3.11  0.3153287050431288  \\
            3.12  0.3164177244343467  \\
            3.13  0.3175058468046057  \\
            3.14  0.3185930828112107  \\
            3.15  0.3196794429563128  \\
            3.16  0.3207649375896078  \\
            3.17  0.32184957691098814  \\
            3.18  0.32293337097313346  \\
            3.19  0.3240163296840537  \\
            3.2  0.3250984628095832  \\
            3.21  0.32617977997581704  \\
            3.22  0.32726029067150764  \\
            3.23  0.32834000425040655  \\
            3.24  0.3294189299335657  \\
            3.25  0.330497076811588  \\
            3.26  0.3315744538468359  \\
            3.27  0.33265106987559356  \\
            3.28  0.3337269336101949  \\
            3.29  0.33480205364109583  \\
            3.3  0.3358764384389188  \\
            3.31  0.33695009635645023  \\
            3.32  0.3380230356306049  \\
            3.33  0.3390952643843397  \\
            3.34  0.34016679062854926  \\
            3.35  0.3412376222639084  \\
            3.36  0.3423077670826875  \\
            3.37  0.34337723277052845  \\
            3.38  0.3444460269081939  \\
            3.39  0.34551415697327403  \\
            3.4  0.3465816303418654  \\
        }
        ;
    \addlegendentry {$P_{22}^{\text{rot}}$} 
\end{axis}
\end{tikzpicture}

%% file: sections/sec6.tex
%

\section{Conclusion}\label{sec:conclusion}
In this work, we presented a novel convexification approach based on hierarchical rank-one sequences for the approximation of the rank-one convex envelope.
First, the required notions and the general variational problem for solid mechanics were introduced.
After this, the algorithm was motivated by first establishing how to convexify along one dimensional lines and then how to convexify for a single laminate.
Furthermore, this procedure was recursively applied to refine the laminates (or the supporting points) of the hierarchical rank-one sequence.
The algorithm was then tested in mathematical benchmark problems, where the convergence with respect to the discretisation parameter $N$ was shown and its limitations were examined.
After that, the performance of the novel approach was shown in the application of a phenomenological scalar-valued continuum damage mechanics model. 
Here, we showed that this algorithm is indeed capable for being used in each integration point at each incremental step in two- and three-dimensional finite element problems at finite strains.
Even if the algorithm in general only delivers an upper bound for the rank-one convex envelope, for many relevant cases, the algorithm suitably approximates the actual rank-one convex envelope and outperforms state-of-the-art algorithms when it comes to computational complexity.
The approximated envelope seems to be sufficient in order to result in mesh-independent results.
Further, we showed the reconstructed deformed microstructures which are described by the laminates occurring at the Gauss points.
In future work, the algorithm needs to be analyzed in the context of anisotropic problems and different dissipative continuum mechanical formulations.
Additionally, the criteria when to split the tree can possibly be enhanced by a more global approach, e.g.~by taking the ellipticity condition into account.

%% file: main.bbl
\newcommand{\etalchar}[1]{$^{#1}$}
\begin{thebibliography}{KWSM18}

\bibitem[ABD{\etalchar{+}}19]{AntBalDesDeeMacKam:2019:mdc}
Eric Anttila, Daniel Balzani, Anastasia Desyatova, Paul Deegan, Jason
  {MacTaggart}, and Alexey Kamenskiy.
\newblock Mechanical damage characterization in human femoropopliteal arteries
  of different ages.
\newblock {\em Acta Biomaterialia}, 90:225--240, 2019.

\bibitem[AFO03]{AubFagOrt:2003:csa}
Sylvie Aubry, Matt Fago, and Michael Ortiz.
\newblock A constrained sequential-lamination algorithm for the simulation of
  sub-grid microstructure in martensitic materials.
\newblock {\em Computer Methods in Applied Mechanics and Engineering},
  192(26):2823--2843, July 2003.

\bibitem[Bar04]{Bar:2004:lca}
S{\"o}ren Bartels.
\newblock Linear convergence in the approximation of rank-one convex envelopes.
\newblock {\em ESAIM: Mathematical Modelling and Numerical Analysis},
  38(5):811--820, September 2004.

\bibitem[Bar05]{Bar:2005:rea}
S{\"o}ren Bartels.
\newblock Reliable and {{Efficient Approximation}} of {{Polyconvex Envelopes}}.
\newblock {\em SIAM Journal on Numerical Analysis}, 43(1):363--385, January
  2005.

\bibitem[Bar15]{Bar:2015:nmn}
S{\"o}ren Bartels.
\newblock {\em Numerical {{Methods}} for {{Nonlinear Partial Differential
  Equations}}}, volume~47 of {\em Springer {{Series}} in {{Computational
  Mathematics}}}.
\newblock {Springer International Publishing}, {Cham}, 2015.

\bibitem[BB{\v S}07]{BerBohSil:2007:rcs}
Albrecht Bertram, Thomas B{\"o}hlke, and Miroslav {\v S}ilhav{\'y}.
\newblock On the {{Rank}} 1 {{Convexity}} of {{Stored Energy Functions}} of
  {{Physically Linear Stress-Strain Relations}}.
\newblock {\em Journal of Elasticity}, 86(3):235--243, February 2007.

\bibitem[BCHH04]{BarCarHacHop:2004:erm}
S.~Bartels, C.~Carstensen, K.~Hackl, and U.~Hoppe.
\newblock Effective relaxation for microstructure simulations: Algorithms and
  applications.
\newblock {\em Computer Methods in Applied Mechanics and Engineering},
  193(48-51):5143--5175, December 2004.

\bibitem[BKK00]{BalKirKri:2000:rqe}
John~M. Ball, Bernd Kirchheim, and Jan Kristensen.
\newblock Regularity of quasiconvex envelopes.
\newblock {\em Calculus of Variations and Partial Differential Equations},
  11(4):333--359, December 2000.

\bibitem[BKN{\etalchar{+}}23]{BalKohNeuPetPet:2023:mrc}
D.~Balzani, M.~K{\"o}hler, T.~Neumeier, M.~A. Peter, and D.~Peterseim.
\newblock Multidimensional rank-one convexification of incremental damage
  models at finite strains.
\newblock {\em Computational Mechanics}, June 2023.

\bibitem[BLJ23]{BhaLarJan:2023:mpm}
Ritukesh Bharali, Fredrik Larsson, and Ralf J{\"a}nicke.
\newblock A micromorphic phase-field model for brittle and quasi-brittle
  fracture.
\newblock {\em Computational Mechanics}, August 2023.

\bibitem[BO12]{BalOrt:2012:riv}
Daniel Balzani and Michael Ortiz.
\newblock Relaxed incremental variational formulation for damage at large
  strains with application to fiber-reinforced materials and materials with
  truss-like microstructures.
\newblock {\em International Journal for Numerical Methods in Engineering},
  92(6):551--570, November 2012.

\bibitem[CCO08]{CarConOrl:2008:man}
Carsten Carstensen, Sergio Conti, and Antonio Orlando.
\newblock Mixed analytical{\textendash}numerical relaxation in finite
  single-slip crystal plasticity.
\newblock {\em Continuum Mechanics and Thermodynamics}, 20(5):275, September
  2008.

\bibitem[CD18]{ConDol:2018:ara}
Sergio Conti and Georg Dolzmann.
\newblock An adaptive relaxation algorithm for multiscale problems and
  application to nematic elastomers.
\newblock {\em Journal of the Mechanics and Physics of Solids}, 113:126--143,
  April 2018.

\bibitem[CHM02]{CarHacMie:2002:ncp}
Carsten Carstensen, Klaus Hackl, and Alexander Mielke.
\newblock Non{\textendash}convex potentials and microstructures in
  finite{\textendash}strain plasticity.
\newblock {\em Proceedings of the Royal Society of London. Series A:
  Mathematical, Physical and Engineering Sciences}, 458(2018):299--317,
  February 2002.

\bibitem[CHO07]{ConHauOrt:2007:cmc}
Sergio Conti, Patrice Hauret, and Michael Ortiz.
\newblock Concurrent {{Multiscale Computing}} of {{Deformation Microstructure}}
  by {{Relaxation}} and {{Local Enrichment}} with {{Application}} to
  {{Single}}-{{Crystal Plasticity}}.
\newblock {\em Multiscale Modeling \& Simulation}, 6(1):135--157, January 2007.

\bibitem[CKBG12]{CoeKouBosGee:2012:mab}
E.~W.~C. Coenen, V.~G. Kouznetsova, E.~Bosco, and M.~G.~D. Geers.
\newblock A multi-scale approach to bridge microscale damage and macroscale
  failure: A nested computational homogenization-localization framework.
\newblock {\em International Journal of Fracture}, 178(1):157--178, November
  2012.

\bibitem[CT05]{ConThe:2005:sem}
Sergio Conti and Florian Theil.
\newblock Single-{{Slip Elastoplastic Microstructures}}.
\newblock {\em Archive for Rational Mechanics and Analysis}, 178(1):125--148,
  October 2005.

\bibitem[Dac08]{Dac:2008:dmca}
Bernard Dacorogna.
\newblock {\em Direct {{Methods}} in the {{Calculus}} of {{Variations}}},
  volume~78 of {\em Applied {{Mathematical Sciences}}}.
\newblock {Springer New York}, {New York, NY}, 2 edition, 2008.

\bibitem[DD02]{DeSDol:2002:mrn}
Antonio DeSimone and Georg Dolzmann.
\newblock Macroscopic {{Response}} of {{Nematic Elastomers}} via {{Relaxation}}
  of a {{Class}} of {{SO}}(3)-{{Invariant Energies}}.
\newblock {\em Archive for Rational Mechanics and Analysis}, 161(3):181--204,
  February 2002.

\bibitem[DH08]{DimHac:2008:mge}
B.~J. Dimitrijevic and K.~Hackl.
\newblock A {{Method}} for {{Gradient Enhancement}} of {{Continuum Damage
  Models}}.
\newblock {\em Technische Mechanik - European Journal of Engineering
  Mechanics}, 28(1):43--52, 2008.

\bibitem[DH11]{DimHac:2011:rfd}
B.~J. Dimitrijevic and K.~Hackl.
\newblock A regularization framework for damage{\textendash}plasticity models
  via gradient enhancement of the free energy.
\newblock {\em International Journal for Numerical Methods in Biomedical
  Engineering}, 27(8):1199--1210, 2011.

\bibitem[Dol99]{Dol:1999:ncr}
Georg Dolzmann.
\newblock Numerical {{Computation}} of {{Rank-One Convex Envelopes}}.
\newblock {\em SIAM Journal on Numerical Analysis}, 36(5):1621--1635, January
  1999.

\bibitem[DW00]{DolWal:2000:ena}
G.~Dolzmann and N.J. Walkington.
\newblock Estimates for numerical approximations of rank one convex envelopes.
\newblock {\em Numerische Mathematik}, 85(4):647--663, June 2000.

\bibitem[FOC98]{FarOliCer:1998:spv}
R.~Faria, J.~Oliver, and M.~Cervera.
\newblock A strain-based plastic viscous-damage model for massive concrete
  structures.
\newblock {\em International Journal of Solids and Structures},
  35(14):1533--1558, May 1998.

\bibitem[For19]{For:2019:maga}
Samuel Forest.
\newblock Micromorphic {{Approach}} to {{Gradient Plasticity}} and {{Damage}}.
\newblock In George~Z. Voyiadjis, editor, {\em Handbook of {{Nonlocal Continuum
  Mechanics}} for {{Materials}} and {{Structures}}}, pages 499--546. {Springer
  International Publishing}, {Cham}, 2019.

\bibitem[GAS07]{GitAskSlu:2007:rve}
I.~M. Gitman, H.~Askes, and L.~J. Sluys.
\newblock Representative volume: {{Existence}} and size determination.
\newblock {\em Engineering Fracture Mechanics}, 74(16):2518--2534, November
  2007.

\bibitem[GM11]{GurMie:2011:edm}
E.~G{\"u}rses and C.~Miehe.
\newblock On evolving deformation microstructures in non-convex partially
  damaged solids.
\newblock {\em Journal of the Mechanics and Physics of Solids},
  59(6):1268--1290, June 2011.

\bibitem[Gra72]{Gra:1972:ead}
R.L. Graham.
\newblock An efficient algorith for determining the convex hull of a finite
  planar set.
\newblock {\em Information Processing Letters}, 1(4):132--133, June 1972.

\bibitem[Hac97]{Hac:1997:gsm}
Klaus Hackl.
\newblock Generalized standard media and variational principles in classical
  and finite strain elastoplasticity.
\newblock {\em Journal of the Mechanics and Physics of Solids}, 45(5):667--688,
  May 1997.

\bibitem[Hol00]{Hol:2000:nsm}
Gerhard~A. Holzapfel.
\newblock {\em Nonlinear Solid Mechanics: A Continuum Approach for
  Engineering}.
\newblock {Wiley}, {Chichester ; New York}, 2000.

\bibitem[JRB22]{JunRieBal:2022:erna}
Philipp Junker, Johannes Riesselmann, and Daniel Balzani.
\newblock Efficient and robust numerical treatment of a gradient-enhanced
  damage model at large deformations.
\newblock {\em International Journal for Numerical Methods in Engineering},
  123(3):774--793, 2022.

\bibitem[JSJH19]{JunSchJanHac:2019:afa}
Philipp Junker, Stefan Schwarz, Dustin Jantos, and Klaus Hackl.
\newblock A fast and robust numerical treatment of a gradient-enhanced model
  for brittle damage.
\newblock {\em International Journal for Multiscale Computational Engineering},
  17(2):151--180, 2019.

\bibitem[KB23]{KohBal:2023:emr}
Maximilian K{\"o}hler and Daniel Balzani.
\newblock Evolving microstructures in relaxed continuum damage mechanics for
  the modeling of strain softening.
\newblock {\em Journal of the Mechanics and Physics of Solids}, page 105199,
  January 2023.

\bibitem[KNM{\etalchar{+}}22]{KohNeuMelPetPetBal:2022:acma}
Maximilian K{\"o}hler, Timo Neumeier, Jan Melchior, Malte~A. Peter, Daniel
  Peterseim, and Daniel Balzani.
\newblock Adaptive convexification of microsphere-based incremental damage for
  stress and strain softening at finite strains.
\newblock {\em Acta Mechanica}, 233(11):4347--4364, November 2022.

\bibitem[KS86a]{KohStr:1986:odr:a}
Robert~V. Kohn and Gilbert Strang.
\newblock Optimal design and relaxation of variational problems, {{I}}.
\newblock {\em Communications on Pure and Applied Mathematics}, 39(1):113--137,
  1986.

\bibitem[KS86b]{KohStr:1986:odr}
Robert~V. Kohn and Gilbert Strang.
\newblock Optimal design and relaxation of variational problems, {{II}}.
\newblock {\em Communications on Pure and Applied Mathematics}, 39(2):139--182,
  1986.

\bibitem[KVK20]{KumVidKoc:2020:ant}
Siddhant Kumar, A.~Vidyasagar, and Dennis~M. Kochmann.
\newblock An assessment of numerical techniques to find energy-minimizing
  microstructures associated with nonconvex potentials.
\newblock {\em International Journal for Numerical Methods in Engineering},
  121(7):1595--1628, 2020.

\bibitem[KWSM18]{KieWafSprMen:2018:gdm}
Bjoern Kiefer, Tobias Waffenschmidt, Leon Sprave, and Andreas Menzel.
\newblock A gradient-enhanced damage model coupled to
  plasticity{\textemdash}multi-surface formulation and algorithmic concepts.
\newblock {\em International Journal of Damage Mechanics}, 27(2):253--295,
  February 2018.

\bibitem[LDR95]{LeRao:1995:qes}
Herv{\'e} Le~Dret and Annie Raoult.
\newblock The quasiconvex envelope of the {{Saint
  Venant}}{\textendash}{{Kirchhoff}} stored energy function.
\newblock {\em Proceedings of the Royal Society of Edinburgh: Section A
  Mathematics}, 125(6):1179--1192, 1995.

\bibitem[LJM18]{LanJunMos:2018:qdm}
K.~Langenfeld, P.~Junker, and J.~Mosler.
\newblock Quasi-brittle damage modeling based on incremental energy relaxation
  combined with a viscous-type regularization.
\newblock {\em Continuum Mechanics and Thermodynamics}, 30(5):1125--1144,
  September 2018.

\bibitem[LKM22]{LanKurMos:2022:hrc}
K.~Langenfeld, P.~Kurzeja, and J.~Mosler.
\newblock How regularization concepts interfere with (quasi-)brittle damage: A
  comparison based on a unified variational framework.
\newblock {\em Continuum Mechanics and Thermodynamics}, 34(6):1517--1544,
  November 2022.

\bibitem[LKM23]{LanKurMos:2023:cdg}
K.~Langenfeld, P.~Kurzeja, and J.~Mosler.
\newblock On the curvature dependence of gradient damage models: {{Control}}
  and opportunities.
\newblock {\em Computer Methods in Applied Mechanics and Engineering},
  410:115987, May 2023.

\bibitem[LMD03]{LamMieDet:2003:ern}
M.~Lambrecht, C.~Miehe, and J.~Dettmar.
\newblock Energy relaxation of non-convex incremental stress potentials in a
  strain-softening elastic{\textendash}plastic bar.
\newblock {\em International Journal of Solids and Structures},
  40(6):1369--1391, March 2003.

\bibitem[LRL18]{LiaRenLi:2018:mmq}
Shixue Liang, Xiaodan Ren, and Jie Li.
\newblock A mesh-size-objective modeling of quasi-brittle material using
  micro-cell informed damage law.
\newblock {\em International Journal of Damage Mechanics}, 27(6):913--936, June
  2018.

\bibitem[Lub72]{Lub:1972:tfn}
J.~Lubliner.
\newblock On the thermodynamic foundations of non-linear solid mechanics.
\newblock {\em International Journal of Non-Linear Mechanics}, 7(3):237--254,
  June 1972.

\bibitem[Lub73]{Lub:1973:sre}
J.~Lubliner.
\newblock On the structure of the rate equations of materials with internal
  variables.
\newblock {\em Acta Mechanica}, 17(1):109--119, March 1973.

\bibitem[MAR16]{MieAldRai:2016:pfm}
Christian Miehe, Fadi Aldakheel, and Arun Raina.
\newblock Phase field modeling of ductile fracture at finite strains: {{A}}
  variational gradient-extended plasticity-damage theory.
\newblock {\em International Journal of Plasticity}, 84:1--32, September 2016.

\bibitem[Mil12]{Mil:2012:1pf}
M.~Militzer.
\newblock 13 - {{Phase}} field modelling of phase transformations in steels.
\newblock In Elena Pereloma and David~V. Edmonds, editors, {\em Phase
  {{Transformations}} in {{Steels}}}, volume~2 of {\em Woodhead {{Publishing
  Series}} in {{Metals}} and {{Surface Engineering}}}, pages 405--432.
  {Woodhead Publishing}, January 2012.

\bibitem[MKPG10]{MasKouPeeGee:2010:chl}
Thierry~J. Massart, Varvara Kouznetsova, Ron H.~J. Peerlings, and Marc G.~D.
  Geers.
\newblock Computational {{Homogenization}} for {{Localization}} and {{Damage}}.
\newblock In {\em Advanced {{Computational Materials Modeling}}}, chapter~4,
  pages 111--164. {John Wiley \& Sons, Ltd}, 2010.

\bibitem[MTL02]{MieTheLev:2002:vfr}
Alexander Mielke, Florian Theil, and Valery~I. Levitas.
\newblock A {{Variational Formulation}} of {{Rate-Independent Phase
  Transformations Using}} an {{Extremum Principle}}.
\newblock {\em Archive for Rational Mechanics and Analysis}, 162(2):137--177,
  April 2002.

\bibitem[Nee88]{Nee:1988:mrd}
A.~Needleman.
\newblock Material rate dependence and mesh sensitivity in localization
  problems.
\newblock {\em Computer Methods in Applied Mechanics and Engineering},
  67(1):69--85, March 1988.

\bibitem[NPPW23]{NeuPetPetWie:2023:cpi}
Timo Neumeier, Malte~A. Peter, Daniel Peterseim, and David Wiedemann.
\newblock Computational polyconvexification of isotropic functions, July 2023.

\bibitem[OR99]{OrtRep:1999:nem}
M.~Ortiz and E.~Repetto.
\newblock Nonconvex energy minimization and dislocation structures in ductile
  single crystals.
\newblock {\em Journal of the Mechanics and Physics of Solids}, 47(2):397--462,
  February 1999.

\bibitem[OR17]{ObeRua:2017:pde}
Adam~M. Oberman and Yuanlong Ruan.
\newblock A {{Partial Differential Equation}} for the {{Rank One Convex
  Envelope}}.
\newblock {\em Archive for Rational Mechanics and Analysis}, 224(3):955--984,
  June 2017.

\bibitem[ORS00]{OrtRepSta:2000:tsd}
M.~Ortiz, E.~A. Repetto, and L.~Stainier.
\newblock A theory of subgrain dislocation structures.
\newblock {\em Journal of the Mechanics and Physics of Solids},
  48(10):2077--2114, October 2000.

\bibitem[OS99]{OrtSta:1999:vfv}
M.~Ortiz and L.~Stainier.
\newblock The variational formulation of viscoplastic constitutive updates.
\newblock {\em Computer Methods in Applied Mechanics and Engineering},
  171(3):419--444, April 1999.

\bibitem[PBMT21]{PlaBarMisTim:2021:med}
Luca Placidi, Emilio Barchiesi, Anil Misra, and Dmitry Timofeev.
\newblock Micromechanics-based elasto-plastic{\textendash}damage energy
  formulation for strain gradient solids with granular microstructure.
\newblock {\em Continuum Mechanics and Thermodynamics}, 33(5):2213--2241,
  September 2021.

\bibitem[RB23]{RieBal:2023:sel}
J.~Riesselmann and D.~Balzani.
\newblock A simple and efficient lagrange multiplier based mixed finite element
  for gradient damage.
\newblock {\em Computers \& Structures}, 281:107030, June 2023.

\bibitem[RRM{\etalchar{+}}24]{RezRieMisBalPla:2024:apf}
Nasrin Rezaei, Johannes Riesselmann, Anil Misra, Daniel Balzani, and Luca
  Placidi.
\newblock A procedure for the experimental identification of the strain
  gradient characteristic length.
\newblock {\em Zeitschrift f\"ur Angewandte Mathematik und Physik {ZAMP}},
  75(80), 2024.

\bibitem[SB16]{SchBal:2016:riv}
Thomas Schmidt and Daniel Balzani.
\newblock Relaxed incremental variational approach for the modeling of
  damage-induced stress hysteresis in arterial walls.
\newblock {\em Journal of the Mechanical Behavior of Biomedical Materials},
  58:149--162, May 2016.

\bibitem[SJH20]{SchJunHac:2020:vrd}
Stephan Schwarz, Philipp Junker, and Klaus Hackl.
\newblock Variational regularization of damage models based on the emulated
  {{RVE}}.
\newblock {\em Continuum Mechanics and Thermodynamics}, May 2020.

\bibitem[TBS16]{TanBalSch:2016:ioi}
Masato Tanaka, Daniel Balzani, and J\"org Schr\"oder.
\newblock Implementation of incremental variational formulations based on the
  numerical calculation of derivatives using hyper dual numbers.
\newblock {\em Computer methods in applied mechanics and engineering},
  301:216--241, 2016.

\bibitem[WLM21]{WilLamMos:2021:pfm}
Hendrik Wilbuer, Henning Lammen, and J{\"o}rn Mosler.
\newblock Phase field modeling with deformation-dependent interface energies.
\newblock {\em PAMM}, 21(1):e202100114, 2021.

\bibitem[Wu17]{Wu:2017:upt}
Jian-Ying Wu.
\newblock A unified phase-field theory for the mechanics of damage and
  quasi-brittle failure.
\newblock {\em Journal of the Mechanics and Physics of Solids}, 103:72--99,
  June 2017.

\bibitem[YZSK22]{YinZhaStoKal:2022:mdm}
Bo~Yin, Dong Zhao, Johannes Storm, and Michael Kaliske.
\newblock A micromorphic damage model based on a gradient extension for robust
  crack deformations.
\newblock {\em Computer Methods in Applied Mechanics and Engineering},
  399:115328, September 2022.

\bibitem[ZHA{\etalchar{+}}22]{ZuoHeAvrYanHac:2022:tfua}
Di~Zuo, Yiqian He, St{\'e}phane Avril, Haitian Yang, and Klaus Hackl.
\newblock A thermodynamic framework for unified continuum models for the
  healing of damaged soft biological tissue.
\newblock {\em Journal of the Mechanics and Physics of Solids}, 158:104662,
  January 2022.

\end{thebibliography}
